\documentclass[11pt]{wlscirep}
\usepackage[utf8]{inputenc}
\usepackage[T1]{fontenc}

\usepackage[page,title]{appendix}
\usepackage{url}
\usepackage{graphicx}
\usepackage{amssymb}
\usepackage{bm}
\usepackage{mathrsfs}
\usepackage{amsmath}
\usepackage{amsfonts}
\usepackage{color}
\usepackage{xspace}
\usepackage{tikz}
\usepackage{slashed}
\usepackage{cancel}
\usepackage{hyperref}
\usepackage[utf8]{inputenc}    
\usepackage{xcolor}
\usepackage{slashed}
\usepackage[colorinlistoftodos,textsize=scriptsize]{todonotes}
\usepackage[square,compress,numbers]{natbib}
\usepackage{graphicx}
\usepackage[capitalize]{cleveref}
\creflabelformat{equation}{#2\textup{#1}#3}
\usepackage{subfigure}
\usepackage[subfigure]{tocloft}

\usepackage[toc]{glossaries}

\makeglossaries

\newglossaryentry{Ai-media}
{
    name={\texttt{Ai-Media}},
    description={An AI/ML-based real-time captioning service. \url{https://www.ai-media.tv/}}
}

\newglossaryentry{Airmeet}
{
    name={\texttt{Airmeet}},
    description={An online meeting platform. \url{https://www.airmeet.com/}}
}

\newglossaryentry{BinderHub}
{
    name={\texttt{BinderHub}},
    description={A cloud service to share reproducible interactive computing environments from code repositories. \url{https://binderhub.readthedocs.io/}}
}

\newglossaryentry{BlackboardCollaborate}
{
    name={\texttt{Blackboard Collaborate}},
    description={Virtual classroom tool. \url{https://www.blackboard.com/teaching-learning/collaboration-web-conferencing/blackboard-collaborate}}
}

\newglossaryentry{Crowdcast}
{
    name={\texttt{Crowdcast}},
    description={Event hosting platform. \url{https://www.crowdcast.io/}}
}

\newglossaryentry{Discord}
{
    name={\texttt{Discord}},
    description={A VoIP, instant messaging, and digital distribution platform. \url{https://discord.com/}}
}

\newglossaryentry{Facebook}
{
    name={\texttt{Facebook}},
    description={A social media and social networking service. \url{https://www.facebook.com/}}
}

\newglossaryentry{Gather.Town}
{
    name={\texttt{gather.\-town}},
    sort={\texttt{Gather.Town}},
    description={An avatar-based meeting platform. \url{https://gather.town/}}
}

\newglossaryentry{GitHub}
{
    name={\texttt{GitHub}},
    description={Version control and software development host. \url{https://github.com/}}
}

\newglossaryentry{GoogleDoc}
{
    name={Google \texttt{Doc}},
    sort={\texttt{Google Doc}},
    description={Collaborative document editing service. \url{https://docs.google.com/}}
}

\newglossaryentry{GoogleMeet}
{
    name={Google \texttt{Meet}},
    sort={\texttt{Google Meet}},
    description={A video communication service. \url{https://meet.google.com/}}
}

\newglossaryentry{Indico}
{
    name={\texttt{Indico}},
    description={An event management platform developed by CERN. \url{https://indico.cern.ch/}}
}

\newglossaryentry{Jamboard}
{
    name={\texttt{Jamboard}},
    description={A virtual whiteboard. \url{https://jamboard.google.com/}}
}

\newglossaryentry{Jitsi}
{
    name={\texttt{Jitsi}},
    description={A collection of free and Open Source multiplatform VoIP, video conferencing, and instant messaging applications. \url{https://jitsi.org/}}
}

\newglossaryentry{Jupyter}
{
    name={\texttt{Jupyter}},
    description={Interactive data science and scientific computing application. \url{https://jupyter.org/}}
}

\newglossaryentry{Mattermost}
{
    name={\texttt{Mattermost}},
    description={An Open Source collaboration platform. \url{https://mattermost.com/}}
}

\newglossaryentry{Mentimeter}
{
    name={\texttt{Mentimeter}},
    description={Interactive presentations with live polls, quizzes, word clouds, and Q\&As. \url{https://www.mentimeter.com}}
}

\newglossaryentry{Mibo}
{
    name={\texttt{Mibo}},
    description={An avatar-based meeting platform. \url{https://getmibo.com/}}
}

\newglossaryentry{MicrosoftTeams}
{
    name={Microsoft \texttt{Teams}},
    sort={\texttt{Microsoft Teams}},
    description={A business communications platform featuring workspace chat and videoconferencing, file storage, and application integration. \url{https://teams.microsoft.com/}}
}

\newglossaryentry{Miro}
{
    name={\texttt{Miro}},
    description={An online collaborative whiteboard platform. \url{https://miro.com/about}}
}

\newglossaryentry{mozillahubs}
{
    name={Mozilla \texttt{Hubs}},
    sort={\texttt{Mozilla Hubs}},
    description={An avatar-based meeting platform. \url{https://hubs.mozilla.com}}
}

\newglossaryentry{Mural}
{
    name={\texttt{Mural}},
    description={A digital workspace for visual collaboration. \url{https://mural.co}}
}

\newglossaryentry{Otter.ai}
{
    name={\texttt{Otter.ai}},
    description={An AI/ML-based real-time speech to text transcription service. \url{https://otter.ai/}}
}

\newglossaryentry{Remo}
{
    name={\texttt{Remo}},
    description={An interactive virtual event platform. \url{https://remo.co}}
}

\newglossaryentry{RemotelyGreen}
{
    name={\texttt{RemotelyGreen}},
    description={AI driven online networking events. \url{https://remotely.green}}
}

\newglossaryentry{Slack}
{
    name={\texttt{Slack}},
    description={A virtual workspace, featuring chat, teams, document sharing, and app integration. \url{https://slack.com/}}
}

\newglossaryentry{Skype}
{
    name={\texttt{Skype}},
    description={A video and voice calling platform. \url{https://www.skype.com/}}
}

\newglossaryentry{Slido}
{
    name={\texttt{Slido}},
    description={An interactive app for hybrid meetings. \url{https://www.sli.do/}}
}

\newglossaryentry{Twitter}
{
    name={\texttt{Twitter}},
    description={A micro-blogging and social networking service. \url{https://www.twitter.com/}}
}

\newglossaryentry{Vidyo}
{
    name={\texttt{Vidyo}},
    description={A real-time video communication platform. \url{https://www.vidyo.com/}}
}

\newglossaryentry{Vimeo}
{
    name={\texttt{Vimeo}},
    description={A video sharing platform. \url{https://vimeo.com/}}
}

\newglossaryentry{Webcast}
{
    name={\texttt{Webcast}},
    description={A CERN-hosted webcast server. \url{https://webcast.web.cern.ch/}}
}

\newglossaryentry{Whova}
{
    name={\texttt{Whova}},
    description={Event management software. \url{https://whova.com/}}
}

\newglossaryentry{WonderMe}
{
    name={\texttt{Wonder}},
    description={A virtual meeting space. \url{https://www.wonder.me/}}
}

\newglossaryentry{YouTube}
{
    name={\texttt{YouTube}},
    description={A video sharing platform. \url{https://www.youtube.com/}}
}

\newglossaryentry{Zenodo}
{
    name={\texttt{Zenodo}},
    description={A general-purpose open-access repository for deposition of research papers, data sets, research software, reports, and any other research related digital artifacts. \url{https://zenodo.org/}}
}

\newglossaryentry{Zoom}
{
    name={\texttt{Zoom}},
    description={A video conferencing platform. \url{https://zoom.us/}}
}

\begin{document}

\newcommand\note[2]{{\color{blue}[#1: #2]}}
\newcommand\inprogress[0]{{\color{red}[In Progress]}}
\newcommand\ready[0]{{\color{green}[Ready for Review]}}
\newcommand\done[0]{{\color{blue}[Done]}}

\title{{\large \textit{\hfill IRIS-HEP Blueprint Workshop Summary}} \\
Learning from the Pandemic: the Future of Meetings in HEP and Beyond}

\author[1]{Mark S. Neubauer}
\author[2]{Todd Adams}
\author[3]{Jennifer Adelman-McCarthy}
\author[4]{Gabriele Benelli}
\author[5]{Tulika Bose}
\author[6]{David Britton}
\author[7]{Pat Burchat}
\author[3]{Joel Butler}
\author[5]{Timothy A. Cartwright}
\author[8]{Tomáš Davídek}
\author[9]{Jacques Dumarchez}
\author[10]{Peter Elmer}
\author[1]{Matthew Feickert}
\author[1]{Ben Galewsky}
\author[7]{Mandeep Gill}
\author[11]{Maciej Gladki}
\author[1]{Aman Goel}
\author[13]{Jonathan E. Guyer}
\author[3]{Bo Jayatilaka}
\author[3]{Brendan Kiburg}
\author[14]{Benjamin Krikler}
\author[10]{David Lange}
\author[3]{Claire Lee}
\author[15]{Nick Manganelli}
\author[16]{Giovanni Marchiori}
\author[4]{Meenakshi Narain}
\author[10]{Ianna Osborne}
\author[10]{Jim Pivarski}
\author[2]{Harrison Prosper}
\author[11]{Graeme A Stewart}
\author[17]{Eduardo Rodrigues}
\author[18]{Roberto Salerno}
\author[19]{Marguerite Tonjes}
\author[20]{Jaroslav Trnka}
\author[21]{Vera Varanda}
\author[10]{Vassil Vassilev}
\author[22]{Gordon T. Watts}
\author[3]{Sam Zeller}
\author[3]{Yuanyuan Zhang}

\affil[1]{University of Illinois at Urbana-Champaign, Champaign, IL, USA}
\affil[2]{Florida State University, Tallahassee, FL, USA}
\affil[3]{Fermi National Accelerator Laboratory, Batavia, IL, USA}
\affil[4]{Brown University, Providence, RI, USA}
\affil[5]{University of Wisconsin--Madison, Madison, WI, USA}
\affil[6]{University of Glasgow, Glasgow, UK}
\affil[7]{Stanford University, Stanford, CA, USA}
\affil[8]{Charles University, Prague, Czech Republic}
\affil[9]{LPNHE - Sorbonne Université - Paris, France}
\affil[10]{Princeton University, Princeton, NJ, USA}
\affil[11]{European Organization for Nuclear Research (CERN), Meyrin, Switzerland}
\affil[12]{University of Delhi, New Delhi, India}
\affil[13]{National Institute of Standards and Technology, Gaithersburg, MD, USA}
\affil[14]{University of Bristol, Bristol, UK}
\affil[15]{University of California at Riverside, Riverside, CA, USA}
\affil[16]{APC Paris, France}
\affil[17]{Oliver Lodge Laboratory, University of Liverpool, Liverpool, UK}
\affil[18]{LLR - Ecole Polytechnique, Palaiseau, France}
\affil[19]{University of Illinois at Chicago, Chicago, IL, USA}
\affil[20]{University of California at Davis, Davis, CA, USA}
\affil[21]{ARISF / LPNHE, Paris, France}
\affil[22]{University of Washington, Seattle, WA, USA}


\maketitle

\begin{abstract}
The COVID-19 pandemic has by-and-large prevented in-person meetings since March 2020. While the increasing deployment of effective vaccines around the world is a very positive development, the timeline and pathway to “normality” is uncertain and the “new normal” we will settle into is anyone’s guess. Particle physics, like many other scientific fields, has more than a year of experience in holding virtual meetings, workshops, and conferences. A great deal of experimentation and innovation to explore how to execute these meetings effectively has occurred. Therefore, it is an appropriate time to take stock of what we as a community learned from running virtual meetings and discuss possible strategies for the future. Continuing to develop effective strategies for meetings with a virtual component is likely to be important for reducing the carbon footprint of our research activities, while also enabling greater diversity and inclusion for participation.

This report summarizes a virtual two-day workshop on \textit{Virtual Meetings} held May~5--6, 2021 which brought together experts from both inside and outside of high-energy physics to share their experiences and practices with organizing and executing virtual workshops, and to develop possible strategies for future meetings as we begin to emerge from the COVID-19 pandemic. This report outlines some of the practices and tools that have worked well which we hope will serve as a valuable resource for future virtual meeting organizers in all scientific fields.

\end{abstract}


\pagebreak
\setcounter{tocdepth}{2}
\tableofcontents
\addtocontents{toc}{~\hfill\textbf{Page}\par}

\pagebreak


\section{Introduction}
\label{sec:introduction}
The COVID-19 pandemic caused by the SARS-CoV-2 virus has had a devastating effect on human health and well-being, and on the global economy. First identified in Wuhan, China in December 2019, SARS-CoV-2 spread rapidly among human population leading to the World Heath Organization (WHO) declaring COVID-19 a global health emergency on January 31, 2020. The continued spread around the world led to global travel restrictions in February 2020 and the WHO officially declaring COVID-19 as a pandemic on March 11, 2020. 

As the alarming scale of the global health crisis became apparent in March 2020, workshop organizers began to cancel or postpone planned in-person events. Many organizers scrambled to re-purpose their workshops for a virtual format where attendees participated remotely through videoconferencing technologies such as \gls{Zoom}. In some cases this re-purposing was done on short order rather than cancelling the event. The \textit{Connecting-the-Dots} Workshop held in April 2020 is one such an example and is described in Section~\ref{sec:experiences_CTD}.

This report summarizes a two-day workshop on the topic of \textit{Virtual Meetings} held over the period of May~5 to 6, 2021. The workshop is part of the "Blueprint" process of the NSF-funded \textit{Institute for Research and Innovation in Software for High-Energy Physics} (IRIS-HEP)~\cite{irishep}.

The report is organized similarly to flow of the workshop. In Section~\ref{sec:overview}, the IRIS-HEP Blueprint process and this workshop is introduced followed by an overview of pre-COVID-19 meeting organization and the ways that COVID-19 rapidly changed the \emph{status quo}. Section~\ref{sec:experiences} provides a summary of the talks given during the workshop by organizers of virtual workshops since April 2020. At the end of Section~\ref{sec:experiences}, common themes and key findings from these workshop experiences are presented. In Section~\ref{sec:tools_techniques} we outline some of the practices and tools that have worked well (or have not but seem like they should have) for virtual workshops. 
In Section~\ref{sec:DIA} we present important considerations for diversity, inclusion and accessibility for virtual workshops. Best practices for virtual workshops at each of the stages of preparation and execution are presented in Section~\ref{sec:best_practices}. The final section (\S\ref{sec:looking_forward}) synthesizes what we have learned about virtual workshops over the last year and presents some ideas for the organization of future meetings in HEP and beyond, such as the "hybrid with hubs" approach.


\section{Overview}
\label{sec:overview}


\subsection{IRIS-HEP and the Blueprint Process}
\label{sec:overview_irishep_blueprint}

The goal of the IRIS-HEP is to address key computational and data science challenges of the High-Luminosity Large Hadron Collider (HL-LHC) experiments and other HEP experiments in the 2020s. IRIS-HEP resulted from a 2-year community-wide effort involving 18 workshops and 8 position papers, most notably a Community White Paper~\citep{Albrecht2019} and a Strategic Plan~\citep{Elmer:2017rej}. The institute is an active center for software R\&D, functions as an intellectual hub for the larger community-wide software R\&D efforts, and aims to transform the operational services required to ensure the success of the HL-LHC scientific program. 

The IRIS-HEP Blueprint activity is designed to inform development and evolution of the IRIS-HEP strategic vision and build (or strengthen) partnerships among communities driven by innovation in software and computing. The blueprint process includes a series of workshops that bring together IRIS-HEP team members, key stakeholders, and domain experts from disciplines of importance to the Institute's mission. This Blueprint meeting on the topic of \emph{Virtual Meetings} is one of a series of workshops that have also included
\begin{itemize}
    \item Analysis Systems R\&D on Scalable Platforms (2019) \vspace{-8pt}
    \item Fast Machine Learning and Inference (2019 \& 2020) \vspace{-8pt}
    \item A Coordinated Ecosystem for HL-LHC Computing R\&D (2019) \vspace{-8pt}
    \item Software Training (2020) \vspace{-8pt}
    \item Sustainable Software in HEP (2020) \vspace{-8pt}
    \item Future Analysis Systems \& Facilities (2020) \vspace{-8pt}
    \item Portable Inference (2020)
\end{itemize}
The Blueprint workshop discussions are captured and inform key outcomes which are summarized in a short report made publicly available, such as this report. 


\subsection{Virtual Meetings Workshop}
\label{sec:overview_meeting}

While the increasing deployment of effective vaccines around the world is a very positive development, the timeline and pathway to “normality” is uncertain and the “new normal” we will settle into is anyone’s guess. Particle physics, like many other scientific fields, has more than a year of experience in holding virtual meetings, workshops, and conferences. A great deal of experimentation and innovation to explore how to execute these meetings effectively has occurred. The IRIS-HEP team has substantial experience with (co-)organizing and sponsoring workshops, conferences, and meetings over the years through the process of establishing the institute and its role as an intellectual hub for software R\&D. 

For these reasons, it was viewed as an opportune time to take stock of what we as a community have learned from running virtual meetings and discuss effective strategies for the future through this virtual meetings workshop. Continuing to develop effective strategies for meetings with a virtual component is likely to be important for reducing the carbon footprint of our research activities, while also enabling greater diversity and inclusion for participation.

The workshop on \textit{Virtual Meetings} was held May~5 to 6, 2021. The aim for the workshop was to bring together experts from both inside and outside of particle physics to share their experiences and practices with organizing and executing virtual workshops, and to develop possible strategies for future meetings as we begin to emerge from the challenging conditions of the COVID-19 pandemic. 

Attendees participated remotely using a variety of videoconferencing and collaborative tools. Eighty-nine people registered for the workshop, though as a virtual workshop, attendees came and went at various times. We estimate that there were about 70 to 80 participants in the workshop at any one time.

The timeline of the workshop presentations and activities are summarized in Table~\ref{tab:agenda}.
\begin{table}[ht]
\centering
\begin{tabular}{rll}
\toprule
\multicolumn{3}{l}{\textbf{Day 1} (via \gls{Zoom})} \\
\addlinespace[0.3ex]
08:30 & Welcome, Blueprint Activity and Workshop Overview & M. Neubauer \\ 
08:50 & A Virtual Hitchhiker's Guide to Virtual Conferencing & B. Krikler \\ 
09:20 & HSF/WLCG Virtual Workshop Experience & G. Stewart \\
09:40 & LHCP 2020 & G. Marchiori, R. Salerno \\ 
10:00 & ICHEP 2020 & T. Davidek \\ 
10:20 & Moriond 2021 & V. Varanda \\ 
10:40 & Connecting the Dots 2020 & D. Lange \\ 
11:00 & LLVM Developers Meeting & V. Vassilev \\ 
11:35 & Snowmass Community Planning Meeting & B. Jayatilaka  \\ 
11:55 & OSG Virtual Meetings & T. Cartwright \\ 
12:15 & PyHEP 2020 Experience and 2021 Plans & E. Rodrigues \\ 
12:35 & SciPy 2020 Experience and 2021 Plans & J. Guyer\\  
12:55 & US ATLAS / Canada ATLAS Computing Bootcamp & M. Feickert \\ 
13:15 & Neutrino 2020 & S. Zeller \\
13:35 & Discussion Session & \\
\addlinespace[1.0ex]
\multicolumn{3}{l}{\textbf{Day 2} (via \gls{Gather.Town})} \\
\addlinespace[0.3ex]
08:30 & Day 2 Topics and Goals & M. Neubauer \\
08:50 & Summary of Community Input from DPF Townhall & M. Narain \\
09:20 & Tools and Techniques for Virtual Workshops Session & B. Galewsky \\
11:00 & Taking Stock of Experiences and Discussion on Future Events & \\
12:00 & Summary Report Discussion and Writing & \\
\bottomrule
\end{tabular}
\caption{Workshop Agenda (times are CDT) Presentations are available from the workshop website\textsuperscript{*}. \\
  \small\textsuperscript{*}\url{https://indico.cern.ch/event/1026363/timetable}
  \label{tab:agenda}}
\end{table}

The first day of the workshop was held using the \gls{Zoom} videoconferencing and chat platform. It was primarily focused on experiences with virtual workshops via talks by its organizational representatives. The workshops were chosen to span a wide range of scales and purpose, from training events with tens of attendees such as the US ATLAS / Canada ATLAS Computing Bootcamp (Section~\ref{sec:experiences_Bootcamp}) to international conferences with thousands of attendees such as the International Conference on High Energy Physics (Section~\ref{sec:experiences_ICHEP}) and Neutrino 2020 (Section~\ref{sec:experiences_Neutrino}). 
At the end of the first day, a discussion session was used to culled out common themes and key finds from the experiences talks. 

\begin{figure}[t!]
    \centering
    \includegraphics[width=\linewidth]{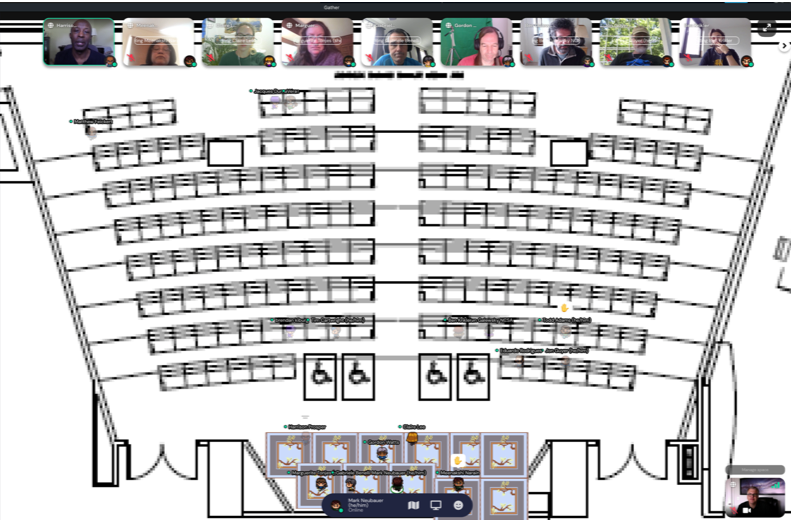}
    \caption{Day 2 of the workshop was held in a virtual Wilson Hall 1 West space within the \gls{Gather.Town} application.}
    \label{fig:1West}
\end{figure}

\begin{figure}[t!]
    \centering
    \includegraphics[width=\linewidth]{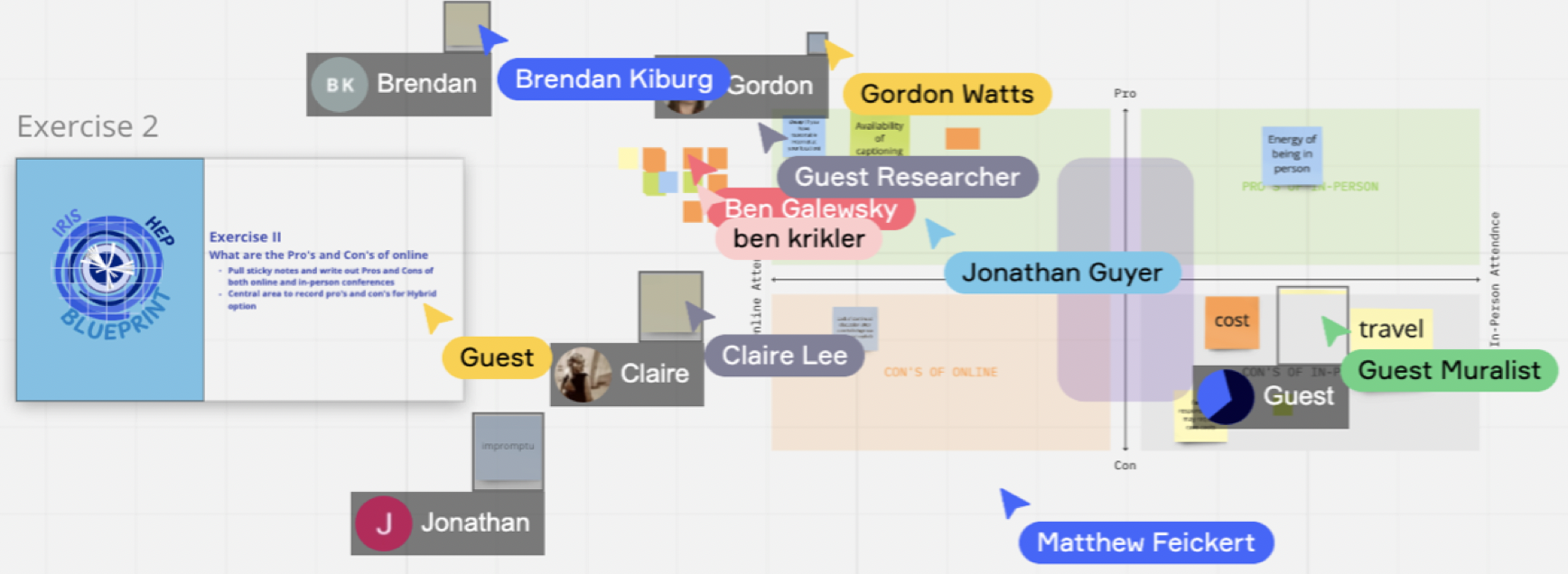}
    \caption{Example use of the \gls{Miro} collaborative whiteboard to work through a set of questions during Day 2 of the workshop. A set of virtual sticky-notes were annotated and placed in the appropriate square.}
    \label{fig:Miro}
\end{figure}

The second day was held using the \gls{Gather.Town} platform. \gls{Gather.Town} is videoconferencing software like \gls{Zoom}, but with the added component of seeing the virtual “room” that you are occupying with other participants represented as avatars, with the ability to move around and interact with others. To demonstrate the \gls{Gather.Town} functionality for the workshop, the Fermilab Wilson Hall floorplan was imported and various areas were defined including a poster area and a private space where participants could arrange for one-on-one or group discussions. The presentations and discussion sessions were held virtually in the Wilson Hall 1 West (WH1W) conference room, as shown in Figure~\ref{fig:1West}.
The Day 2 focus was on community input from an American Physics Society Division of Particles and Fields Townhall event, demonstration of tools \& techniques for virtual meetings and forward-looking discussion of possible meeting strategies as we begin to emerge from the challenging conditions of the COVID-19 pandemic. This discussion was facilitated by the use of a collaborative whiteboard tool called a "\gls{Miro} board" where participants answered questions in the form of five "exercises":
\begin{enumerate}
    \item \textbf{\textit{Conference Essence}}: What are elements that are core to the conference? What are important elements that also happen at a conference? \vspace{-8pt}
    \item \textbf{\textit{Online/In-Person Format Comparison}}: What are the pros and cons of an online meeting format? Of an in-person format? Of a hybrid format? \vspace{-8pt}
    \item \textbf{\textit{Attendance Composition}}: Who is most likely to attend in-person conferences? \vspace{-8pt}
    \item \textbf{\textit{Attractors/Repellers}}: What might attract someone to move from one mode of conference to another? What might dissuade them?
 \vspace{-8pt}
    \item \textbf{\textit{Tools and Techniques}}: Which are important to achieving the goals of a given conference? Which might attract people to our attend our conference? \vspace{-8pt}
\end{enumerate}
The input provided via this collaborative whiteboard along with the notes and \gls{Zoom} recording of the discussions are represented in this summary in the appropriate sections. An example of the \gls{Miro} board in action from the workshop is shown Figure~\ref{fig:Miro}.


\subsection{Conferences and Workshops Before COVID-19}
\label{sec:overview_preCOVID}

Most readers of this document will have experience with attending in-person conferences, workshops, and/or meetings as part of their research or other professional pursuits. Therefore, we do not attempt to discuss all elements of in-person meetings in this report. Rather, we use this section to touch on some of the key aspects from the virtual meetings workshop as a baseline for the discussion that follows.

The status quo in HEP before COVID-19 was, especially for the larger workshop and conferences, was to travel to the venue to either deliver a talk on one's research of research on behalf of a collaboration or attend in-person to hear talks and participate in discussions. Of course, videoconferencing has been around for a long time and remote participation has been component to events, but these were more likely than not due to specific constraints by individuals rather than prominent element generally embraced by the organizers.

Travel to and from in-person events most often involves 
a substantial commitment of time and funds (most likely from research grants for which principle investigators typically budget as part of their research program). Depending on distance from the researcher's home location, its common to have one or two days on each side of the event consumed with traveling through a network of airports, train stations, etc. This level of time commitment away from home can be challenging for those with children, take care of others, or have other personal constraints. The transportation itself also involves a carbon load on the environment which can be substantial in the aggregate when one considers the scale of global research travel and the distances involved to participate in-person at international conferences and collaboration events.

Outside of the presentations themselves, in-person meetings provide opportunities for chance (or planned) encounters with other participants (e.g. during coffee breaks for conference dinners) to discuss and develop new research ideas or directions, foster new collaborations and engage in deep technical discussions about talks that were presented. These encounters and informal interactions can be career-changing for young scientists, both in terms of networking and access to leading scientists in their respective fields. These types of interactions are difficult to replicate in a remote format given the current state of technology for virtual presence.

Participants in the virtual meetings workshop noted other benefits of the in-person format, including an opportunity to focus on a specific topic for a period of time that is often difficult given responsibilities and distractions in their home/work environment, meet collaborators in-person (sometimes for the first time!), and obtain an expanded viewpoint on science, society and cultures that comes with traveling to somewhere new.



\subsection{COVID-19 Effects on Meetings}
\label{sec:overview_COVID_effects}

As mentioned in Section~\ref{sec:introduction}, the move from largely in-person meetings to almost exclusively virtual events due to the COVID-19 pandemic was abrupt and unprecedented. Of course, it was not only the organizers and participants of conferences and workshops that had to pivot -- schools and universities had to move quickly to remote or "hybrid" learning environments and businesses and people alike had to adapt to the challenges of COVID-related lock-downs and restrictions. Such a rapid transition to a virtual presence in personal and professional lives at scale was only possible due to the general availability of personal computers (including laptops, tablets and smart phones) coupled to high-speed networks and software tools like \gls{Zoom}.

One clear effect that COVID-19 has had on our community is to expose inefficiencies in the \emph{status quo} model for conferences, workshops and meetings and bring to the surface concerns around equity, access and environmental aspects of this model. A constructive rethinking of this model informed by experiences since April 2020 and emerging technologies is the first step to addressing some of the pre-existing deficiencies toward a more inclusive, environmentally-conscious and effective approach to future meetings. 



\section{Experiences with Virtual Meetings}
\label{sec:experiences}

This section summarizes the experiences drawn from a dozen virtual scientific workshops with a variety of sizes and purposes since April 2020. Not surprisingly, given that this report is a summary of a Blueprint workshop for a HEP software institute, the virtual events presented here are either software-focused, HEP-focused or both.

\subsection{HSF/WLCG Virtual Workshops}
\label{sec:experiences_HSF}

Since the HSF Community White Paper~\cite{Albrecht2019} the HEP Software Foundation (HSF) and the Worldwide LHC Computing Grid (WLCG) have held joint workshops to advance R\&D in software and computing for HEP.

A workshop had been planned for May 2020, in Lund, Sweden, but with the degenerating COVID-19 situation in March it became clear that an in-person workshop was impossible to hold and we started to plan a virtual event~\cite{HSFWLCG2020} that could replace it. This was one of the first reasonably large meetings in the community that moved to a virtual format, but the lessons learned here and at the follow up workshop in November 2020~\cite{HSFWLCGNov2020} helped refine and inform the community on how to hold successful online events \cite{HSF2020-05-Feedback,HSF2020-11-Feedback}. This section is a summary of the most important lessons learned from these workshops. 

\subsubsection{Focused Topic and Scheduling}

It was immediately obvious to the organisers that simply moving the planned five full-day face-to-face meeting to a video conference setup would not work. Instead, the May workshop~\cite{HSFWLCGNov2020} was shortened and refocused to tackle a specific topic: \textit{New Architectures, Portability, and Sustainability}. 

Narrowing the scope meant also that scheduling the event would be easier. Given the relative centres of gravity in high-energy physics of Europe and the Americas, the `golden hour' in which to schedule meetings is later afternoon in Europe and morning in the Americas. e.g., 16:00 CERN time is a comfortable 9:00 at Fermilab and 7:00 in the Pacific timezone, which is early, but possible. In May we went for fairly short sessions of two hours only, 16:00 to 18:00 CERN. Finshing at 18:00 avoided being too disruptive for European participants. However, it does mean that only a limited amount of time is available -- again emphasising the need to focus on key topics during each session.

Unfortunately these times make it really difficult for Asia-Pacific colleagues to participate. In larger events, a mixture of sessions that start in the European mornings (thus afternoon in Asia) work well, e.g., starting at 9:00 CERN time, as was chosen for the recent conference~\cite{vCHEP2021}. Unfortunately, it is very difficult to have an event during working hours for North America and Asia: 17:00 in New York is still 5:00 in Tokyo as the Pacific Ocean is really large.

Bear in mind that concentrating on online events is fatiguing and many participants will be experiencing multiple distractions (e.g. emails arriving, phone calls, children needing attention, etc.). It is imperative to schedule regular breaks to allow people to have a mental rest, stretch out, make a cup of tea, etc. Practical experience suggests that one break every two hours is a minimum. If one insists on running an activity during the break, light-hearted polls can keep people amused.

\subsubsection{Material Upload and Videos}

\begin{figure}[t!]
    \centering
    \includegraphics[width=0.75\linewidth]{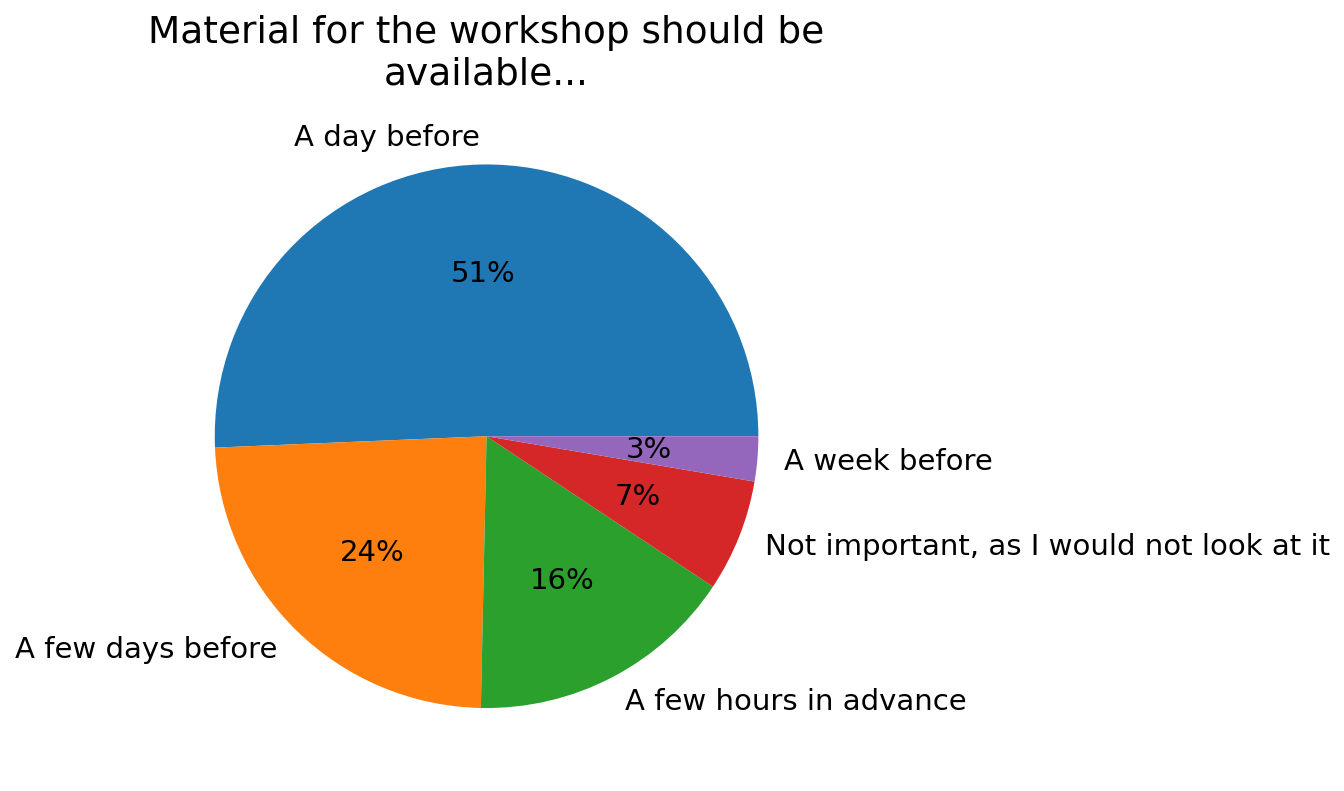}
    \caption{Participant survey responses from the May HSF-WLCG workshop on their preference for when presentations should be uploaded.}
    \label{fig:hsf_talk_upload_time}
\end{figure}

Given the difficulties of scheduling a convenient time for presentations that can suit all participants, one compensating factor is to make sure that material is uploaded in advance. For the first workshop we attempted to get material a week in advance. However, this was hard for speakers to manage and, in fact, feedback from participants was that one day in advance is sufficient, Figure~\ref{fig:hsf_talk_upload_time}.

\begin{figure}[t!]
    \centering
    \includegraphics[width=0.45\linewidth]{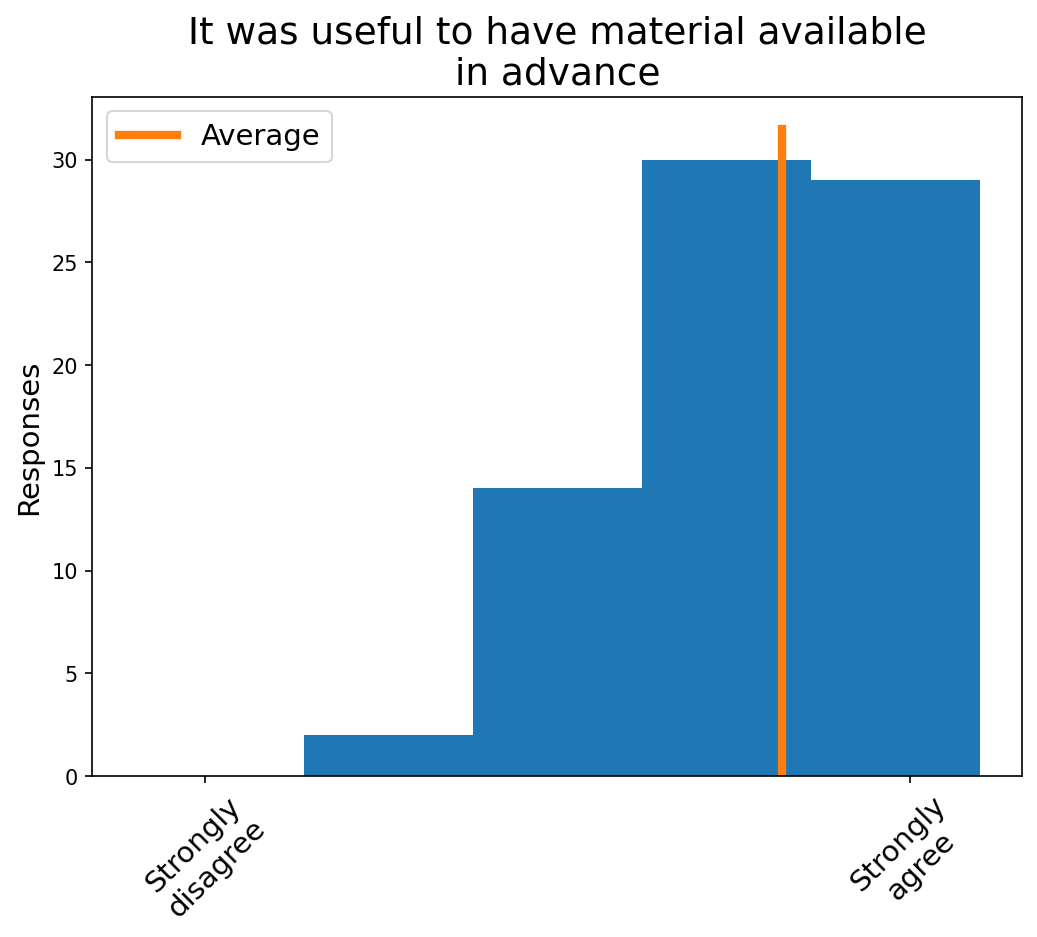}    \includegraphics[width=0.45\linewidth]{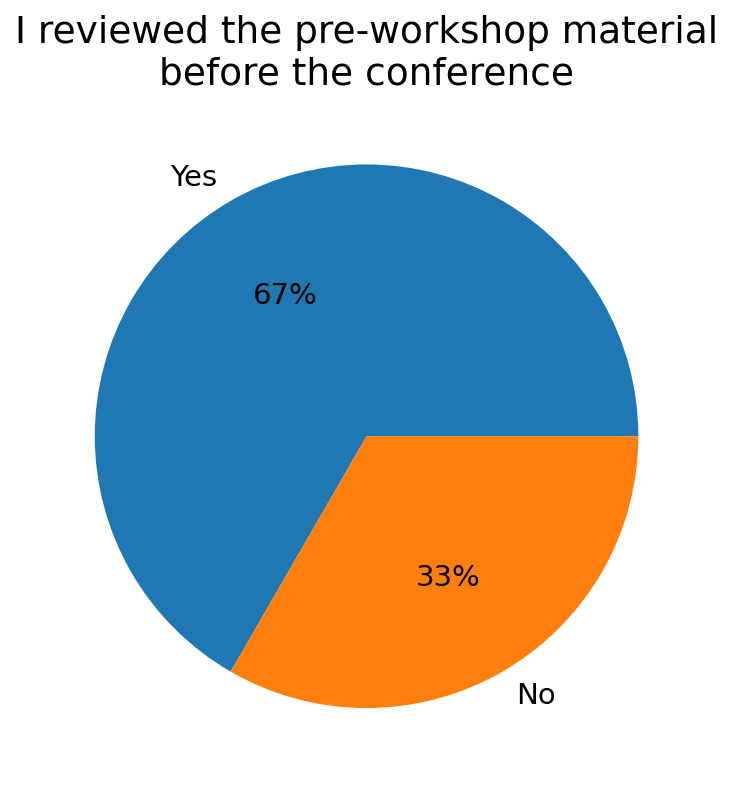}
    \caption{Survey response on if participants thought that material should be available in advance and if they viewed it.}
    \label{fig:hsf_pre_review}
\end{figure}

The usefulness of this advanced availability of the material was borne out by the fact that 2/3 of workshop attendees reported viewing the material before the sessions, as shown in Figure~\ref{fig:hsf_pre_review}.

If the session is really at an inconvenient time for some people, the best solution is to record the video of the talk and make it available \emph{post-facto}. If \gls{Zoom} is being used for the presentations, then its built-in recording facilities are ideal for this purpose. Depending on the organisers familiarity with video editing tools, session videos can be posted unedited, or split by talk or even given fancy introductions and epilogues. Splitting by talk is relatively easily managed using open source tools like \texttt{ffmpeg}.

Be aware, especially at CERN and in Europe, that data privacy laws apply. It is strongly advised, in fact even mandatory, that you gather specific consent from all participants to not only record the meeting, but also to make it available publicly on platforms such as the CERN Document Server or \gls{YouTube}. 

\subsubsection{Hosting the Event}

At a minimum, a video conferencing service will be needed to host the event. Almost every university and laboratory now has a solution for this. More often than not this is the \gls{Zoom} platform, which at the moment is a \emph{de facto} standard in HEP. This is for good reason: the clients are stable, the infrastructure scales well and features such as recording work well. It also has the advantage of being extremely familiar to most participants, thus technical hurdles are minimised. Back in May 2020, quite a number of participants had problems screen sharing, "hand" raising, muting/un-muting, but by now most people have worked these out. Still, it is a wise idea, especially for larger events to:
\begin{itemize}
    \item Run some practice sessions in advance for speakers (and session chairs) to test their setup \vspace{-8pt}
    \item Schedule one or two minutes between speakers to change sharing \vspace{-8pt}
    \item Have a copy of the slides downloaded locally by the chairs, in case the speaker has a slide sharing issue
\end{itemize}

During presentations, the usual recommendation is that the speaker shares slides and that they have their camera enabled to give a slightly more intimate feel. Everyone else should be muted and have their camera off, unless they are speaking. Having a participants' guide for the event makes it clear to people how they should contribute to the meeting discussion. E.g., in larger meetings using \gls{Zoom}'s \emph{raise hand} feature is a very good discipline to adopt.

In addition to these preparatory steps for speakers the hosts of the event need to be well prepared. At the very least there should be a chair and a co-chair for each session. The co-chair can monitor the meeting chat (\S \ref{sec:chat_tools}) while the chair is managing the speaker and contributors.

The hosts of the meeting should have an out-of-band communication mechanism that they can use to discuss issues as the meeting is ongoing -- private chat channels on instant messaging tools like \gls{Mattermost}, \gls{Slack}, or \gls{Skype} work well for this. The co-chair can warn if a speaker looks badly set to go over time, for example, or perhaps the organisers feel there is an important question that should be prioritised.

Good timekeeping in online meetings is very important. Like it or not, participants will multiplex with other meetings and events and knowing that a presentation will happen at the scheduled time is vital for people to join for the sessions they desire. Family or other personal commitments are a reality for many people and it is not acceptable to exclude people by throwing the schedule out the window when the meeting starts. Make sure that when organising the meeting there is sufficient discussion time scheduled and that it is stressed to the speakers how much time they have for their presentation. Do not be afraid to remind them during their talk if it looks like they will overrun, or even insist they finish up if they exceed their time budget.

Finally, although "\gls{Zoom}-bombing" (unwanted remote attendance by uninvited individuals) is quite rare in our community, it can happen. Make sure that \gls{Zoom} rooms are password protected and only distribute the passwords to people registered for the event (CERN's \gls{Indico} allows \gls{Zoom} links to be nicely protected, or email them to the list of registered participants). Knowing what to do if a \gls{Zoom}-bombing occurs is really important to minimise disruption if there is an incident -- \gls{Zoom} has good tools for managing this now and even a `big red button' that will allow the hosts to suspend all participant activities in the case of a concerted attack. Know how you will proceed with the meeting in case something happens: e.g., keep the waiting room active, the (co-)chair shares slides, keep chat and rename disabled, don't allow participants to unmute except by invitation, etc.

All of this preparation will help the event to run smoothly.

\subsubsection{Chat Tools}
\label{sec:chat_tools}

Having a channel by which discussion can happen on presentations is a very useful thing for online meetings. It allows participants to ask questions in a non-disruptive way during the talk (and some people even just feel more comfortable asking a question in text chat, so it also improves inclusivity). It also allows for discussions to continue after the time slot for the presentation has finished; it can also allow people to ask questions in advance (as we noted, slides should go up well before the meeting starts).

The chat channel of the video system itself is rather disfavoured for this. Usually it is ephemeral and lacks many useful features, so save that only for technical issues during the meeting (it can be the co-chair's job to redirect questions if they do get asked there).

A number of options are used in the community, each with pros and cons:

\begin{itemize}
    \item Publicly writable \glspl{GoogleDoc} work well at a small to medium scale -- make sure that a skeleton layout is in place in advance so that comments and questions go to the correct place. A few of the drawbacks here are that contributors need some discipline to identify themselves in the comment they make (anonymous tapirs are cute, but not helpful) and that comments are easily misplaced as people can write anywhere in the document at any time.
    \item \gls{Slack} and \gls{Mattermost} are chat tools that impose some restrictions, in the sense that everything is a serial stream. That discipline can be useful to prevent chaos and more easily manage a discussion. However, it can also be confusing if multiple discussions are happening at the same time interleaved. This is mitigated if people use `Reply' functionality that threads a discussion; success depends on how familiar people are with these kind of tools. If multiple sessions are running, use a different channel for each one to cut down on cross-talk.
    \item \gls{Discord} is a popular platform (albeit less well known in HEP) that allows for multiple chats, breakouts and even video. It is very popular in the gaming community and seems to work particularly well for training and tutorial events. This platform was used successfully in the US ATLAS / Canada ATLAS Bootcamp described in Section~\ref{sec:experiences_Bootcamp}.
\end{itemize}

\begin{figure}[ht]
    \centering
    \includegraphics[width=0.5\linewidth]{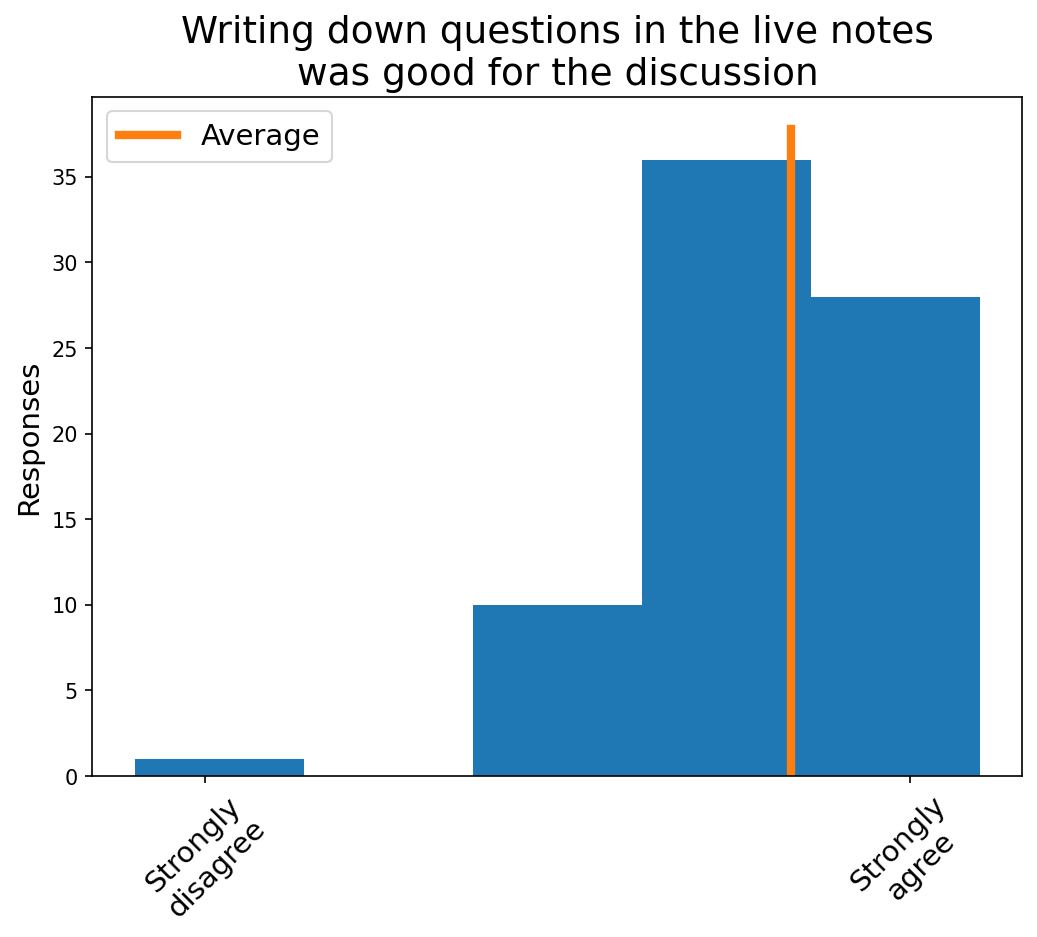}
    \caption{Participant survey responses towards asking questions in the \gls{GoogleDoc} live notes.}
    \label{fig:hsf_live_notes}
\end{figure}

\begin{figure}[ht]
    \centering
    \includegraphics[width=0.5\linewidth]{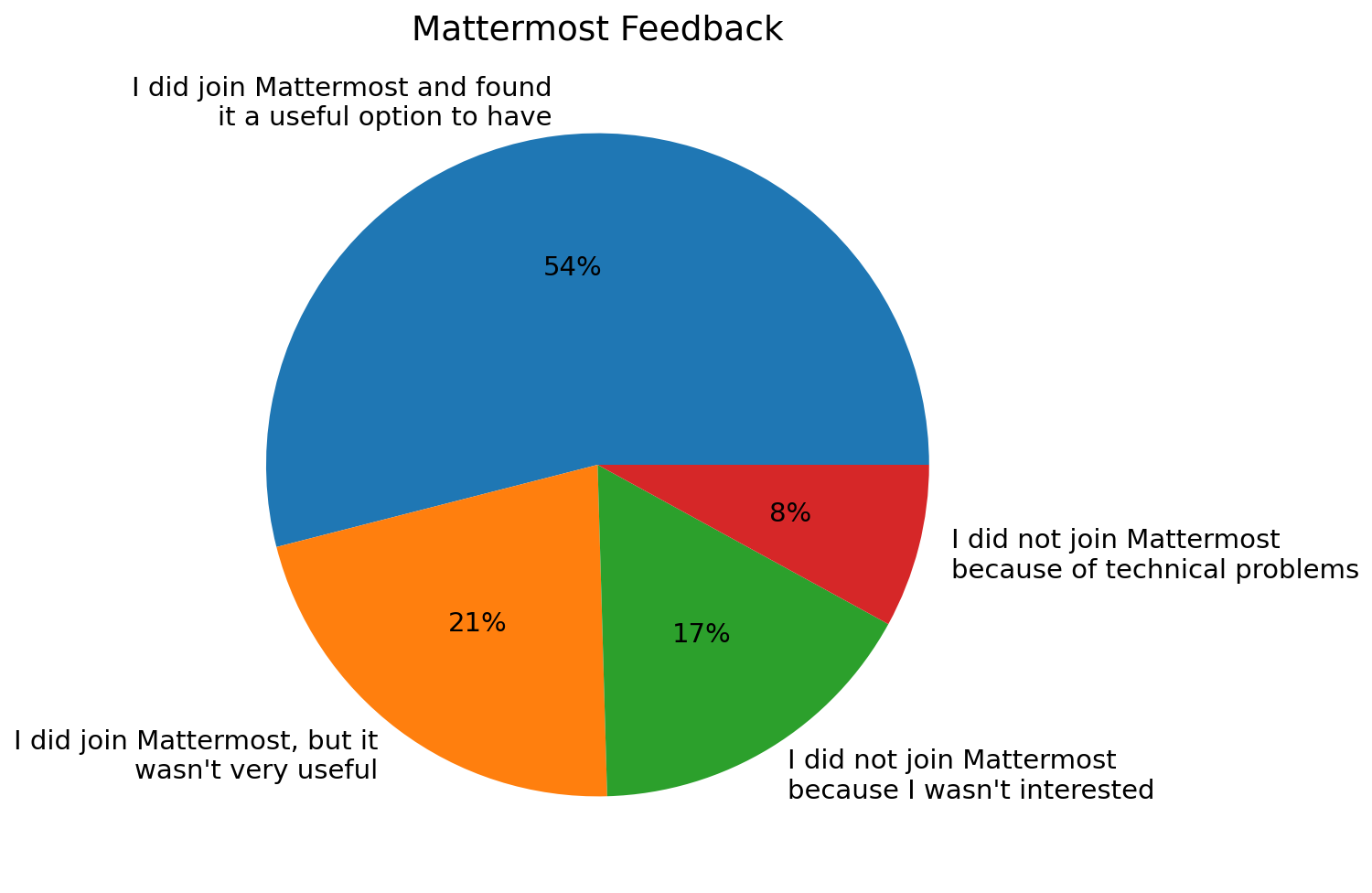}
    \caption{Participant survey responses at vCHEP about the \gls{Mattermost} chat tool.}
    \label{fig:hsf_vchep_mattermost}
\end{figure}

For the HSF-WLCG workshops, the \gls{GoogleDoc} solution was adopted and people thought that it did help the discussion, Figure \ref{fig:hsf_live_notes}. Likewise, \gls{Mattermost}, as used at the vCHEP2021 conference was view positively by 54~\% of participants, Figure \ref{fig:hsf_vchep_mattermost} (this was despite a definite tension between security and convenience that caused some participants technical problems).

\subsubsection{Virtual Social Platforms}

One aspect of workshops and conferences that is much harder to replicate online is the social and informal human contact that we have during coffee breaks, lunches and social events. The opportunity to have less structured discussions, to branch out to other topics or just to enjoy the company of colleagues is not yet easily replicated through any virtual platform. There are no lack of platforms here (\gls{Gather.Town}, \gls{WonderMe}, \gls{Mibo}, to name a few), but it is clear that barriers to normal human interaction do remain. Perhaps in the future this will change, with technical advances (e.g., virtual reality) or with us becoming more accustomed to such platforms. For now, though not without their merits, do not expect them to be a substitute for face-to-face interactions.

\subsubsection{Conclusions}

The HSF and WLCG experience of running virtual events demonstrates that for the presentation of scientific content a virtual event functions rather well. The same is true of direct discussions, which can be supported in a variety of ways through live notes or text chat as well as live chat. Running a successful virtual event (just like an in-person one) requires careful planning and a lot of work from the organisers, from scheduling the event at the best times, running the event properly and then posting videos of the event after the fact. Virtual social platforms can supplement the presentations and discussion, but still fall far short of the in-person experience.

One meta-consideration when organising events is that as the barrier to entry for organising a virtual event is rather low, there was a tendency to have more events during 2020 than we would have had if it had been a normal year. \texttt{Zoom}-fatigue arose for more reasons that just what was happening in the HEP universe, but it affected many of our colleagues in a real way during difficult times. So we should endeavour to make any virtual events focused, meaningful and useful and, as ever, be sympathetic and understanding to the pressures that people have found themselves under during these extraordinary times.

\subsection{LHCP 2020}
\label{sec:experiences_LHCP}

\subsubsection{Origin of the LHCP 2020 online conference}
\label{LHCP2020:sec:intro}

The LHCP conference series started in 2013 after a successful fusion of two international conferences, "Physics at Large Hadron Collider Conference" and "Hadron Collider Physics Symposium". 
It consists of a  series of yearly conferences where the latest experimental and theoretical results on the Large Hadron Collider (LHC) physics are presented. 
They include research areas such as the Standard Model Physics and Beyond, the Higgs boson, Supersymmetry, Heavy Quark Physics, and Heavy Ion Physics as well as recent progress in the high luminosity upgrades of the LHC and future colliders developments.
The LHCP conference typically attracts between 300 and 400 participants every year from all over the world, discussing together in parallel, plenary and poster sessions spanning a full week.

The 8$^{th}$ episode of the series, LHCP 2020, was due to take place in Paris, France, in the International Conference Centre of Sorbonne Universit\'e, the week of May~25 to 30, 2020.
At the beginning of March 2020 due to the outbreak of COVID-19 it became clear that an in-person conference would not be feasible, and it was decided by the conference Steering Committee (SC), in consultation with the conference International Advisory Committee (IAC), to postpone the Paris conference by one full year, to become LHCP 2021.
However, the LHC experimental collaborations strongly suggested that an online "LHCP-like" event would be still held, in a similar period of the year and with a similar format to the planned LHCP 2020 conference, so that physicists could still join together -- though virtually -- to discuss the new results and keep the community together. 
The LHCP 2020 organisers with the assistance of the CERN IT department for technical details embarked in the project of setting up such online event with only two months to go before the start of the conference.

The LHCP 2020 online conference took place the week of May~25 to 30, 2020 as originally planned for the in-person conference.
It was the first online HEP conference with a large (>1000) audience and as such the organisers had to take decisions based on limited previous experience. The success of the LHCP 2020 online conference and the lessons learned from it were useful for the setup of later HEP conferences.

\subsubsection{LHCP 2020 online: the preparation}

The organisation of LHCP 2020 online revolved around two main points: the \textit{programme} and the \textit{technical} organisation.

The programme had already been setup for the in-person conference, and the main issue here concerned the modification of the schedule in order to be accessible to a community that would be scattered over a large number of continents and timezones rather than being all sitting together in the same site. 
Two alternative scenarios were considered:
\begin{itemize}
    \item keep the structure proposed for the in-person conference, with at most four parallel sessions taking place at the same time, and similar duration of the pauses (30${}^\prime$ breaks), but with fewer blocks every day (typically one parallel and one plenary session with a pause in between), for about four hours/day but spanning two full weeks (Monday to Friday);
    \item increment the maximum number of concurrent parallel sessions, shorten the breaks, and keep three sessions/day, for a total of about six hours/day from Monday-Friday plus a shorter concluding 4-hour session on Tuesday.
\end{itemize}
It was not considered to shorten the programme due to the large amount of material that the experimental collaborations and the theory community had already prepared to release and discuss at the conference.

\begin{figure}[t!]
\centering
\includegraphics[width=\textwidth,trim=0 250 0 90,clip]{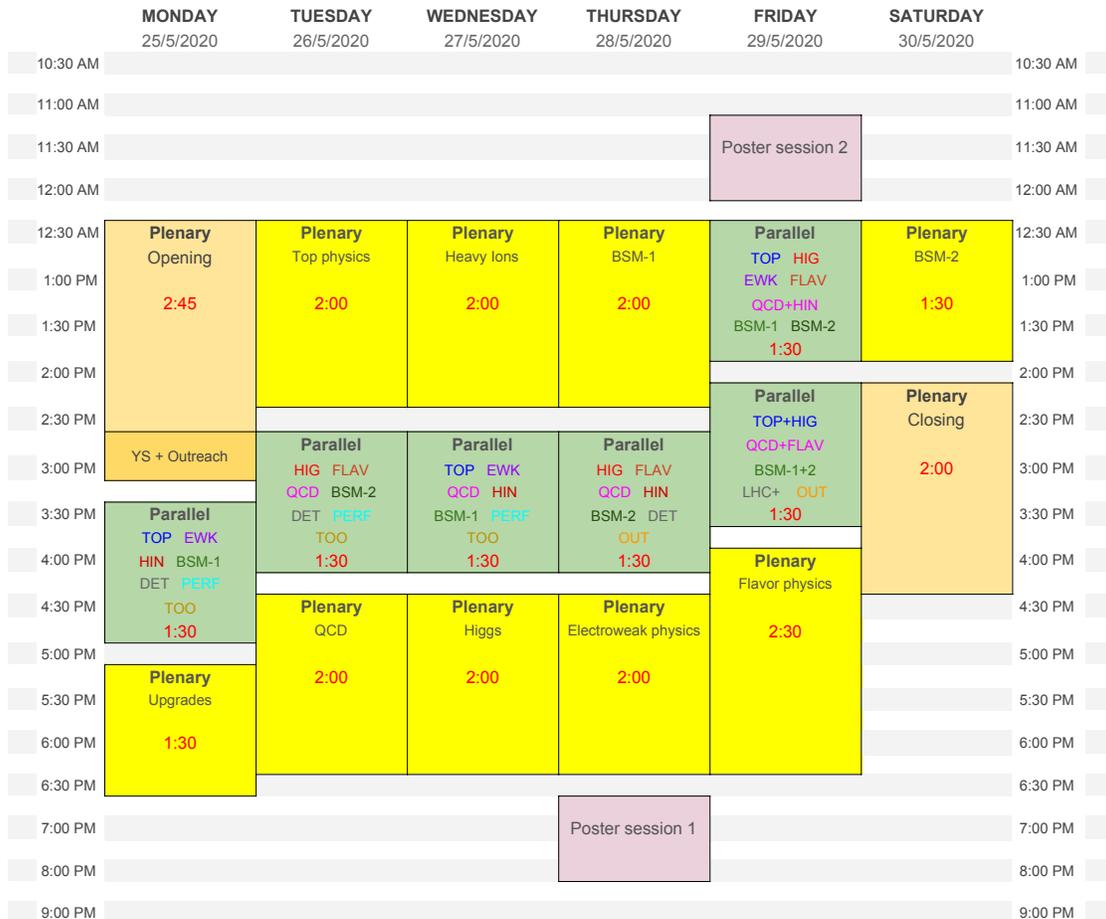}
\caption{\label{fig:ii} The final LHCP 2020 timeline (CEST time zone) and block structure.}
\end{figure}

The second scenario (see Figure~\ref{fig:ii}) was chosen, with sessions between 12:30 and 18:30 CEST (18:30 to 00:30 CST, 5:30 to 11:30 PDT). This would make a little fraction of the session less accessible to our colleagues in the countries on timezones far away from CEST, but would have the advantage of keeping the event within the same week as initially planned and advertised, so that participants would have less difficulties to find the time to interrupt their teaching and research activities and join the virtual event. 

Concerning the technical and practical organisation, the main decisions taken were the following:
\begin{itemize}
    \item to keep the poster sessions, which would be held in virtual meeting rooms and to keep the awards for the best poster; 
    \item to publish the conference proceedings as initially planned, giving to the speakers and poster presenters the possibility to have a short citeable write-up on their presentation after the conference;
    \item  to waive completely the fees for all participants making the conference accessible to the widest possible option and avoiding the need to handle processing small amounts of money from a large number of participants;
    \item to extend the deadlines for registration as well as poster abstract submission until a few days before the start of the conference to maximize the potential audience.
\end{itemize}

\subsubsection{Technical setup}

When the online LHCP 2020 conference was announced, participants started enthusiastically to register, at a pace of order of 50 new participants per day, increasing with the approach of the starting day of the conference. It soon became clear that the number of participants would reach several hundreds and that it would require an online videoconferencing tool that would be able to handle well such a large number of simultaneous connections and that would be intuitive to use for the participants, the speakers, and the session conveners. At the same time, when the world locked down and the meetings of the large LHC collaborations moved online, the videoconferencing solution adopted until then by CERN, \gls{Vidyo}, started to show saturation issues with several persons unable to join meetings. The recommendation from CERN IT was therefore to move to \gls{Zoom}, which proved to be extremely scalable and intuitive to use. CERN established, in collaboration with \gls{Zoom}, a pilot program that provided professional licences of the tool to CERN users during an evaluation period. Some \gls{Zoom} features such as recording of the meetings on the cloud were disabled for privacy concerns.

In order to control access to the online meetings and restrict it to authorised people, to avoid "gatecrashing", and even to avoid that participants by accident would unmute their microphone or start broadcasting their video during the presentations and disrupt the normal flow of the conference (not an unlikely event with more than 1000 participants), we decided to choose the "Webinar"-style format for the plenary and parallel sessions, rather than the regular \gls{Zoom} meeting style. The webinar format would give more control to the conveners ("Hosts" in \gls{Zoom}'s terminology) on who could (or could not) turn on their microphone, after having virtually raised their hand; it would provide a practice session for speakers ("Panelists") where they could join the meeting before its actual start, to test their connection and the sharing of their screen, and to avoid the problem of not being able to connect to the meeting in case they by accident tried to connect only after 500 participants (the limit of the \gls{Zoom} CERN pilot) had already connected. On the negative side of this choice, the participants could not see who the other persons attending the same event (apart from the speakers and conveners) were, reducing the sense of community that is an important, integral part of in-person conferences.

A few days before the beginning of the conference, when the deadline for the registration was reached, all registrations were reviewed and those from participants not from the HEP community nor with an institutional e-mail address (for instance, an address provided by a University to a student) were rejected (only a handful of registrations would not be accepted). The confirmed registered participants then received by e-mail the connection details and the passwords for the \gls{Zoom} webinars.

The webinars for the plenary sessions were created by the conference organisers. On the other hand, due to the impossibility for a single \gls{Zoom} user to create several meetings running in parallel in the same time slot, the webinars for the parallel sessions were created by the conveners of the session, who were instructed on how to setup and password-protect the meeting by the conference organisers. To allow for off-line access to the conference presentations and Q\&A, the option to record the meetings (on the local PC of the host) was enabled. The consent of each speaker to have their talk recorded was collected through the \gls{Indico} webpage of the conference beforehand (in case of a speaker not accepting to be recorded, recording would have been paused). The participants connecting to the \gls{Zoom} webinar would have to accept by clicking on the "Accept" button of a \gls{Zoom} pop-up window or otherwise disconnect.

For the "online" poster presentations, it was decided to have two sessions of one hour each, in different times of the day to allow poster presenters from each part of the world to find at least one suitable time slot for them to join it. Poster sessions took place in \gls{Zoom} regular meeting rooms, one for each poster presenter. The speakers were asked to create their own \gls{Zoom} meeting rooms and upload the link to their meeting room on the conference website, while passwords were sent separately to the participants by e-mail. For presenters without a \gls{Zoom} license, the CERN IT department created a lightweight CERN account for the duration of the conference, giving them access to the CERN \gls{Zoom} pilot program. Poster presenters were also asked to pre-record a 3 min short presentation of their poster and upload it, together with the poster itself and the connection details, to the conference agenda before the poster sessions.

Finally, since the number of registered participants in the end (1300) exceeded the maximum capacity of the CERN \gls{Zoom} license (1000 participants connected at the same time), for the plenary sessions it was put in place by the CERN IT department a streaming of the \gls{Zoom} webinars to the CERN \gls{Webcast}. In order to allow participants connected to the plenary sessions through \gls{Webcast} to ask their question, a \texttt{plenary-qa} channel was created on a dedicated \gls{Discord} online server created for the LHCP 2020 conference.

\subsubsection{Accessibility and inclusion}
An effort to make the conference as widely accessible as possible was done by the organisers by compressing the schedule into an agenda that would allow participants from all over the world to connect to most of the sessions
at a reasonable time of the day. 

Arrangements had also been made by the local organisation, before the conference was moved online, to support the stay of young participants from developing countries with the creation of a dedicated budget for waiving fees and booking of low-cost accommodation solution in Paris, as well as - for instance - to provide dedicated quiet rooms for mothers needing to breastfeed.

When the conference was announced in its online version, four weeks before the starting date of the conference, a request was brought forward by one participant to provided full real-time captioning for the whole conference (plenary and parallel sessions). The organisation looked around for potential solutions that would suit the user requesting them as well as fit in the very limited budget that the online conference had, from the CERN and EPJC sponsorships (that were also to cover for other expenses such as the poster awards, the publication of the conference proceedings, and communication activities). As the option to caption only the Q\&A sessions was deemed not acceptable by the participant, it was proposed to caption in real time only a few, selected sessions chosen by the participant. For the other sessions, we decided to propose off-line captioning of all the talks, as obtained by running the recordings through an automatic AI-powered tool, \gls{Otter.ai}. However, the agreement on this option came only a few working days before the starting of the contract; meanwhile, the company that had already been chosen for the captioning was not available any longer, and an alternative one had to be found quickly (including finding extra budget to cover for the price increase compared to the other one). This introduced some unfortunate delays, that we regret, in the setting up of the contract and we were only able to arrange live captioning for the sessions attended by the participant during the second half of the conference. 

For the future conferences we recommend, based on this experience, that users with special requests like this one get in contact with the organisers well in advance with respect to the start of the conference, possibly at least two months before, and that they share with the conference organisers any knowledge they might have of captioning companies providing sufficiently accurate transcripts.
For LHCP2020, the transcripts from the company that was hired can be found on the conference website; the quality of the transcript is not close to 100~\%, due to the jargon/communication ubiquitous in our field. Another possibility would be that the speakers record their presentations and prepare a transcript of the talk in advance, before the start of the conference, and upload it to the conference website when their talk is scheduled. In this case only the question time would require to be captioned.

On a longer term, the CERN IT department is launching a project on an automated speech recognition (ASR) system for high-energy physics in collaboration with Universita Politecnica de Valencia, which has a very advanced system for automated transcription with a very efficient training system. The goal is to train an ASR on HEP talks with captions validated by HEP physicists, so that such an ASR could be used in the future to caption conferences as well as other online meetings (at CERN during most of the working days there are hundreds of meetings running in parallel on \gls{Vidyo} or \gls{Zoom}, which would translate into tens of millions of euros per year if all meetings were captioned). Some of the LHCP2020 participants kindly volunteered to edit the transcripts (automatic or from the 3rd party company) of some LHCP2020 presentations, and thus provide a high-quality training dataset for this tool, consisting of 40 presentations with transcripts validated by high-energy physicists. Hopefully this will become an accurate, and at the same time cost effective, solution, that could have large application (transcribed talks can also be searched for and are more accessible not just to hearing-impaired physicists but also to those with difficulties in understanding the speaker).

\subsubsection{Running of the conference}

Each day the conference would alternate plenary and parallel sessions, with short breaks between each session. 
Each session of the conference was chaired by two or more conveners.

The \gls{Zoom} webinars for each session were started by the "host" (one of the session conveners or the conference organisers) typically 30 minutes before the starting time of the session, in practice mode, to allow the speakers to join the webinar and test their audio and video setup as well as the slide sharing. The employee of the captioning company providing real-time subtitles would also join at this time and test the captioning functionality.

Typically 10 minutes before the start of the session, the webinar were broadcast to all the participants, who would then be able to connect. At this point, streaming of the plenary sessions to \gls{Webcast} would also be started.

People joining the webinar would have their microphone muted and their video turned off by default. Video was reserved for speakers and session chairpersons.

The session would then begin with the chairpersons starting the recording of the session, turning on their video, saying a few introductory words, with a brief welcome statement on the scope of the session, and a reminder to the speakers to mute themselves unless they are speaking, and to the participants to use the "raise hands" feature of \gls{Zoom} to ask questions at the end of each presentation (or, alternatively, to post their questions in  the \gls{Discord} chat if connected via \gls{Webcast}).

The chairpersons would then introduce the first speaker, who would turn on their video and microphone, share their screen, and start their presentation, while the chairpersons would mute their own microphone. During the presentation, from time to time, the chairpersons would unmute temporarily to remind the speaker of the time left. At the end of the presentation, the chairpersons would unmute and handle the question time, unmuting the participants who had "raised their hand" in \gls{Zoom} and would thus be able to ask their own question, or directly forwarding to the speaker the question posted in the \gls{Discord} chat. Then, the chairpersons would repeat the previous steps for each of the following presentations, until the end of the session, when the host would close the meeting.

Overall this setup worked very well, with no particular hiccups, with the exception of one afternoon in which, due to a broader outage of various IT services at CERN, it was not possible to start any \gls{Zoom} webinar. The parallel session meetings of the afternoon were thus reverted to \gls{Vidyo} meeting rooms that had been created on purpose as backup solutions, while the CERN IT department worked on a fix of the outage, that arrived in time for the following plenary session. The sessions went on as planned, the only problem being that some of the recordings of the parallel sessions of that day were not saved due to a glitch in the \gls{Vidyo} server and are thus lost forever.

\subsubsection{(Virtual) coffee breaks}

In an in-person conference, coffee (and other meal) breaks serve two important purposes (in addition of course to provide the participants the time for a much needed pause or to get their favourite beverage): they allow the participants to socialise, as well as to discuss with the speakers of the session that just finished and asked them more questions on their presentations. 
To recreate these possibilities in the online conference experience, two solutions were adopted:
\begin{itemize}
    \item various virtual rooms, on the same \gls{Discord} online server used to allow plenary session attendees on \gls{Webcast} to ask their questions, were created, one for each parallel topic as well as a more general "coffee-and-tea" room, where people could, if they wish, interact with each other, on physics or more general topics. The same channel were also provided to allow the participants connected to the same parallel session, if they wished, to say "hello" to their fellow participants, and thus enforce the feeling of community that might be partially lost with the \gls{Zoom} webinars, in which the regular participants cannot see who the other participants are.
    \item It was suggested to the speakers, on a voluntary basis, to be available, after the end of their session, in a \gls{Zoom} meeting room ("breakout room") where they could meet directly the participants that were interested in their presentations and wanted to ask more questions about them.
\end{itemize}
While the \gls{Discord} channels were barely used, the breakout room after the sessions had a reasonable success. Not all the speakers provided them nor all the participants took advantage of them, but those who did were largely satisfied.

\subsubsection{Communication}
A small, but dedicated communication team created content that highlighted events of the day which they posted on the \gls{Twitter} channel of the conference, \texttt{@LHCPConference}, re-posting on this social medium slides - with a short associated comment - on the hot topics being shown in the plenary sessions.

\subsubsection{Statistical information about the participants and the organising committees}
\begin{figure}[t!]
\centering
\includegraphics[width=\textwidth,trim=0 100 0 100,clip]{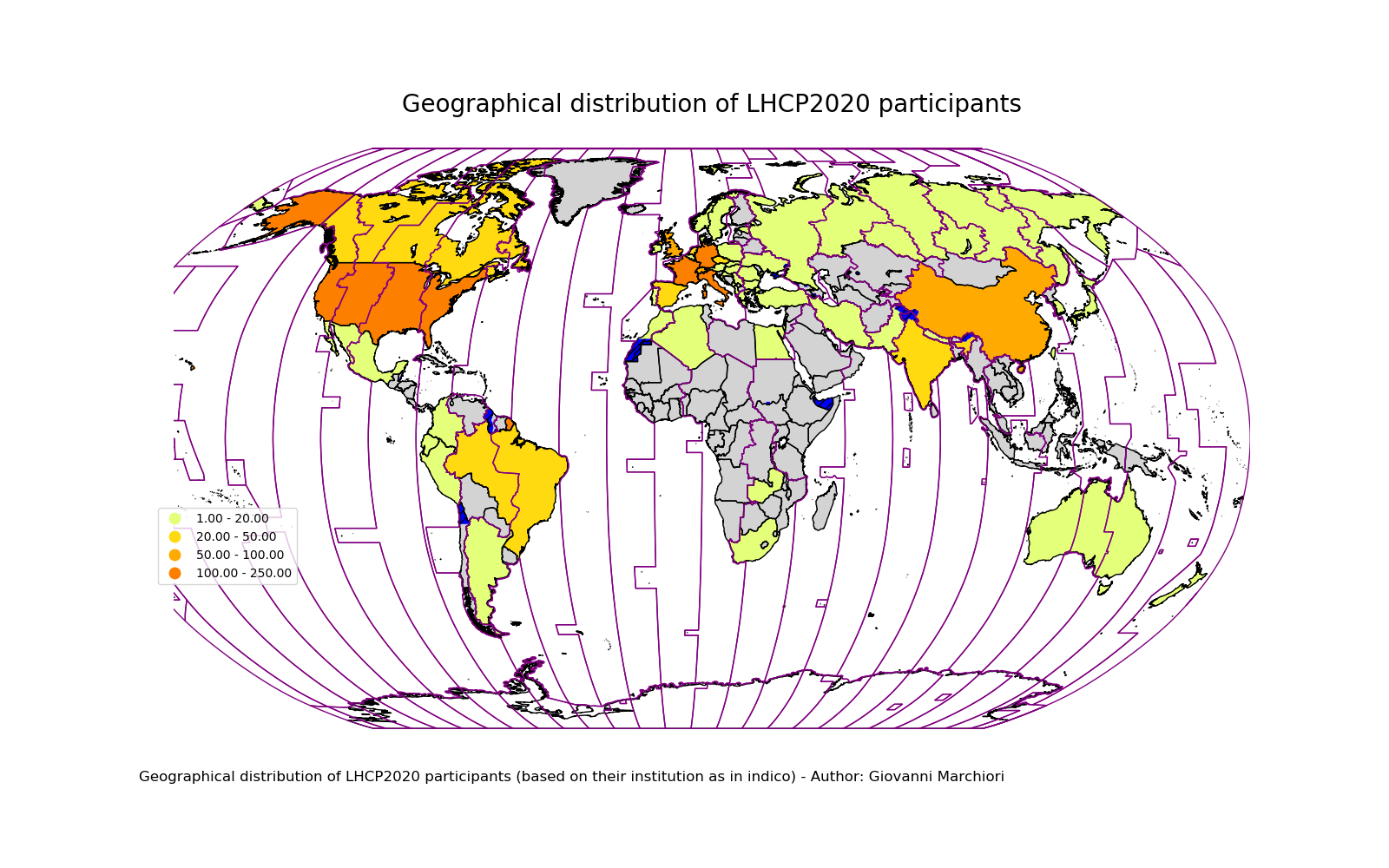}
\caption{\label{fig:iii} The geographical distribution of the participants of the LHCP2020 conference, based on the institute they work for (or university they study at), with timezones overlaid.}
\end{figure}

The final number of registered participants to the conference was 1301, coming from institutions based in 56 countries and spanning 17 time zones. The geographical distribution of the participants is shown figure~\ref{fig:iii}.

\begin{figure}[t!]
\centering 
\includegraphics[width=0.7\textwidth,trim=0 50 0 250,clip,page=1]{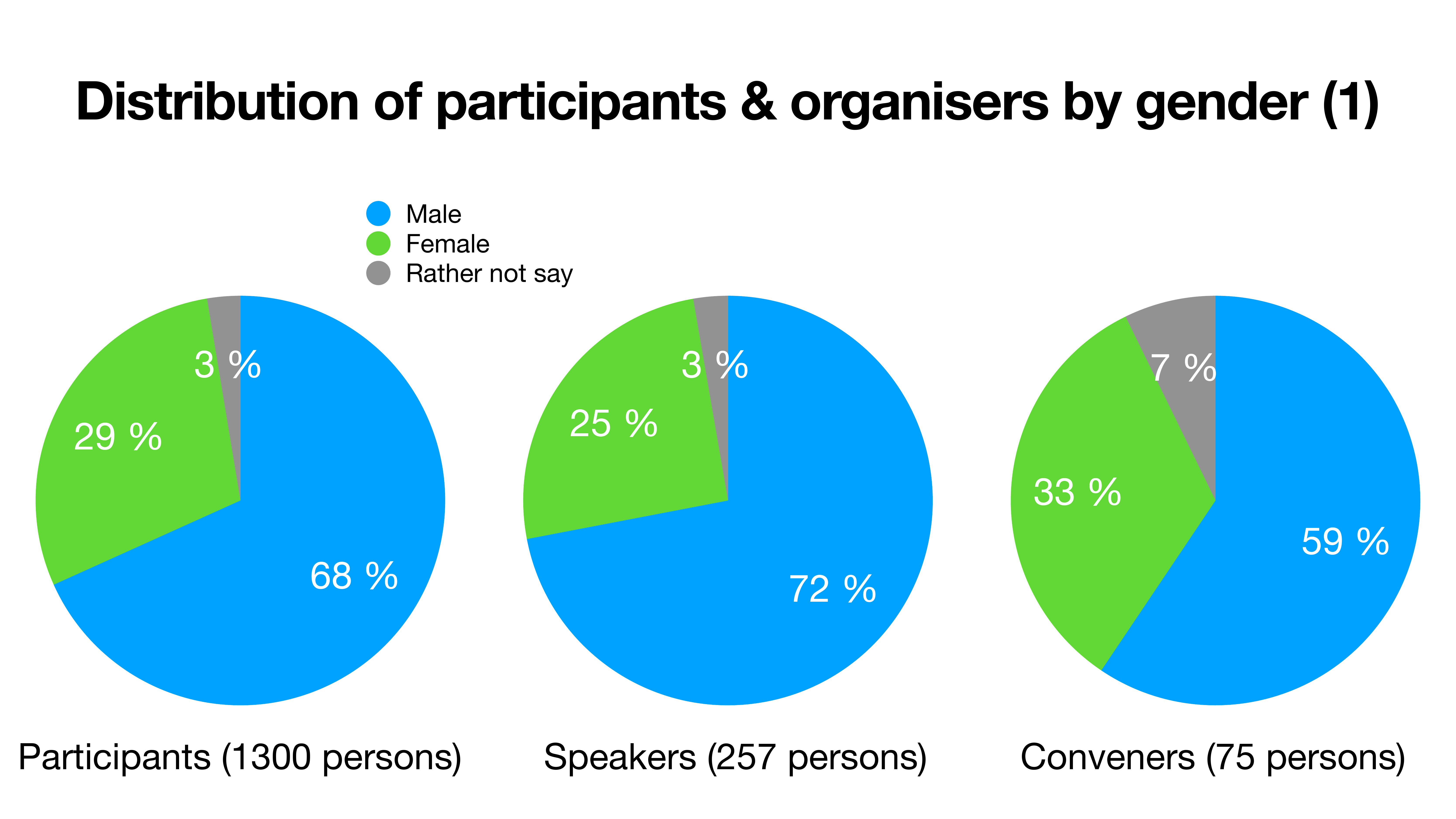}
\includegraphics[width=0.7\textwidth,trim=0 50 0 370,clip,page=2]{LHCP2020_stats.pdf}
\caption{\label{fig:iv} The distribution of the participants and organisers of the LHCP2020 conference according to their self-declared gender.}
\end{figure}

The distributions of the gender of the conference participants and of the members of various organising committees, as defined by the participants themselves at registration time when they were given the choice between 'Female', 'Male' and 'Rather not say', are shown in figure~\ref{fig:iv}.
Among the organising committees (PC, SC, IAC) the male:female ratio was varying between 63:37 and 65:35, never exceeding 2:1. Among the session conveners, who were appointed by the SC for the plenary sessions and by the PC for the parallel ones, a similar ratio is found (64:36). Among the participants a higher male:female ratio of 70:30 was observed, and even slightly larger among the speakers (74:26), that were partially invited, partially selected by the experimental collaborations, or otherwise volunteered for a poster presentation by submitting an abstract to the conference website.

The logs of the videoconferencing sessions indicate that there were up to 520 unique participants in the plenary sessions connected on \gls{Zoom} as well as up to 377 on \gls{Webcast}. In the parallel sessions there were typically several tens of connections, with some sessions reaching up to 150 participants.

\subsubsection{Participant satisfaction survey}

After the conference, an online survey was circulated to the conference participants to collect their feedback about the logistics, organisation, format of the conference, and potential ideas on how to improve it. The questionnaire was left open for about two weeks, and two reminders were sent to the conference participants before the deadline. We collected a large sample size ($N$) of 354 people. The survey participants are well distributed in current position, age, continent in which their institution is located, and between experimental and theoretical physicists. The sample does not differ from the true population average attending a high energy conference as one of the LHCP conference series.
The information about the respondents are collected in an online PDF file~\cite{LHCP2020_stats}.

The sample is composed 44.4~\% active conference participants (namely participants that had at least one of the following roles: speakers, poster presenters, conveners, members of the organising committee) and 55.6~\% simple attendees. Most of the respondents attended either the plenary or the parallel or both the sessions while the poster sessions were less attended by 7.1~\% of the respondents.

\paragraph{Time format}

The first set of questions that was asked in the survey concerned the choice of the time format of the conference.
Most of the participants were satisfied with the choice of concentrating the programme in the hours between 12:30 and 18:30 Central European Summer Time (CEST), with two short 15 minutes breaks and several parallel sessions taking place at the same time. On a 0--10 scale, the average satisfaction was 8.7, with a drop to 7.6 for participants in North America.
The choice of scheduling the parallel sessions in the "prime time" of the day, typically between 14:30 and 16:00 CEST, in order to make them as much accessible as possible to participants from all over the world, was also appreciated, with an average satisfaction of 8.6/10.

The participants were also asked about possible alternative formats:
\begin{itemize}
    \item a 2 week long conference, with less parallel sessions running in parallel, longer breaks, and about 4 hours of conference per day.
    \item a 1-week long conference, with "premiere" sessions whereby talks are live  followed by 1 or 2 "replay" sessions to cover all time zones and during which recorded talks would be re-transmitted.
\end{itemize}
Only 23~\% to 24~\% of the respondents would have preferred these alternative options to the format that was chosen for LHCP2020.


As an optional request, the respondents were asked to provide additional feedback on the time format. The few received suggestions were on reducing the numbers of sessions in parallel, on having less compressed sessions, and on having longer time for discussions. These would unavoidably affect the conference time schedule forcing to have either longer conference days or going to a two-week long conference, solutions that as shown before are disfavoured by the majority of the respondents.


\paragraph{Technical setup}

The choice of \gls{Zoom} as the videoconferencing platform was highly appreciated, with an average satisfaction of 8.8/10. A few respondents also proposed alternative platforms such as \gls{Vidyo} or \gls{MicrosoftTeams}. It was also suggested that CERN develops its own videoconferencing system based on \gls{Jitsi}. Among the participants who filled the survey, only 17~\% connected to the plenary sessions through \gls{Webcast}, either because they could not use \gls{Zoom} (10.5~\%) or they preferred \gls{Webcast} over \gls{Zoom} (6.2~\%). The average satisfaction of the \gls{Webcast} users is 8.5/10. No alternative solution was proposed.

Concerning the format of the sessions, we asked the participants which one would be more appropriate between a Webinar and a Meeting for plenary and parallel sessions, poster sessions and breakout rooms. The webinar format is clearly preferred for the plenary sessions, while the meeting format is favoured for the poster and breakout rooms among the respondents who have an opinion on this topic. For the parallel sessions the respondents with an opinion are split almost evenly between the webinar and meeting formats. 


The satisfaction of the participants concerning the way the question time was technically handled can be summarised as follows. The use of the raise-hand feature of \gls{Zoom} in order to be able to turn on the microphone and ask the question, as would happen in in-person conferences, was appreciated by a large majority or participants, with an average satisfaction of 8.4/10. About a quarter of the respondents of the survey would prefer to use a chat-based system in which questions can be entered in a text window and can be up or down voted. The \gls{Discord} chat that was provided to users following the plenary sessions on \gls{Webcast} in order to ask questions was essentially not used. The respondents that used \gls{Webcast}, however, found it more useful than not (average vote 5.6/10) to have the option available.

The satisfaction of the participants concerning how the feeling of community, lost in a online conference, was recreated, can be summarised as follows. Most participants did not actually use the \gls{Discord} chat rooms created for the parallel sessions and for the virtual coffee breaks; the minority who did, found them of limited utility (average vote 5.3/10). Concerning the breakout rooms where to meet the speakers and discuss with them during the virtual coffee breaks, a bit more than one thirds of the respondents answered that they used them and the idea was deemed quite satisfactory (average vote 6.7/10). On average, all the respondents said that providing such rooms is rather important for a virtual conference (average vote 7.0/10), and future conferences could thus consider to request it to all the speakers.


The satisfaction of the participants concerning the recordings and live transcriptions can be summarised as follows. The respondents found the recordings of great utility (average vote 8.5/10), few of them already watched the recordings (17.8~\%) and about half of the participants will watch them in the future.

When asked (optionally) to provide further feedback about the technical organisation of the conference, the respondents provided a few comments such as:
\begin{itemize}
    \item find a better way for the participants to see who else is attending the same event. A possible solution for this, at least for parallel sessions, could be moving from webinar-style to meeting-style \gls{Zoom} meeting rooms
    \item having a clock running during a speaker's presentation, visible to the speakers themselves, rather than the chair interrupting to tell the speaker how much time is left. Of course the speakers themselves are encouraged to use a clock on their desk or on the computer that they use during the presentation to keep track of the time, but solutions integrated into the technical tool used for the videoconference would be a welcome addition.
    \item breakout rooms are an interesting idea for virtual conferences and should be further encouraged
    \item having a system to upvote questions (but no downvoting)
    \item releasing the recordings of the talks the same day as the presentation, otherwise people could lose interest afterwards. It should be noted that post-processing the recordings of the whole sessions (splitting them into single contributions, adding captions) takes time and person-power and even with a dedicated IT team it took 2 to 3 weeks after the end of the conference to have all the recordings online on the LHCP2020 conference website. A potential alternative would be that speakers themselves provide pre-recorded, self-transcribed talks. This would also solve the problem of the quality of the transcripts, which was deemed terrible by some respondents and acceptable (though not perfect) by others.
\end{itemize}



\paragraph{Funding}

The third set of questions of the survey concerned questions related to the conference funding.
Considering that the budget is very severely constrained for an online conference without a fee, the participants assigned to the items to be funded the following importance, ordered from higher to lower:
\begin{itemize}
    \item communication, average vote 7.8/10; \vspace{-8pt}
    \item publication of conference material (recordings, post production, ...), average vote 7.5/10; \vspace{-8pt}
    \item outreach, average vote 6.6/10; \vspace{-8pt}
    \item poster awards, average vote 5.9/10; \vspace{-8pt}
    \item proceedings, average vote 5.9/10; \vspace{-8pt}
    \item others, average vote 1.3/10.
\end{itemize}
We had explicitly asked any other possible items to spend money and we have received few suggestions. Three suggestions were related to assuring the accessibility of the conference material to (visually, hearing, etc.) impaired participants and one suggestions was to have a solid IT infrastructure.

The majority of the respondents would prefer to maintain a "fee free" registration (average vote 7.4/10). However, when asked about a reasonable fee for the the (major) online conferences, 45~\% of the participants would find it acceptable to pay a fee of between 20 and 50 euros, with 43~\% favoring lower (26~\%) or no fees (17~\%) at all. The respondents found that their institutes/funding agencies should contribute financially to support the organisation of a fully online conference with an average vote 6.4/10.



\paragraph{Outreach}

While outreach activities had been foreseen for the in-person conference in Paris, no outreach programme was included in the online conference, since the organising committee focused, in the limited amount of time that was left for preparing the online conference, on the scientific programme and the technical setup. However, we asked the conference participants how important would it be to propose a public online outreach program during an online conference and, optionally, who could be the target public and to propose some ideas for the future. 
The average importance of proposing outreach activities in an online event is 6.6/10.


Such outreach activities should mainly target students of various levels (high-school, undergraduates) as well as general public (ranging from kids to retirees) with an interest in science. Attention should be paid also to propose activities that can reach out to students and post docs from nations without much budget for travelling around the world.
Such activities could be useful to remind the general public why fundamental science is important and how the public money that supports it is used. Plenty of ideas are already being developed on potential online outreach activities, by the outreach teams of the LHC Collaborations as well as by the International Particle Physics Outreach Group (IPPOG). 
They include webinars, virtual tours of the detectors, virtual labs, hackathons, physics-inspired quizzes, and summary talks of the main highlights of the conference explained in a way that is accessible to everybody.


\paragraph{Overall satisfaction and feedback}

\begin{figure}[t!]
\centering 
\includegraphics[width=0.9\textwidth,trim=0 50 0 100,clip,page=36]{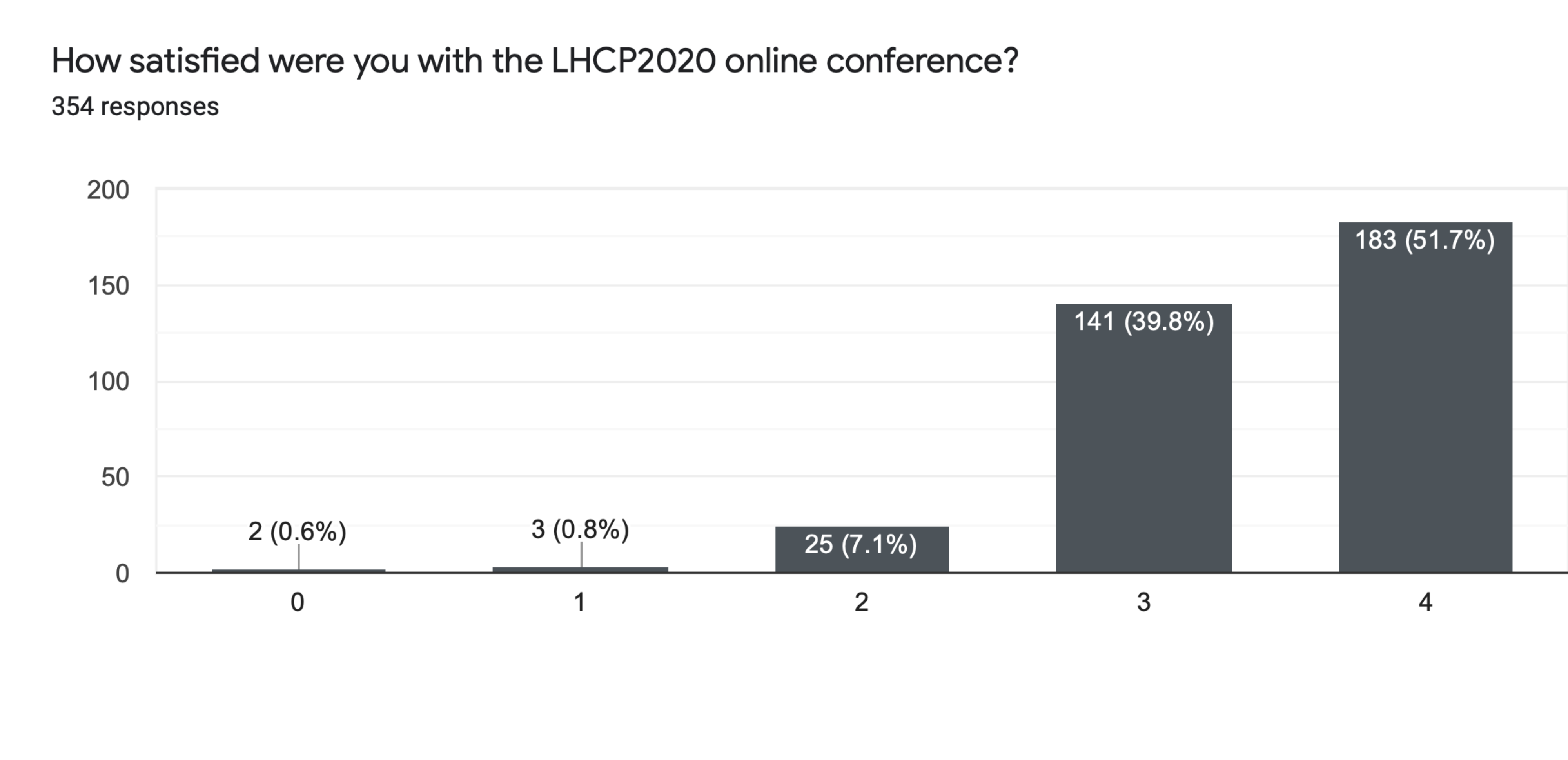}
\caption{\label{fig:xv} LHCP2020 online survey: overall satisfaction of the conference participants.}
\end{figure}

The overall satisfaction of the conference participants is shown in figure~\ref{fig:xv}. The average satisfaction was found to be high at 8.6 out of 10.


\subsubsection{Conclusion}

Because of the extraordinary conditions related to the COVID-19 pandemic, the LHCP 2020 conference, the eighth in the series of annual conferences on the physics of the Large Hadron Collider, was exceptionally organised online, in the week of May~25 to 30, 2020. It was the first fully online HEP conference with more than 1000 participants. The conference was a success with a very high level of satisfaction based on participant surveys. We believe the experience gained in the preparation of the LHCP2020 online conference, as well as the feedback received by the participants, will be useful not only in the organisation of online events, but can also be beneficial for low-cost, inclusive, and delocalized conferences.

\subsubsection{Acknowledgments}
We would like to thank all the people involved in the organisation of the conference, the members of the international advisory for their suggestions and guidance, the programme committee for setting up such an exciting program and the speakers for delivering high quality talks and adapting to the new schedule, the local organising committees for the preparatory work for the Paris conference that was partially reused for LHCP2020.

We would also like to thank the CERN IT department, and in particular Thomas Baron and Jonathan Coloigner, for all the technical support in the setup of \gls{Zoom}, \gls{Webcast}, \gls{Discord}, the recordings of the presentations and their captioning. Many thanks to Connie Potter and Dawn Hudson who helped with the preparation of the conference \gls{Indico} webpage, the conference advertisement, and took care of reimbursing the fees already paid when the in-person conference was cancelled. An outstanding work of communication on the social media was done during the conference by our \gls{Twitter} team: kudos to Yasmine Amhis, Zaida Conesa del Valle, and Laure Marie Massacrier.

Finally, we would like to thank all the participants of the conference, who all together made its success, and in particular those that volunteered to edit the transcripts of the recordings -- hoping that future online events might benefit from this work -- and those who filled the survey and gave us the precious feedback presented in this document.
\subsection{ICHEP 2020}
\label{sec:experiences_ICHEP}

\subsubsection{Introduction}
The International Conference on High Energy Physics (ICHEP) is organized every two years. Since its first edition in Rochester 1950 it has grown to the leading conference in particle physics worldwide, where the latest experimental results and theory achievements are
presented. The usual format is three full days of parallel sessions and another three days of plenaries. The typical attendance exceeds 1000 participants. 

The Local Organizing Committee (LOC) in Prague started the preparations in 2016, foreseeing a standard conference format. In early spring of 2020 it became clear that the conference could not be held as originally planned due to the COVID-19 pandemic. Several discussions within LOC and with C11 IUPAP Commission took place, the final decision was to move the conference wholly to a virtual format. 

\subsubsection{Conference format}
Since the conference participants are scattered over the whole world, the sessions were organized in shorter time slots, alternating between morning (8:00 to 13:00 CEST) and afternoon (15:30 to 20:30 CEST) slots that reasonably match people's working hours in Asia and America, respectively. Speakers were asked during the registration about their preferred time slots. 

\begin{figure}[ht!]
  \includegraphics[width=0.49\linewidth]{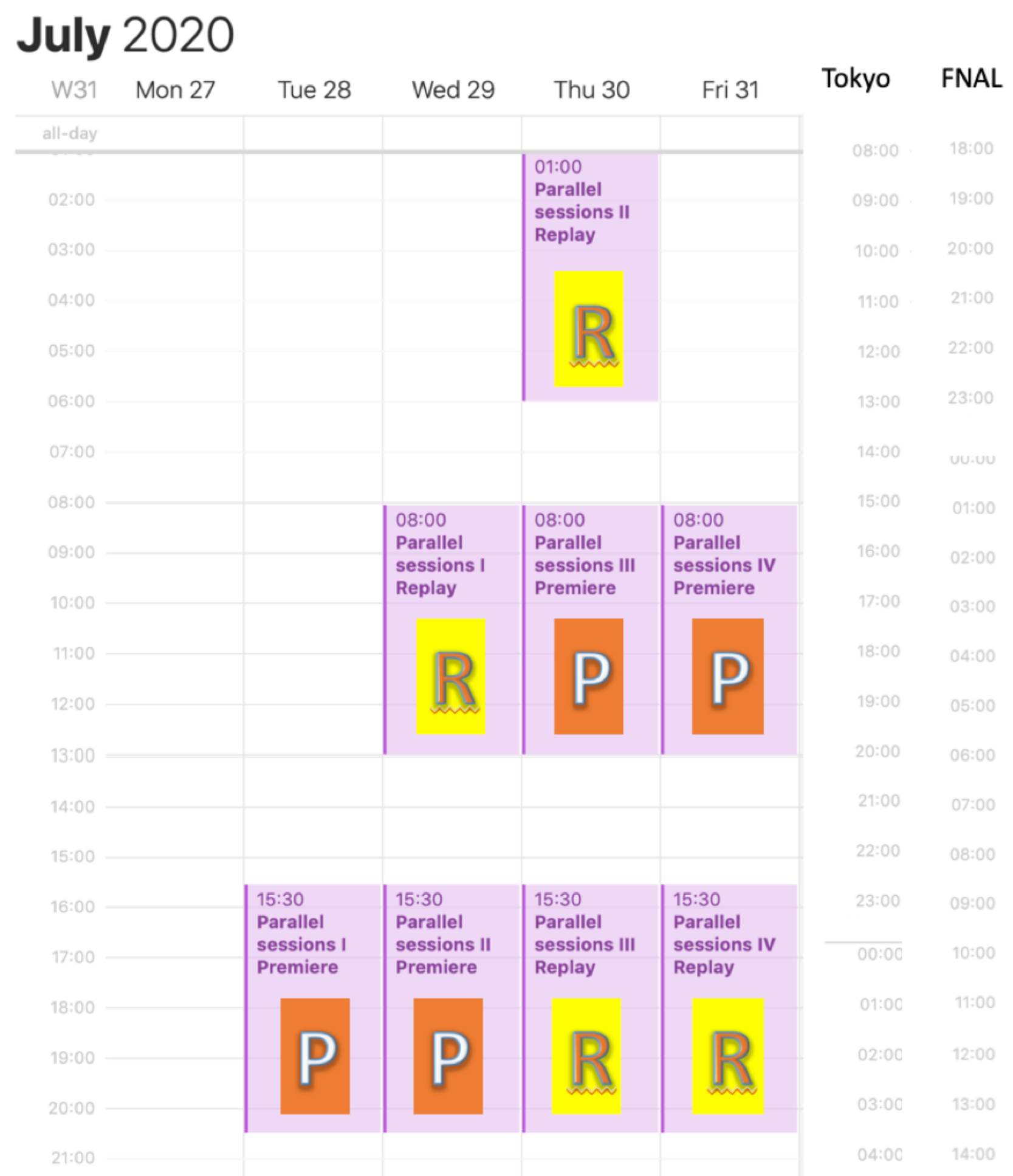}
  \includegraphics[width=0.42\linewidth]{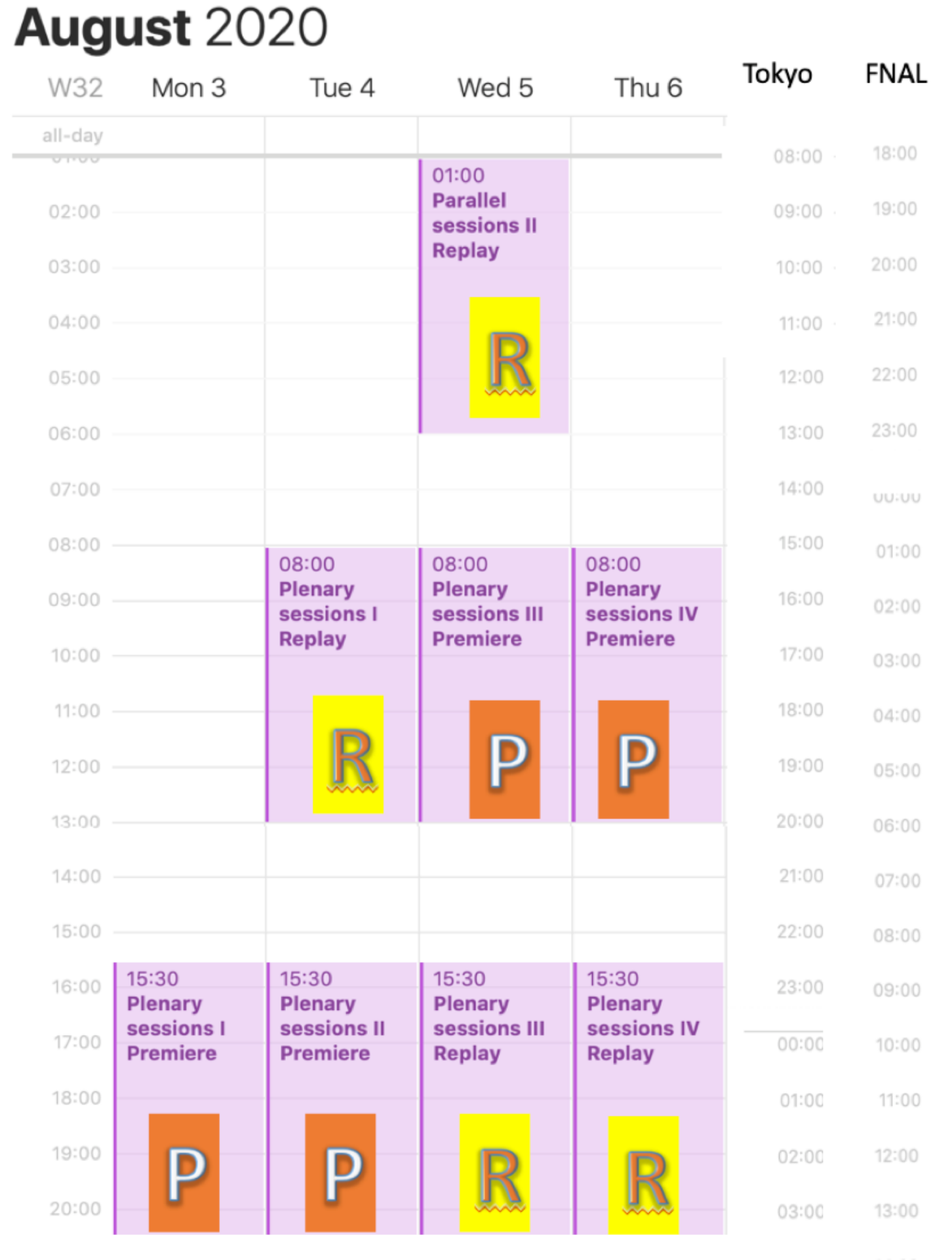}
  \caption{The layout of ICHEP2020 parallel (left) and plenary (right)
    sessions. ``P'' stands for the premiere session, ``R'' indicates
    the replay session.}
  \label{fig:ichep_sessions}
\end{figure}

In order to accommodate the large number of accepted contributions (total 800~talks in 17~parallel sessions, 44~plenary contributions, 150~posters), the conference was extended to four days of parallel sessions and four days of plenary sessions. In addition to these so-called "premiere" sessions involving presentations, specific ``replay'' sessions were organized to further ease the attendance to participants from East/West time zones. During the replay sessions, the recordings on \gls{YouTube} were streamed. The timeline and sessions are summarized in Figure~\ref{fig:ichep_sessions}. All
sessions were recorded and the videos were available on \gls{YouTube} and later also in CERN Document Server (CDS). 

\subsubsection{Technical solutions}
The \gls{Zoom} platform was chosen as the technical solution for the ICHEP2020 virtual conference. All talks were given live during the premiere sessions rather than being pre-recorded. A dedicated \gls{Zoom} room with a technical assistant from LOC was available few days before the conference to allow participants to test their audio/video. This proactive approach minimized problems during the premiere sessions. As already mentioned above, the premiere sessions were
recorded in \gls{Zoom} platform by technical assistants from LOC as well as by the corresponding session convenor. This solution was adopted as precaution of potential internet connection problems on the organizer’s side. Nevertheless, these backup recordings were never needed, as everything wound up running smoothly. 

A \gls{Zoom} Webinar up to 3000~participants was used for all plenary sessions and Neutrino parallel session (where the largest audience was expected). For the remaining parallel sessions \gls{Zoom} large meeting
(maximum 500~participants) was used. All sessions were simultaneously streamed to \gls{YouTube}, which provided a backup solution for those who could not eventually join \gls{Zoom} directly. 

Apart from talks, young researchers and students were also invited to present their work as posters. Poster sessions were organized during the parallel session week. Posters were turned into mini-talks (maximum 5~slides) and presenters were offered to upload into the
agenda additional material (e.g.\ a standard poster) with further details. Plenary sessions were close-captioned by Whitecoat Captioning company that was hired specifically for this purpose.

\subsubsection{Interactions}
Opportunity for discussion -- both within each session as well as informal discussion afterwards -- is a key ingredient of every conference. The discussion
between participants was stimulated by several means:
\begin{itemize}
\item Usual questions and answers at the end of each talk or poster presentation. Additionally, \gls{Zoom} rooms remained open for typically one hour after the respective session finished in order to enable further discussion \vspace{-7pt}
\item A dedicated \gls{Mattermost} channel was setup for continuous discussion in each parallel session \vspace{-7pt} 
\item Topical discussion sessions were organized during the plenary session week. Senior scientists were asked to lead and stimulate discussion on individual topics \vspace{-7pt}
\item A virtual tour through Prague via drawings by one of our colleagues was offered
\end{itemize}

\subsubsection{Other program elements}
Other events originally planned to accompany the conference were also moved to a virtual format:
\begin{itemize}
\item Outreach and public relations activities were performed on \gls{Facebook} (Czech) and \gls{Twitter} (English) \vspace{-7pt} 
\item An art competition called "BeInspired" was organized\vspace{-7pt}
\item A public lecture was given online by Barry Barish, the Nobel Prize winner 2017, in the evening of the first day of plenary session week \vspace{-7pt}
\item A European Research Council (ERC) workshop was organized during the weekend between the parallel and plenary sessions
\end{itemize}

\subsubsection{Statistics \& feedback}
Table~\ref{tab:statistics} summarizes the attendance for the last three ICHEP conferences. The virtual format of ICHEP2020 allowed for a much larger number of registrants as compared to the previous in-person conferences. Since the ICHEP2020 conference fee was waived, the number of registrants does not necessarily reflect the number of active participants. Nevertheless,
we encountered 2835 unique participants (connected more than 15~min) on \gls{Zoom} and approximately 200 on \gls{YouTube}. Parallel sessions were typically attended by 100 to 200~people and plenary sessions had typically 500 to 1200~participants. The discussion panels were attended by 20 to 100~people.
\begin{table}
  \centering
  \begin{tabular}{|c|r|r|r|r|r|r|}
    \hline
    & \multicolumn{2}{c|}{ICHEP2020} & \multicolumn{2}{c|}{ICHEP2018}
    & \multicolumn{2}{c|}{ICHEP2016}\\ 
    \cline{2-7}
    Gender & Regs. & Fraction & Regs. & Fraction & Regs. & Fraction \\
    \hline
    Male & 2178 & 72~\% & 893 & 77~\% & 1144 & 80~\% \\
    Female & 772 & 26~\% & 267 & 23~\% & 286 & 20~\% \\
    Rather not say & 54 & 2~\% & & & & \\
    Other & 6 & 0~\% & & & & \\
    \hline
    Total & 3010 & 100~\% & 1160 & 100~\% & 1430 & 100~\% \\
    \hline
  \end{tabular}
  \caption{Number of registrants over the last three ICHEP conferences}
  \label{tab:statistics}
\end{table}

\begin{figure}[ht!]
  \centering
  \includegraphics[width=1.0\linewidth,bb=50 115 790 480,clip]{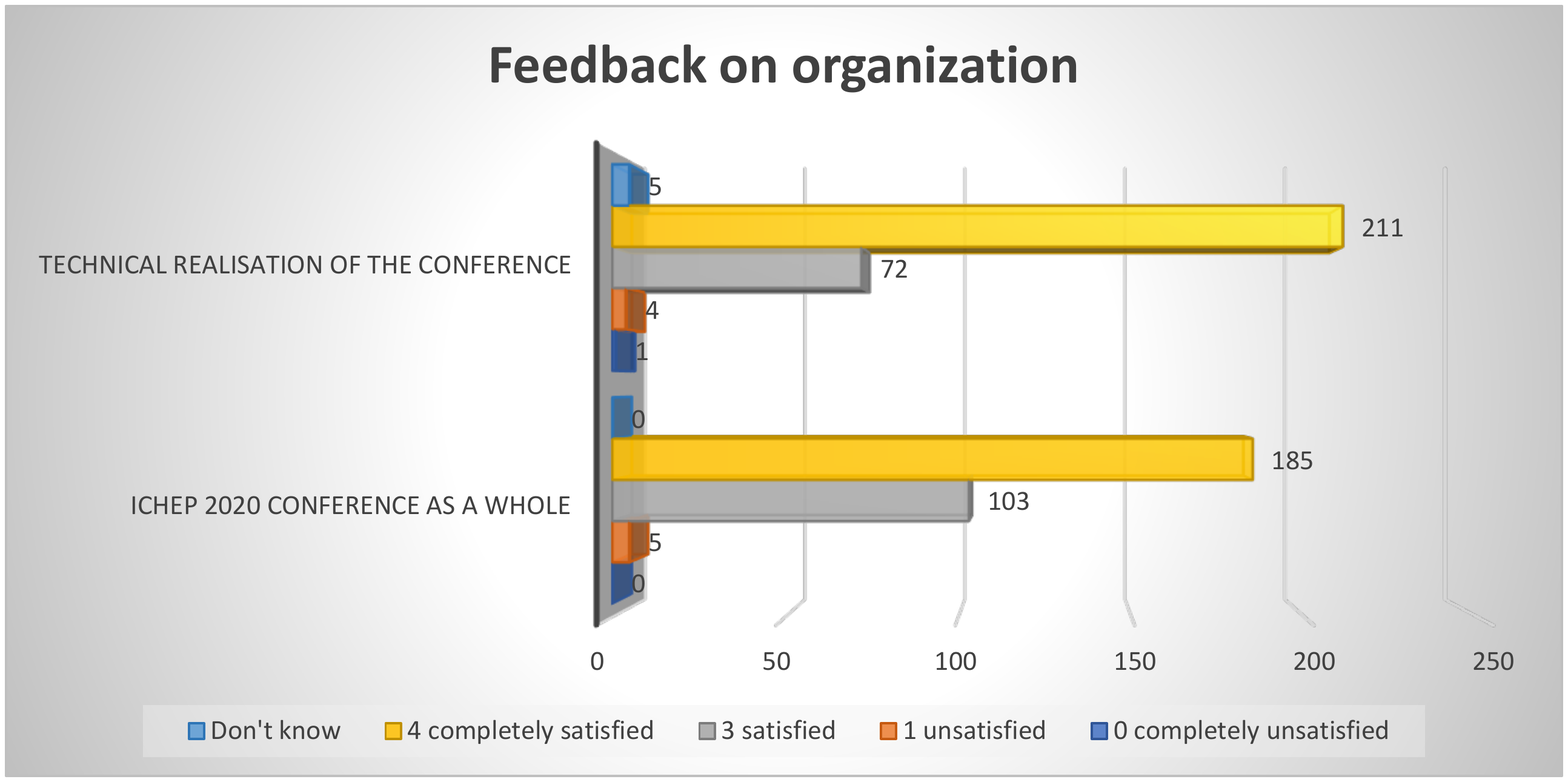}
  \caption{The results the satisfaction survey, 300 participants
    responded.}
  \label{fig:feedback}
\end{figure}

The effort involved in organization of the virtual conference was very appreciated and we received many congratulations. The positive feedback is illustrated in
Fig.~\ref{fig:feedback}, which shows the response to the
post-conference survey.

\subsubsection{Concluding remarks}
The ICHEP2020 was a great success for a virtual conference as evidenced by the high-level of satisfaction from the surveys. Many more people participated in the conference and the lack of travel burden helped make the conference more inclusive. Nevertheless, more people were needed to run the conference (\gls{Zoom} technical assistants, management of recordings, etc.) compared to a standard in-person event. The limitation on attendees' interactions were at least partially offset by topical discussion sessions and live chats on \gls{Mattermost}. Strong institutional support for ICHEP2020 is acknowledged, which allowed for waiving of the conference fee, for example.

\subsection{Moriond 2021}
\label{sec:experiences_Moriond}

\subsubsection{The spirit of Moriond}

Since 1966, physicists have been meeting in the snowy mountains of the Alps for a week of discussions. It was thought to be a place where all persons were sharing a time and place, with no hierarchy. Everybody has the same speech duration, whether they be theorists or experimentalists, graduate students or Nobel Laureates. People from all over the world live for one week in a family-like environment, presenting and discovering the latest results in the field, but more importantly, \emph{interacting}. Full attendance is required and specific grants are proposed for those with difficulties to pay for the living expenses.

This concept has been working for 54 years, never failing. Until the one we do not name arrived and got us all stuck at our homes. Again, no hierarchy applied. But the ``meeting'' was seriously compromised and with a knife in our hearts we had to cancel the 2020 edition only two weeks before its launch.

One year later, responding to a certain demand, but also to our own need to pursue, we had to face that the 2021 edition of the {\it Rencontres de Moriond} would be virtual or would again not be at all, at the risk of falling into a definitive breach where results would be presented elsewhere, finances would fall into a serious situation and inactivity would lead us to different paths. After all, there was a calendar to maintain, results kept being obtained and released, students kept working on their theses and needed to be presented to society in view of postdocs. The show must go on, as Freddy would say.

\subsubsection{Three sessions, Three experiences}

Over the years, the {\it Rencontres de Moriond} have split into half a dozen different sessions, dedicated to different subjects and communities. Some are annual, others biennial, others even quadrennial. In 2021, for this {\it Moriond@home} we decided to organise three sessions
\begin{itemize}
    \item Electroweak Interactions (annual)\vspace{-7pt}
    \item QCD and High Energy Interactions (annual)\vspace{-7pt}
    \item Gravitation (bi-annual)
\end{itemize}

\noindent Each session had a very different structure with different tools -- we were definitively trying different things!

\paragraph{Gravitation}

The Gravitation session taking place every odd year would have to jump into 2023 if we skipped 2021. This would have made it impossible for a large number of graduate students to present results during their thesis time, which motivated us to go forward with a virtual event dedicated to them -- a full poster session.

It was a short three half-day format (Tuesday to Thursday, from 13:00 to 17:00 local time), divided into two daily poster sessions of sixty minutes each preceded by a thirty minute plenary introduction to the theme of the session. The existence of the \gls{Gather.Town} platform was discovered by a scientific organizing committee member and we thought that, as it was addressed mainly to a young public, it was worth trying.

The result was better than the organizers could have expected. It required, on the organising side, a thorough learning on how to use and customise the \gls{Gather.Town} platform for our purposes. We ended up having one conference room, six poster rooms, one meeting room, a bar and a garden where people could meet informally.

We had 124 participants, mostly graduate students and postdocs, but also quite a few ``regulars'' that were happy to meet the community, even if virtually. The possibility to present in small groups and navigate freely from group to group made it really close to an on-site conference. The selective activation of microphones and cameras inside a well-defined area without interfering with neighbouring areas is a real bonus. Also, the fact that the venue was always available and that participants could stay and meet at any time outside the official hours made the attendance quite lively. From a technical point of view, we created a short User's Guide with basic hints and proposed daily ten minute meetings during the prior week so that people could test the platform. Overall, there were no major difficulties except for really senior users, who needed to be individually guided for the first five minutes or so. 

At the end it was almost frustrating not to continue using the venue, as it appeared so full of possibilities remaining unexplored.

\paragraph{Electroweak Interactions and Unified Theories}

The EW session was a hybrid between what we usually do and the possibilities that virtual meetings offer. It consisted of short half-day live sessions (14:00 to 17:00 from Sunday to Saturday local time) over \gls{Zoom}. Regular talks were pre-recorded using also \gls{Zoom} and stored using \gls{Vimeo}. Links to the talks were accessible from the program on the website and released 24 hours before the dedicated live session. Each of these sessions started with a Young Scientists Forum (five minute talks on the session's theme) followed by the discussion on the recorded talks and questions. In parallel, there was a conference \gls{Slack} account, with one channel per talk and one channel per live session. Participants could then watch the talks at their convenience and take the time to write questions on the corresponding \gls{Slack} channel. These questions could be answered by the speaker but would also be considered by the session Chairperson, along with live questions that were included in the discussion. Speakers could also follow-up with a discussion on \gls{Slack} after the live session.

This organisation was intended to limit the problem of people in significantly different time zones, making shorter live sessions and allowing each person to watch and comment on talks at their own pace and availability.

As questions were expected on \gls{Slack}, we chose to disable the chat on \gls{Zoom} in order to avoid multiple sources of input for the session Chairpersons. Indeed exchanges worked rather well summing up with a peak of 758 messages in a single day for an attendance of 216 persons. Of course, things do not always go well by themselves -- we had to make a real effort for these exchanges to work properly. Each session was moderated by a dedicated member of the Scientific Organizing Committee, who appointed two dedicated Chairpersons (one theorist and one experimentalist) who in turn chose secret agitators. A whole system!

\paragraph{QCD and High Energy Interactions}

The QCD session was the closest one to the classical format -- seven days, from 14:00 to 20:00 local time with live talks and discussion, split by a 30 min break. It was held fully over \gls{Zoom}, each 15 min talk was followed by five minutes of questions and discussion. For some themes, a general discussion time was reserved at the end of the program.

Virtual coffee breaks and after-meeting discussions were expected to happen on \gls{Gather.Town}, but participants did not seem to find an interest in using this extra space. The truth is that sessions were quite long and intense and breaks were used as ... well, real breaks! You can't stretch your real legs in \gls{Gather.Town}.

There was an attendance of 120 persons, which is quite similar to usual in-person participation.

\subsubsection{Technical Aspects}

\paragraph{Costs}

We thought there would be an important investment in subscriptions to all the different tools and platforms we wanted to use. But in the end it was surprisingly cheaper than expected, as all tools used have a special plan for education that we benefited from after a few email exchanges and documentation provided. The reduced price was about 20~\% of the regular one, which was very affordable. Furthermore, there was no special need for hardware or supplies.

\paragraph{Human Resources}

Regarding the secretariat, we had more or less the same need as for a in-person event -- dealing with the website, registrations, participant requests, accounting, etc. was a little less demanding than for an on-site event. On the other hand, we did not need any specific computer support. The managing of the various platforms did require a large amount of preparation time, but could be dealt with by the Organising Committee -- creating the necessary spaces and channels, checking and uploading each poster and video in due time, creating all necessary links and reacting to other various details that one could not foresee. Also, a constant human presence at-the-ready in the background was required, making sure everything was working from the technical point of view and being prepared to jump into a back-up solution if anything failed. Practically speaking, this really meant an enormous amount of time during the week.

Also, as already mentioned, an increased participation from the members of the scientific committee was required as compared to a traditional in-person event where building the program is their main concern. Things needed to run like clockwork - checking speakers sound, screen sharing and image, preparing questions, avoiding silences, handling the discussion threads, etc.

\paragraph{Fees}
We chose to charge a limited fee of 50 euros, which was an amount that was thought to be accessible to all. This was not meant to fully cover organising expenses since we have a permanent secretariat, but it served as a kind of engagement, avoiding inactive registrations and limiting paperwork and mail exchanges for a null participation. We also hoped it would maximise the presence of participants, since they did pay for the thing after all.
In the very few cases of participants explaining that they could not pay for this fee, for budgetary or administrative reasons, we did waive it, although we did not open a webpage for grant applications in these cases.

\subsubsection{Participation}

We had a similar number of participants in the virtual conference to what we usually have. This relates both to the amount of talks (that was the same) and to the fee 
(that might have discouraged a few visitors). We were satisfied with this result, as we were not aiming for a very large or different event, but rather a substitute to what we usually do. It stayed between the limits of what is comfortably handled by our team. 

We could see different trends of the daily participation amongst the three sessions. While Gravitation was quite stable around 80 persons for 124 registrants, the EW and QCD instances had a general tendency to decrease along the week. Also, on a daily analysis, we could see that in QCD there were fewer participants per day relative to registrants, and clearly decreasing towards the end of the day, while in EW it was a bit more stable. 
Only 1/4 of the participants in EW and 1/6 in QCD attended the full week, as measured in terms of connection time, and it does not mean people were actually listening all the time. This result is quite understandable -- it is very difficult to stay connected during a full week while at work or at home. One may stay focused on a few subjects but hardly fully participating throughout.

\subsubsection{Lessons Learned}

A very detailed preparation is needed, otherwise interactions fall into a background noise. To achieve this, more human presence is needed before and during the events. We also noticed that interactions are often based on prior acquaintances. This was quite obvious in \gls{Gather.Town} and \gls{Slack}, where people seemed happy to ``see'' each other. The discussion relies greatly on the fact that people know each other and hence dare to talk, but it is difficult to create new connections under these circumstances.

Sessions need to be short since participants cannot make themselves available for the whole day. Sessions of three hours seem to be a good compromise. As mentioned before, there is no real cut-off from the home/work environment as when one is isolated in a hotel in the mountains, sometimes with a poor internet connection (those were the days...). Also it is difficult to include all participant time zones comfortably, so for someone having to wake up very early or go to bed very late, making things short is generally appreciated.

Regarding the platforms used, \gls{Zoom} is the easiest to use, but by far not the most interactive.  \gls{Gather.Town} can do it all, but might not be intuitive for some users and can be limiting in resources available -- not all web browsers supported it equally well at the time of the conference and smartphone features are reduced. \gls{Slack} seems to be a must have for communication in the background, allowing an organised and long-lasting exchange. Combining the two functionalities into a single application would seem to be good practice, as too many tools or spaces can get people confused and/or information can get lost. But this requires taking the time to structure and adapt these tools to specific needs, so participants will find a time and place for everything.

Proposing a dedicated time slot for testing is a really useful activity. This could happen either the week before, or simply 1/2 hour before each daily session where organizers make themselves available for individual testing of the various features, navigation, document sharing, sound and image, resources, etc. This really helped to make people comfortable and confident that everything would work smoothly when the time comes to start the conference.

\subsubsection{Conclusion}

Virtual meetings can work, but cannot efficiently replace an in-person meeting where the purpose is to discuss and where as much happens outside the conference room as inside, if not more. Some things cannot happen online and in the Moriond case, the virtual is not at all the preferred option. But in many other situations, it is clearly a field where things are improving and many intelligent solutions are being found. 

Remains the difficulty of setting the world to rights while having a beer virtually...

\subsection{Connecting the Dots 2020}
\label{sec:experiences_CTD}

The Connecting The Dots (CTD) workshop series brings together experts on track reconstruction and other problems involving pattern recognition in sparsely sampled data. The CTD scientific program is nominally 2.5 days with a fully plenary format, including three types of presentations: Plenary talks; young-scientist talks; and posters. An in person workshop had been scheduled for April of 2020 with about 80 people expected to attend. 

CTD2020 was moved to a virtual conference in mid-March once it was clear that the local and international restrictions due to the COVID-19 pandemic would prevent an in-person event. By this time, the workshop had accepted abstracts for contributions and had published an initial timetable for its event. Essentially six weeks remained before the workshop, so only a limited time was available to determine the format and overall details of a virtual event. Both the local organizers and international committee were strongly in favor of a virtual conference rather than no conference. 

A virtual format was formulated in less than two weeks. The organizers wanted to preserve the full scientific program as planned for the in-person meeting – talks and posters, the planned length of oral presentations, and the planned opportunities for interacting with all presenters (oral and poster). We judged it to be not possible to schedule the entire scientific program during the overlap between the EU extended-working day and the US extended-working day. Asia-pacific participation was not prioritized as there were no abstracts from that region contributed to the workshop. Finally, the organizers felt that we had limited flexibility in rescheduling as speakers needed time to get results approved (e.g, could not move presentations forward).

In the end, the organizers settled on a two-phase workshop: “recording sessions” followed by dedicated question \& answer sessions with about one week in between for viewing recording contributions. Posters were allocated as short talks as we had insufficient time to converge on a more creative solution appropriate for a poster presentation. 

The recording sessions happened outside of the EU/US overlap time period to make scheduling easier, had an audience, and had a short question \& answer period for each time. There were six sessions over three days. Recordings were done via \gls{Zoom}, which was new to the CERN community at the time, and used \gls{Zoom}’s cloud recording service. Speakers were asked to share their video during their presentation if they could (most but not all could). Sessions had twenty participants on average and went essentially without technical issues. 

The cloud recordings were available to the organizers about one hour after each recording session. The organizers automated the process of splicing each session into talks and managed to upload recordings to both the \gls{Indico} workshop agenda and \gls{YouTube} before the start of the next recording session. Some aspects were done by hand, including finding the time break for each talk and uploading to both \gls{Indico} and \gls{YouTube}. \gls{YouTube} does have an API to upload videos but its free API quota is sufficient to upload only a few videos per day. 

A \gls{Mattermost} channel was set up to host discussion for each contribution as videos were viewed. In the end, this went essentially unused, suggesting that something more integrated with other infrastructure was needed instead of a standalone discussion platform.

The question/answer sessions happened about one week after the recording sessions. There were three sessions, each two hours. Sessions were held from 10:00 to 12:00 Eastern time and were recorded. Each speaker had a one-minute per one-slide introduction to their talk to get the discussion started. More than 200 people had registered for these sessions. In general, the question \& answer sessions had good (50+) attendance, however the discussion was largely driven by a few “experts”. The short introduction to each talk worked well. Speakers followed the format and generally successfully conveyed the important conclusions of their work. 

Conclusions that the organizers drew from this workshop included: 
\begin{itemize}
    \setlength\itemsep{-0.7pt}
    \item  There was a clear audience preference towards real-time interactivity rather than recordings.
    \item Free/open discussion was clearly more difficult in a virtual setting. Perceived higher barrier to speaking up and coffee/lunch/evening discussion time was lost
    \item \gls{Indico} timetables did not handle events without events every day very well. Days without contributions were shown in the agenda making it more difficult to find active days.
    \item It is important to engage, remember and thank the audience. Workshop photographs are one good approach (enabled by \gls{Zoom}).
    \item Often conferences and workshops are organized by a small team having limited experience as event organizers. For CTD, the organizers did not consider the potential impact of up-front cost commitments when planning CTD2020. As HEP starts to organize in-person or hybrid events, it is also a good opportunity to think about best-practices for in-person event organization. 
\end{itemize}
\subsection{LLVM Developers Meeting}
\label{sec:experiences_LLVM}
\subsubsection{Introduction}
The LLVM Project~\cite{llvm} is a collection of modular and reusable compiler and toolchain technologies. The LLVM Developers' Meeting is a bi-annual gathering of the entire LLVM Project community. The conference is organized by the LLVM Foundation and many volunteers within the LLVM community. Developers and users of LLVM, Clang, and related subprojects will enjoy attending interesting talks, impromptu discussions, and networking with the many members of our community. 

One of the event's main goals is to provide a "venue" where the geographically distributed developers community can interact and exchange ideas. The canonical event format includes:
\begin{itemize}
    \setlength\itemsep{-0.7pt}
    \item Technical Talks -- These 20 to 30 min talks cover all topics from core infrastructure talks, to project's using LLVM's infrastructure. Attendees will take away technical information that could be pertinent to their project or general interest.
    \item Tutorials -- Tutorials are 50 min sessions that dive down deep into a technical topic. Expect in-depth examples and explanations.
    \item Lightning Talks -- These are fast 5 min talks that provide a "taste" of a project or topic. Attendees will hear a wide range of topics.
    \item Panels -- Panel sessions are guided discussions about a specific topic. The panel consists of $\approx3$ developers who discuss a topic through prepared questions from a moderator. The audience is also given the opportunity to ask questions of the panel.
    \item Birds-of-a-Feather -- Large round table discussions with a more formal directed discussion.
    \item Student Research Competition -- Students present their research using LLVM or related subprojects. These are usually 20 min technical presentations with Q\&A. The audience will vote at the end for the winning presentation and paper.
    \item Poster Session -- An hour long session where selected posters are on display for attendees to ask questions and discuss.
    \item Round Table Discussions -- Informal and impromptu discussions on a specific topic. During the conference there are set time slots where groups can organize to discuss a problem or topic.
    \item Evening Reception -- After a full day of technical talks and discussions, attendees gather for an evening reception to continue conversations and meet other attendees.
\end{itemize}

The type of attendance is active project developers, novice and advanced users, students and researchers, programming language enthusiasts, and anybody interested in compilers. The usual in-person attendance is around 400 people.

\subsubsection{Event Format}

In 2020 due to the COVID-19 pandemic the event was fully virtual~\cite{llvm-event}. During its duration of three days there were keynote, technical talks, lightning talks, tutorials, poster sessions, bird-of-a-feather, round tables, student research competition (SRC), breaks, social event. The keynote, technical talks, lightning talks, tutorials and SRC were pre-recorded with a moderated live Q\&A session. The SRC also included voting for the three best talks. The talks relied on the \gls{Whova} app for content delivery and \gls{Zoom} for the live Q\&A. The rest used the \gls{Remo} platform which allowed attendees to virtually walk around and network.

The event was aided by a professional event organization team~\cite{bline}.

The registration deadline was one week before the event and then extended until the end of the event. There were two types of tickets: a free registration subsidized by sponsors; and a US\$50 supporter ticket which supports the LLVM Foundation activities. The total amount of registrants exceeded 800, but the real attendance was around 400.

The event had a code-of-conduct and special program encouraging diversity and inclusivity. The event organized branded merchandise such as T-shirts and other apparel.

\subsubsection{Interactions}

The conference materials were made available at the registration deadline -- a week before the event. The discussions happened in the \gls{Whova} platform as well as the live session questions. There were channels for public discussions. The talks were made available in \gls{YouTube}.

\subsubsection{Conclusion}

The LLVM Developers' Meeting spans at least at three continents which makes picking a convenient time for speakers and attendees very challenging. It also seemed difficult for many people to dedicate the necessary amount of time while being in their usual work/home environment. The virtual platforms served their purpose well on a technical level, however did not achieve the usual attendance at breaks and the social event.

The event happened relatively early in the pandemic when the experience organizing virtual events was not substantial. Virtual events pose specific set of challenges in terms of time zones, networking and technical solutions. Overall, the event achieved its mission due to the extra efforts by the organizers and the community.
\subsection{Snowmass Community Planning Meeting}
\label{sec:experiences_Snowmass}

\subsubsection{Origin of the Snowmass Community Planning Meeting}
\label{SCPM2020:sec:intro}
The Snowmass Planning process \cite{Snowmass} gathers the U.S. particle physics community to begin defining the most important questions in our field. Over the course of $\approx$1 year, various frontier working groups identify opportunities to address those questions and produce a report that is used as input for the Particle Physics Project Prioritization Panel (P5). The output of P5 defines the roadmap for the field over the subsequent $\approx$10 years. This exercise is repeated approximately every 7 to 8 years. A Snowmass Community Planning Meeting (CPM)~\cite{snowmassCPM} was scheduled to bring the community together and commence this cycle.

The nature of the meeting involves a significant amount of planning discussions, debate, and interactions with physicists from a wide set of working groups. The CPM represents the beginning of a process, rather than a summary or conclusion of work that has already been performed. There are nominally ten frontiers that reflect physics drivers as well as technology and infrastructure. Naturally, many proposed topics span multiple frontiers. Thus there are needs for both plenary lectures as well as parallel discussions.

The goals of the CPM were stated on the event page at~\cite{snowmassCPM}, and highlighted communication:
\begin{itemize}
    \item Inspire the community about the field, and encourage them to engage broadly in the Snowmass process\vspace{-7pt}
    \item Inform the community about plans from other regions and from related fields and planned Snowmass activities\vspace{-7pt}
    \item Listen to the community\vspace{-7pt}
    \item Provide space for members across the field to talk to each other and to discuss, promote, and develop new ideas\vspace{-7pt}
    \item Establish cross working-group connections and identify gaps
\end{itemize}

The 2013 Snowmass cycle began with a similar CPM meeting at Fermilab in the fall of 2012. This was a fully in-person meeting, and approximately 400 community members participated in the meeting. The 2020 CPM initially attempted to replicate that successful kickoff meeting, and aimed for a 2.5 day meeting in November of 2020. In March 2020, a local organizing committee (LOC) was formed to begin planning for the meeting. Throughout the spring of 2020, it became clear that the hopes of an in-person meeting were inconsistent with the reality imposed by COVID-19, and by June of 2020, the decision was made to move to a fully virtual meeting format. The LOC chosen for the in-person meeting continued and began the process of adapting for a fully virtual meeting.
 
\subsubsection{Meeting Organization/Technical Setup}
\label{SCPM2020:sec:technical}

As the CPM focus was on U.S. planning, the meeting times were prioritized to fall during the traditional 9~am to  5~pm workday for all continental U.S. time zones. Thus, the meetings started at 11~am US Central time and concluded by 4~pm US Central. To accommodate the broad program the community needed to discuss, the meeting was extended to four days. New dates were selected to avoid major conferences, holidays, and the US general election. The virtual Snowmass CPM was scheduled for October of 2020, and the meeting organization benefited from the conferences, meetings, and workshops that were forced into a virtual format on short notice during the spring and summer of 2020. Registration fees were eliminated to encourage participation and improve accessibility. 

The meeting structure consisted of two days of plenary meetings and two days of parallel sessions. The plenary agenda focused on interactions with the representatives of other global HEP programs, the views from funding agencies, and a community town hall to introduce new initiatives. On the final day, the plenary sessions focused on summaries from the frontier working groups and motivational visionary talks from experienced physicists. 

The plenary sessions utilized a \gls{Zoom} webinar with a capacity of 5000 participants. A decision was made to keep the security settings fairly strict to avoid unwanted incidents. Chat was disabled in the public-facing portion of the webinar, and the question \& answer feature was enabled for the participants to communicate with the speakers and moderators. To improve accessibility during the meeting, captioning services were procured from \gls{Ai-media} to provide real-time description of the presentations via the \gls{Zoom} application. To encourage participant engagement, multiple \gls{Slack} channels were created within the Snowmass \gls{Slack} space. These channels were communicated to the participants and provided means to contact the organizers, discuss the content of the plenary program and ask questions of the speakers.

Plenary sessions speakers were provided with unique "panelist" links to access the private side of the \gls{Zoom} webinar. In the weeks leading up to the meeting, the LOC held multiple rehearsals to practice the mechanics for the plenary sessions, such as webinar content projection, recording and the question \& answer interface. These rehearsals also provided speakers with opportunities to become familiar with the interface.  Additional technical and code-of-conduct training were required for all session chairs.

During the parallel sessions, up to 18 \gls{Zoom} "rooms" were provided in the format of a traditional \gls{Zoom} meeting. Since these were smaller groups and focused on discussion and interaction, the security settings enabled chat and participants could unmute themselves and share their video. Presenters were promoted to co-host to share their slides. Typically the program was scheduled such that each room contained similarly themed topics. This structure was chosen to prevent excessive fragmentation of the meeting, and to ensure participants could locate the relevant topics. Between sessions, the rooms were held open to provide space for carryover conversations and to encourage the development of new ideas and relationships. We note these rooms were often underutilized.  Each parallel session room had a designated "host" that focused on the \gls{Zoom} meeting logistics and security, while the physics content of the meeting was handled by the session conveners. 
 
The overall structure of the meeting and instructions on how to navigate the meeting was collected in a guide for participants, which was modeled on a similar conference guide provided for ICHEP 2020 (\S \ref{sec:experiences_ICHEP}). This included specific instructions for \gls{Zoom} hosts, conveners, moderators, and speakers, and was posted to the \gls{Indico} agenda.

\subsubsection{Statistics and lessons learned}
\label{SCPM2020:sec:statistics}
About 2000 people registered to attend the meeting prior to the start of the CPM. Registration was required to receive the link to the \gls{Zoom} webinar. By the final day of the meeting over 3000 people had registered. Statistics were collected on meeting participants from within the \gls{Zoom} software. Unique connections (identified by \gls{Zoom} display name) were tracked as a function of time, and the total integrated unique connections were recorded (see \href{https://gordonwatts.github.io/snowmass-cpm-attendance/}{here} for more information). 
\begin{figure}
     \centering
     \begin{subfigure}
         \centering
         \includegraphics[width=0.49\textwidth]{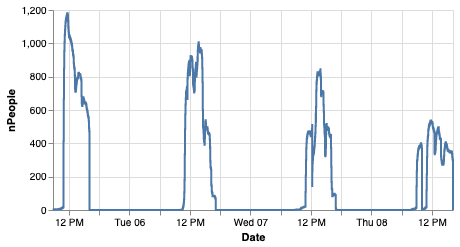}
     \end{subfigure}
     \begin{subfigure}
         \centering
         \includegraphics[width=0.49\textwidth]{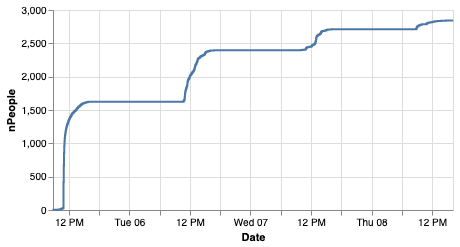}
     \end{subfigure}
        \caption{Snowmass CPM participation statistics as determined by self-identified \gls{Zoom} user display name. (a) Unique connections vs time (b) Integrated unique participations vs time}.
        \label{fig:cpm_stats}
\end{figure}

Since individuals have the freedom to enter different names into the \gls{Zoom} software, it is estimated that these statistics over-count the actual attendance by up to 10~\%.  

An examination of Figure~\ref{fig:cpm_stats} reveals several features associated with this meeting. First, more than 2500 unique individuals connected to the meeting at some point during the week. This represents about 85~\% of registrants. This attendance represents a factor of $\approx$6 times the participation of the in-person meeting in 2012. Removing the barriers of cost and ensuring the content was only a click away greatly enhanced the accessibility of the meeting.  

On the flip side, engagement was a challenge each day and over the course of the week. Attendance tended to peak early in each day and dropped off anywhere from 20~\% to 50~\% from the peak. Attendance also dwindled from a peak of nearly 1200 participants early on the first day to 400 at the conclusion of the meeting

Because security was prioritized during the plenary webinar sessions to avoid \gls{Zoom} "bombing" or other disruptive experiences, participants reported a lack of sense of community. They were unable to clearly see other participants and directly chat through the \gls{Zoom} interface. The \gls{Slack} channels were used at some level, but the traffic was relatively low. Many of the in-person brainstorming interactions were not replicated by this setup. An investigation into tools such as \gls{Slido} which allows an integrated, moderated wrapper to the \gls{Zoom} webinar is warranted for future efforts.

\subsection{OSG Virtual Meetings}
\label{sec:experiences_OSG}

\subsubsection{Introduction}
The OSG is a distributed organization that advances the state of the art of distributed High-Throughput Computing in the United States and worldwide.  With staff and users scattered around the globe, the OSG has relied on virtual meetings for a long time, especially for smaller groups and events.  However two longstanding and larger in-person kinds of events---the OSG All-Hands Meeting and the OSG User School---were significantly affected by the COVID-19 pandemic, and they are the focus of this section.

\subsubsection{OSG All-Hands Meeting}
The OSG All-Hands Meeting (AHM) is the premier annual meeting for the OSG, bringing together campuses, science users, site administrators, developers, and anyone else interested in the latest news and technologies.  Historically, the format has been a mix of conference presentations, workshop segments for discussion, training, and time for the OSG Council to meet.  In 2020, the usual in-person AHM had been scheduled for March.  In early spring, the then-nascent pandemic caused OSG management to postpone the meeting until September 2020, to see how things played out.  Of course, by autumn, travel and mass gatherings in the United States were largely not an option so the meeting was made virtual~\cite{osg2020}. And the regularly scheduled AHM in March 2021 was never even considered as an in-person event~\cite{osg2021}. Because the focus of the AHM had always been to bring the community together in person, using a virtual format was a significant change.

For the switch to an all-virtual format, the program changed in a few critical ways.  First, the schedule was extended to a whole week (as opposed to 3 and a half days) but with daily content reduced to a single track totalling 3 hours.  The net result of these changes was significantly fewer hours of presentations.  Further, workshop-like and training elements were dropped in favor of other, more focused meetings.  However, open discussion time was added to the end of each day in hopes of recapturing a bit of the feeling of being in a shared space.  Thus, the resulting daily schedule was 1.5~hours of talks, 1~hour of break, 1.5~hours of talks, and then 1~hour of discussion time, for a total of 5~hours of allocated time each day.  Especially for the AHM 2021, the talks for each day were organized around a topic or two with a specific audience in mind; that way, attendees with specific interests could focus on certain days.  There were usually 3 to 5 talks per session, so about 10~min to 20~min apiece with a bit of time for questions during the transition times.

\begin{figure}[t!]
  \centering
  \includegraphics[width=\linewidth]{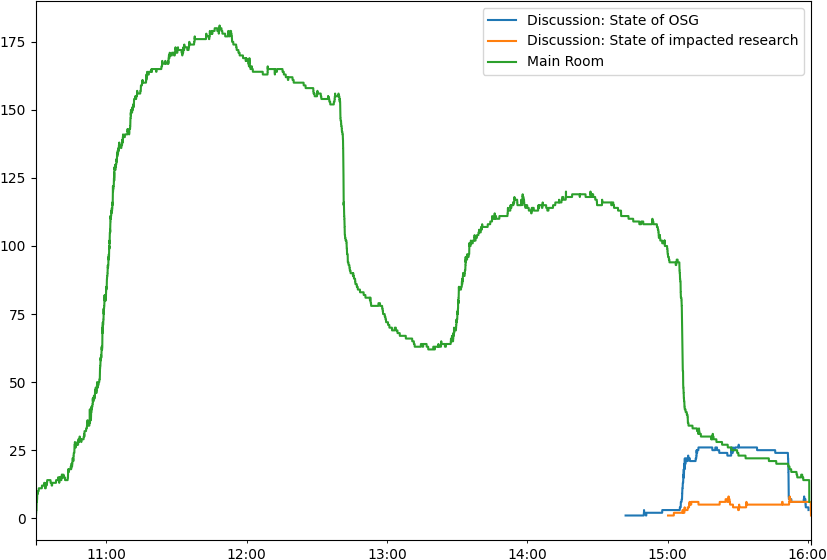}
  \caption{Sample of OSG All-Hands Meeting 2021 live attendance.}
  \label{fig:osgahm21}
\end{figure}

The main technologies used for the AHMs were the Fermi National Accelerator Laboratory instance of the \gls{Indico} scheduling system and \gls{Zoom} for the meetings themselves.  \gls{Indico} has been used for a long time in the OSG community, so that was nothing new. \gls{Zoom} was chosen due to its already ubiquitous nature in March 2020---most people in the community already had and used it, plus several of the institutions containing OSG staff had institutional access which made planning easy.  Each day, one main \gls{Zoom} meeting was home to the talk sessions, the break between, and one of the discussion areas afterward (dubbed ``the hallway''); separate \gls{Zoom} meetings existed for the one or two parallel, topical discussion areas.  Registration in \gls{Indico} was required and the \gls{Zoom} links with embedded passwords were emailed to all registered participants daily.  OSG staff had roles throughout the event, with a session moderator on duty at all times, helpers during sessions to deal with \gls{Zoom} issues and chat questions, and so on.  There was also a \gls{Slack} workspace channel for the meeting, but it was used lightly.
Looking at participants for the AHM 2021, a total of 336 people registered.  By analyzing detailed \gls{Zoom} logs, it seems that just under 300 unique attendees joined at least once during the week, which is easily double the participation of any prior in-person OSG AHM.  However, peak attendance during any day or session was about half of the total registration count.  Thus, it can be misleading to look just at peak attendance statistics, which are easy to see while the meeting is in progress.  From the peaks, one might conclude that a significant fraction of registrants did not attend, when in fact over 85~\% of registrants attended at some point.  Attendance varied considerably by topic, as might be expected, and discussion times were attended lightly.  Figure~\ref{fig:osgahm21} shows an example of OSG AHM 2021 live attendance numbers.

In sum, both virtual OSG AHMs went smoothly and participants who provided feedback (there was no post-event survey) seemed happy.  Having fewer total hours of content meant that the program committee had to focus carefully on invited talks and the structure of sessions and days, and this extra focus seemed to result in a high-quality program that appealed to many.  A real benefit of having virtual meetings was people from around the world who could not attend in person, due to travel constraints, were able to participate.  Attendees came and went as schedules allowed and topics appealed, and it seemed that the hour-long break between sessions helped some people fit the AHM into their busy schedules.  Discussion times were lively and mostly on topic for those few who joined.  Of course, there are always ways to improve!  By far the biggest request was to record and post videos of the talks, which will be done for future virtual events.  And the staff identified a number of smaller ways in which to try to engage the community more and keep participants interested in the content longer.

\subsubsection{OSG User School}
The OSG User School is by far the largest and most important training event that OSG offers to its community of researchers around the world.  For 10 years prior to the pandemic, the School was run at the University of Wisconsin--Madison as a week-long in-person synchronous training program.  In many respects, this event was like the Moriond conference described in Section~\ref{sec:experiences_Moriond}: The focus was on camaraderie and the shared, intensive learning experience, with everyone from advanced undergraduates to graduate students, post-docs, staff, and faculty all treated the same and working together. In 2020, the pandemic began just as the application period for the School was ending, throwing all plans for the year into disarray.

After much consideration, the plan for 2020 evolved into an offering with a curriculum similar to but pared back from previous in-person events.  Then to take advantage of some benefits of remote learning, the OSG Virtual School Pilot 2020, as it became known, added a strong focus on personalized learning and one-on-one and small-group learning and consulting opportunities.  The primary goal for participants morphed into a very personal one: Learn enough about distributed High-Throughput Computing to apply it to at least one research project, and have that project running by the end of the event.  Of the accepted applicants, twenty chose to participate in the re-imagined Pilot~\cite{osgv2020}.

For technology, the venue for lectures was \gls{BlackboardCollaborate} and then \gls{Zoom}, \gls{Slack}, and regular email were used heavily for individual and small-group interactions.  For each lecture day---and there were only five such days---live lectures occurred twice each day to provide flexibility in participant scheduling; lectures were recorded and posted after the event.  Exercises were posted on the public website, and all participants were given logins to Access Points at the University of Wisconsin--Madison and OSG Connect.

Overall, the Pilot School went very well and most participants reached their individual goals for the event.  The staff learned a great deal, too, and so the OSG Virtual School 2021 will follow a similar pattern bit with improvements throughout~\cite{osgv2021}. Nonetheless, the OSG User School will return to its in-person format as soon as possible.

\subsubsection{General Observations}
Based on OSG's experience with virtual events in 2020 and 2021, there are a few observations to make about such events and about what the future might hold.

As others in this report have noted, virtual conferences and open meetings have a number of advantages and disadvantages.  The primary advantage is the opportunity to reach more members of the community who would otherwise be unable to attend due to time constraints, travel costs (financial and other), and so forth.  Both OSG All-Hands Meetings had far more total unique participants---double or more---than previous in-person events.  Accordingly, without travel and other in-person costs, OSG was able to offer its events without charge, expanding access and simplifying logistics.

There are many disadvantages, all fairly obvious at this point: lack of in-person connections and interactions, time zone disparities that make attendance difficult for some, a greater sense of fatigue that necessarily shortens events, and seemingly reduced engagement due in part to still being ``in the office'' and the ease of leaving and rejoining at will.  Nonetheless, perhaps because of the pan- in pandemic, everyone seemed to understand the limitations and did their best to engage and enjoy.

One might be tempted to say that an all-virtual event is easier to organize than an in-person one.  However, the OSG experience was that the effort savings from having no travel, accommodation, food, and venue logistics were largely offset by effort expenditures on preparing virtual venues, documenting and practicing key procedures and back-up plans, preparing staff, and so forth.  Simply put, it takes a great deal of effort to run a successful virtual event.  That being said, preparations for a virtual event need not start as far in advance (by calendar days) as for an in-person one.

Another lesson learned was that it can be useful to completely rethink the approach to and goals of an event when it must become virtual.  For OSG, this was especially true of the User School, where the in-person approach simply did not translate into a virtual one.  But even for the All-Hands Meetings, it was expected that interactions among participants would be severely reduced and hence the common AHM elements that depended on it---such as training events---were best omitted and handled via other, focused events.

Looking forward, the next great challenge will be the hybrid event.  While no specific plans are in place yet, it is generally accepted within OSG that future AHMs will have to include a significant online component.  What does such an event look like and how will it work?  The key will be to identify the elements of virtual events that worked the best and preserve those.  For example, it may be good to keep the condensed, single-track schedule of the virtual AHMs, even for the in-person participants, thereby acknowledging the fatiguing, come-and-go nature of virtual events.  The extra time made available by a condensed schedule could be used for in-person activities that emphasize the benefits of being together for those who choose to do so---that is, more workshops, training events, discussions, and loosely structured collaborative work time.  Much work remains to be done in this space.  Given the uncertainty of the future, it might even be helpful to have another workshop in a couple years to discuss hybrid meetings!

\subsection{PyHEP 2020 Experience and 2021 Plans}
\label{sec:experiences_PyHEP}

\subsubsection{Origin and format of the PyHEP series of workshops}
\label{PyHEP:sec:intro}

The PyHEP, "Python in HEP", workshop series started in 2018, recognizing the increasing importance of Python in Particle Physics. The workshops are organised as part of the activities of the PyHEP Working Group of the HSF, the HEP Software Foundation~\cite{PyHEP:WG}. Under the support of the HSF, the aim from the onset has always been to provide an informal environment to discuss and promote the usage of Python in the Particle Physics community at large.
Furthermore, diversity and inclusion aspects have also been taken seriously into consideration, in terms of the set of participating communities, and cultural backgrounds, gender, ethnicity, disability, sexual orientation of participants.

The first workshop, PyHEP 2018, was run as a pre-CHEP 2018 conference event, profiting from the presence of a large community in Sofia, Bulgaria. Slightly over 10~\% of the CHEP 2018 attendees participated in PyHEP 2018.

The workshop format remained unchanged until the time of the COVID-19 pandemic, that is with an in-person format, given the spirit and goals of PyHEP:
\begin{itemize}
    \item Only plenary, topical, sessions\vspace{-7pt}
    \item Bring together users and developers\vspace{-7pt}
    \item Very informal, with significant time for (lively) discussions\vspace{-7pt}
    \item Educative, not just informative\vspace{-7pt}
    \item A mix of keynote presentations, tutorials and "standard" 20 + 10 minute presentations
\end{itemize}

The COVID-19 pandemic constrained the workshop organization into a decision to either cancel the event or run it fully virtually. With a few months ahead of us we opted for a virtual event and we learned a great deal from this experience.

\subsubsection{PyHEP 2020 organisation and running}
\label{PyHEP:sec:pyhep2020}

To adapt to the requirements and constraints of a virtual event, the duration of the workshop was extended from 2.5 days as in the 2019 in-person format to 5 days, with shorter sessions. Because of the global nature of the event we organized the sessions in two different times zones -- a "Europe-friendly" session and an "Americas-friendly" session. The former was by default 3-hour long whereas the latter only lasted 1 to 1.5 hours. We had approximately a third of the time devoted to tutorials and two thirds devoted to standard presentations.

PyHEP 2020 became a truly global event with participants from all over the world. Without travel and/or budget constraints, and no registration fees,
the level of interest increased significantly, from about 50 to 70 participants in in-person events to an incredible 1000 registrations, which we had to limit to for technical reasons related to the video conferencing system.
Understandably, we observed a significant increase in the fraction of students registered, including undergraduates. In fact, almost 40~\% of the participants were not members of an experiment or a collaboration (e.g. students and theory, simulation or instrumentation colleagues).

The workshops try to timely address topics in the spotlight as well as important topics specific to Particle Physics. PyHEP 2020 had sessions on the following topics: analysis fundamentals, analysis platforms and systems, automatic differentiation, performance, fitting and statistics, and the HEP analysis ecosystem.

We strongly encouraged all presentations to be prepared as (\gls{Jupyter}) \textit{notebook presentations}, with all material made publicly available on \gls{GitHub}.
To enhance the interactive experience we also encouraged the preparation of a "Binder launch button" so that any participant could follow and experiment in real time the notebooks being presented by simply launching on the browser the material.
To ensure a smooth run we used both the Binder Federation and the CERN \gls{BinderHub} resources (for those with CERN accounts),
and made sure that resources on Binder were (kindly) allocated by the Binder Team for the relevant repositories at the time of the presentations.
All material got posted onto the workshop agenda, including slides, \gls{GitHub} repository links, and the links to the recordings. The latter were captioned -- thanks to our sponsors -- and uploaded on to a dedicated playlist of the HSF channel on \gls{YouTube}. 

As mentioned above, PyHEP workshops allow for a fair fraction of time for discussion, which is paramount. The online platform \gls{Slido} was used to crowdsource questions from the audience: via a web page each participant has the opportunity to post a question, even anonymously, and also "upvote" or "downvote" any question so that, effectively, by the beginning of the Q\&A session, the session chair sees a prioritising set of questions that got selected as the most popular and interesting ones.
At the end of the Q\&A sessions all questions got copied to \texttt{Slack} in the appropriate topical channel, where speaker and participants could continue to discuss and exchange. A few polls were also run via \gls{Slido} as a fun means to socialize. The use of \gls{Slido} turned out to work extremely well; about 40~\% of the participants joined and used the platform at least once.

\subsubsection{PyHEP 2021 planning}
\label{PyHEP:sec:pyhep2021}

The PyHEP 2021 workshop~\cite{PyHEP2021} will again be held as a virtual event, in early July. At the time of writing we have approximately 950 registrations, which shows the continuous interest from a large fraction of the community in these series of workshops.

Unlike last year's edition, this year's will not be run in two different time zones but nominally between 14:00 and 17:00 CEST, a time slot that has proven appropriate, or at least acceptable, to a large fraction of the attendees. The format is otherwise largely kept the same, though live streaming to \gls{YouTube} will be experimented for the first time atop the set-up of the usual video conferencing room (with \gls{Zoom}).

\subsubsection{Lessons learned}
\label{PyHEP:sec:lessons}

As expected, virtual events are far more inclusive
and truly global, with participation from all over the world, even if time zone constraints imply that some participants are either attending (very) early in the day or rather late in the day. Workshops such as PyHEP, which are meant to foster exchanges among experts and learners, with plenty of time devoted to discussions, are nevertheless made more challenging when running virtually, though the organisation burden is less.
In the future we may consider alternating the workshops as virtual/hybrid and in-person.
\subsection{SciPy 2020 Experience and 2021 Plans}
\label{sec:experiences_SciPy}

SciPy, the Scientific Computing with Python Conference, is a community conference dedicated to the advancement of scientific computing through open source Python software for mathematics, science, and engineering. The annual SciPy Conference allows participants from all types of organizations to showcase their latest projects, learn from skilled users and developers, and collaborate on code development.

The first SciPy meeting was held at CalTech in 2001 for a few dozen attendees. By 2009, attendance had reached 150 and, exceeding the capacity of the CalTech facilities, the meeting moved to the campus of UT-Austin in 2010. The most recent in-person meeting had roughly 800 registrants.

\subsubsection{Typical in-person SciPy meeting}

A SciPy meeting is organized around three main elements:
\begin{description}
    \item[Tutorials] half- and full-day, interactive classroom setting. Instructors propose topics, and the Tutorials Committee chooses those that make the most compelling case for their pedagocial approach and appeal to the widest audience of SciPy attendees.
    
    \item[Conference] single- and multi-track sessions for audiences ranging from small groups to all attendees.
    
    \begin{description}
        \item[Keynotes] 45 min to 1 h plenary talks, typically one focused on a significant scientific advance enabled by Python, one focusing on the ecosystem of scientific python packages, and one addressing one of the broad themes of that year's meeting, not necessarily from a Python perspective. In recent years, we have also held a keynote on diversity, equity, and inclusion.
        \item[Themes and Minisymposia] 30 min technical talks in parallel sessions. Themes are two or three cross-cutting sessions identified for each year. Minisymposia are topical sessions organized by different scientific communities.
        \item[Birds of a Feather (BoF) Sessions] self-organized sessions where communities of interest can meet, talk, and plan.
        \item[Lightning Talks] 5 min talks (no going over!) for an hour every afternoon on late-breaking news and whimsical projects.
        \item[Posters] traditional static and semi-interactive presentation \emph{en masse}.
    \end{description}
    \item[Sprints] self-organized, intense development of Python packages in the SciPy ecosystem.
\end{description}
In a typical year, tutorials run in two or three parallel sessions on Monday and Tuesday; the conference is held Wednesday through Friday, and sprints run through the night from Saturday to Sunday.

\subsubsection{Virtual SciPy 2020}

\begin{figure}[ht!]
  \includegraphics[width=\linewidth]{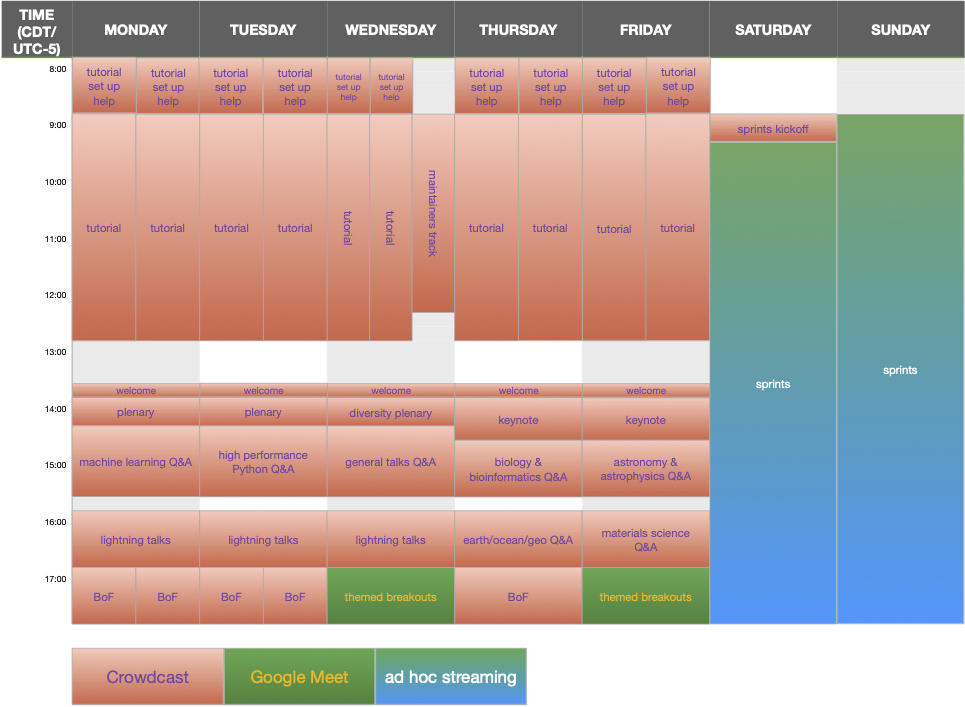}
  \caption{Schedule for SciPy 2020, showing virtual platforms used for different portions.}
  \label{fig:scipy_2020_sessions}
\end{figure}

After briefly debating delaying the meeting, SciPy 2020 was held, as scheduled, from July~6 to 12, 2020. In the transition to a virtual format, we kept the key elements of a traditional meeting, but rearranged the schedule, shown in Fig.~\ref{fig:scipy_2020_sessions}, 
to hold live tutorials in the mornings, followed by live keynotes, and then live question \& answer sessions, lightning talks, and BoFs in the afternoons, all over the course of five days. Sprints were held after the conference, on Saturday and Sunday, as usual. For this first, purely virtual meeting, we reduced registration fees, inclusive of tutorials, to US\$75 standard/US\$25 student (fees for in-person SciPy 2019 were US\$500 standard/US\$225 student, with the same charges again for tutorial attendees). As SciPy traditionally draws a North American audience, meeting times were kept in Central Daylight Time (there is a long history of other regional meetings, e.g., EuroScipy, SciPy India, SciPy Latin America, and SciPy Japan).

SciPy talks have been recorded for many years and distributed via \gls{YouTube} since 2013 \cite{scipy-youtube}; the difference for 2020 being that Theme and Minisymposia talks were pre-recorded and made available via the SciPy \gls{YouTube} channel several days in advance; speakers were given a few minutes to recapitulate their talk at the beginning of their respective question \& answer session. Tutorials and Birds of a Feather were held in two parallel sessions; all other elements were held sequentially. Poster authors were responsible for hosting their own content asynchronously. After an initial kickoff using the \gls{Crowdcast} conference platform, individual sprint leaders set up their preferred mode of communicating and streaming with their team. 

Tutorials and live sessions were hosted on the \gls{Crowdcast} platform. This platform afforded four "hosts" to accommodate speakers and session chairs. For conventional sessions, this worked well, but for non-traditional elements like BoFs and lightning talks, with the potential for many speakers in quick succession, more hosts would have been beneficial. 

SciPy meetings have had dedicated \gls{Slack} workspaces for many years, with both official and self-organized channels. The \gls{Crowdcast} platform included a chat feature that was used heavily, but it did not offer any sort of direct messaging capability and it was challenging to correlate participants between \gls{Crowdcast} and \gls{Slack}.

Due to the limited number of channels and hosts per channel allowed by \gls{Crowdcast}, informal themed sessions were planned for \gls{Zoom} breakout rooms; due to technical difficulties, these were switched to \gls{GoogleMeet} at the last minute.

More than 1400 attended at least one live session and more than 1200 attended at least one tutorial. This substantial increase can likely be attributed to attendees not needing to travel and to the significantly reduced registration fees. Many attendees from overseas emphasized that the virtual format was what enabled them to attend and they expressed the hope that SciPy would retain a virtual component in the future.

\subsubsection{Plans for Virtual SciPy 2021}

With more time to plan and a better idea what to expect than in 2020, we strove to recapture as much of the feeling of the in-person meetings as possible. To achieve this,
\begin{itemize}
\item we returned to a conventional schedule, illustrated in Fig.~\ref{fig:scipy_2021_sessions}, of 2 days of tutorials, 3 days of conference, and 2 days of sprints.
\begin{figure}[ht!]
  \includegraphics[width=\linewidth]{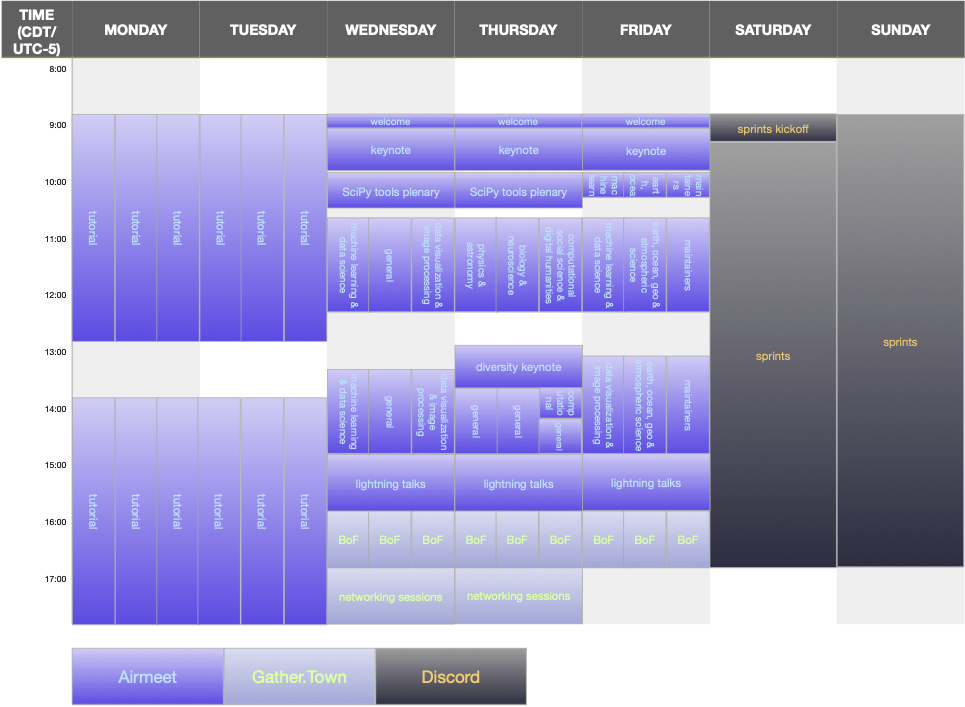}
  \caption{Schedule for SciPy 2021, showing virtual platforms used for different portions.}
  \label{fig:scipy_2021_sessions}
\end{figure}

\item all talks will be live. Talks will still be recorded and made available afterwards on \gls{YouTube}, as they have been for more than a decade, but we will not be posting pre-recorded talks.

\item we are changing the primary meeting platform to \gls{Airmeet}. This change was motivated by scaling considerations, as the new platform allows up to 10 concurrent sessions with 10 presenters, each. A "green room" for presenters to prepare before going live and live technical support were additional attractions of this platform.

\item we will host the more interactive/social elements, e.g., BoFs, posters, and general networking on the \gls{Gather.Town} avatar-based virtual meeting space. This will enable us to have both large and small meeting spaces, as well as a nostalgic recreation of SciPy's original home at CalTech.

\item sprints will be hosted on \gls{Discord}. This platform was favored for collaborative coding due to its combination of video streaming, screen sharing, and chatting. Sprint organizers are free to choose a different platform if it works better for their team.

\item we will again have a \gls{Slack} workspace for SciPy 2021, in addition to chat features in \gls{Airmeet}, \gls{Gather.Town}, and \gls{Discord}.

\item we increased registration to US\$125 standard/US\$50 student in order to support the improved attendee experience via a wider range of platforms this year. These rates are still well below the in-person rates, as well as the rates charged by many domain-specific society virtual conferences. Some sponsor funds are available to support attendees for whom these rates are a hardship.

\end{itemize}

SciPy 2021 is scheduled for July~12 to 18, 2021.

\subsubsection{Lessons learned}

\begin{itemize}
    \item Many more people could attend a virtual meeting who would otherwise be limited by time, cost, travel policies, or family constraints. Several appreciated the reduced environmental impact of virtual attendance. Overseas attendance was up significantly.
    
    \item Conversely, work and family tend to be less forgiving for those who are not really ``away'' at a meeting.
    
    \item Many people miss the informal and social interactions of the traditional in-person meeting.
    
    \item Pre-recorded talks need to be ``dropped'' far enough in advance for people to view them prior to any live discussion/Q\&A session. For attendees already experiencing "Zoom-fatigue", the prospect of spending all night watching videos as soon as the live meeting is over can be exhausting.
    
    \item Interactive, dynamic, or social elements of a meeting either require more people with "hosting" privileges or a platform that allows all participants to engage equally.
    
    \item Quasi-anonymity makes managing conduct issues a challenge. Lack of direct-messaging capability on some platforms and correlating identities between different platforms can make it difficult to address concerns in private.
\end{itemize}

\subsection{US ATLAS / Canada ATLAS Computing Bootcamp}
\label{sec:experiences_Bootcamp}

\subsubsection{Introduction}\label{subsec:computing_bootcamp_intro}

The US-ATLAS Computing Bootcamp is an ongoing annual bootcamp designed to educate newcomers to the ATLAS Collaboration --- in particular graduate students --- on the core technical computing concepts and tools that are used throughout ATLAS and the broader HEP community.
The goal of the bootcamp is to have all students have basic proficiency in the computing tool-chain for a typical ATLAS analysis and an understanding of how to use them. The scope and format of the bootcamp is modeled off of the widespread software workshops by The Carpentries and the training formats of the High Energy Physics Software Foundation (HSF). A similar workshop was planned by Canada ATLAS in 2020, but given the pandemic was canceled. The Canada ATLAS instructor team contained guest instructors from US-ATLAS, and so the Canada ATLAS instructors joined with the US-ATLAS instructor team to form a joint bootcamp that allowed for a total of 44 students from both countries to participate.

\noindent The 2020 Bootcamp covered:
\begin{itemize}
    \setlength\itemsep{-0.7pt}
    \item Version control with \texttt{Git}
    \item Use of CERN’s Enterprise Edition version of \texttt{GitLab}
    \item Building software with \texttt{CMake} and ATLAS \texttt{CMake}
    \item Containerization with \texttt{Docker}
    \item Continuous integration of code and delivery of analysis software with \texttt{GitLab} Pipelines
    \item Analysis preservation and reinterpretation with \texttt{RECAST}
    \item Fundamentals of machine learning
    \item Sustainable software development
\end{itemize}
and was held entirely online with all the resources being publicly available from the bootcamp website~\cite{USATLAS-Bootcamp:2020-Indico,USATLAS-Bootcamp:2020-website}.

\subsubsection{Bootcamp Format}\label{subsec:computing_bootcamp_format}

\begin{figure}[t!]
    \centering
    \includegraphics[width=\linewidth]{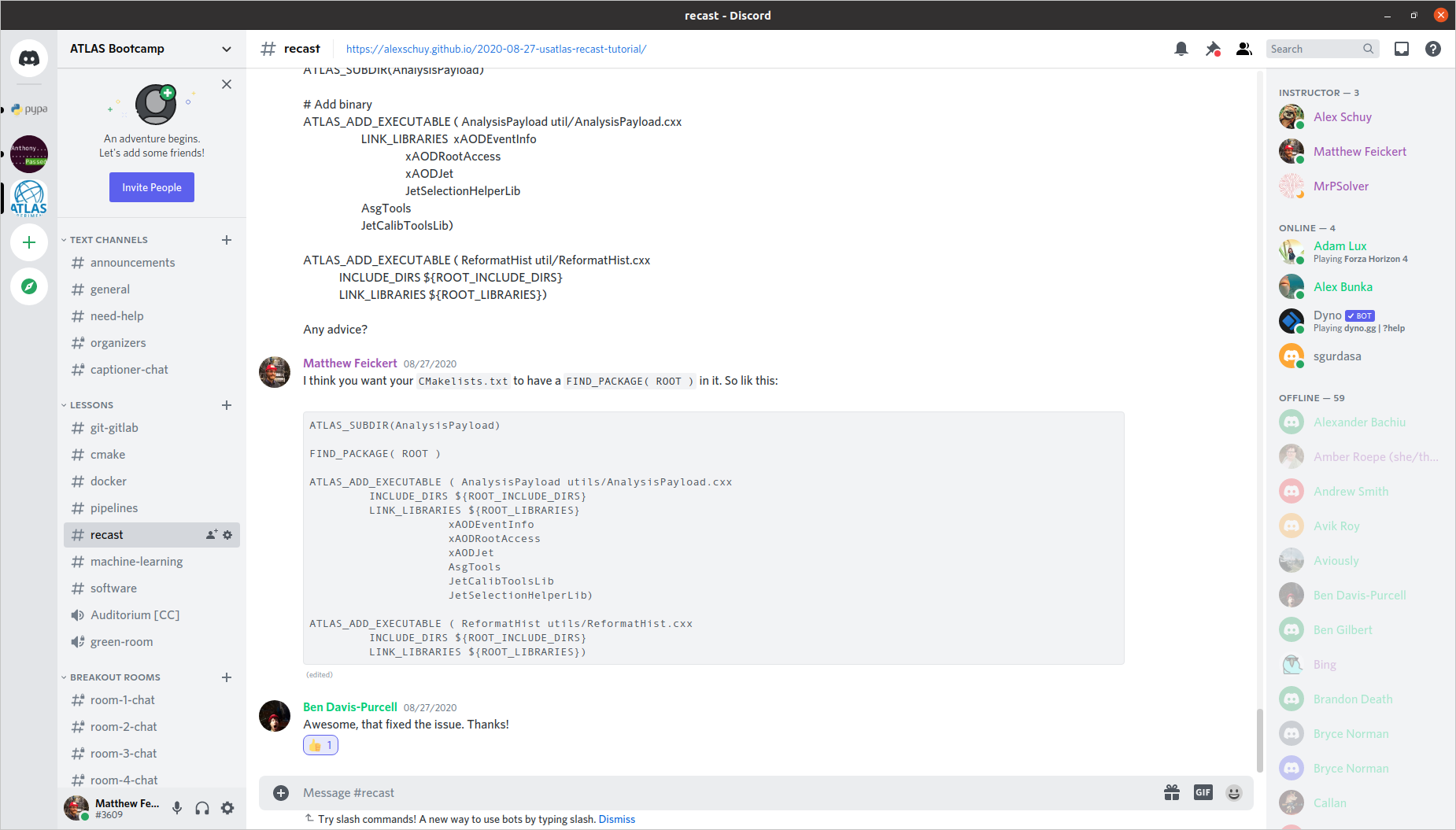}
    \caption{View of the US-ATLAS Computing Bootcamp \gls{Discord} server where students and instructors are interacting on questions related to ATLAS \texttt{CMake} and \texttt{RECAST}.}
    \label{fig:discord_server}
\end{figure}

In addition to having all of the bootcamp materials be publicly available online, the 2020 bootcamp had all instruction and interactions occur through a \gls{Discord} server that was setup for the bootcamp, seen in Figure~\ref{fig:discord_server}. \gls{Discord} offered a unified platform for instruction, discussion, and problem solving across text, audio, and video communication.
This makes for a very effective platform for a teaching environment like a bootcamp, where the video feed of the instructor team and technical discussion are able to coexist on a single platform.
Additionally, unlike \gls{Zoom}, the text support for \gls{Discord} supports a flavor of Markdown that provides syntax highlighting, which is useful tool when trying to communicate subtle differences in code that could otherwise be difficult to see. An example of this is also seen in Figure~\ref{fig:discord_server} where a student has posted code with no syntax highlighting at the top of the screen and an instructor has replied with syntax highlighting enabled. Syntax highlighting exists across many communication platforms, like \gls{Mattermost} and \gls{Slack}, but the advantage of it in \gls{Discord} is the ability to have it integrated into live discussions with video.

Another strong advantage of \gls{Discord} is the ability to move seamlessly between existing video and audio channels or ``rooms``. This is critically useful when teaching, as it allows for a student in the main instructional video stream to signal an instructor that they need assistance, have an instructor and the student move to a breakout room with audio/video and text support to iterate on the student problem, and then move back to the main stream without needing to request permissions or support from an administrator.
Lowering the barrier for students to interact with an instructor and get help without disrupting the main discussion improves the learning experience, and is similar to the kind of interaction an instructor and student would have at an in-person bootcamp.

\begin{figure}[t!]
    \centering
    \includegraphics[width=0.4\linewidth]{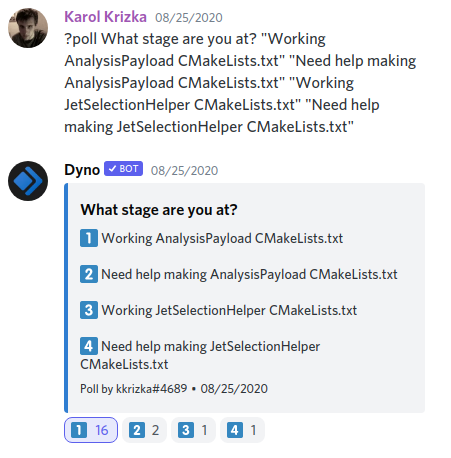}
    \caption{A member of the instructor team uses Dyno's markup syntax to generate an interactive poll to gauge student progress.}
    \label{fig:dyno_poll}
\end{figure}

\gls{Discord} is designed to be extensible and has a rich plugin and ``bot'' ecosystem for adding support and services to the platform. For the 2020 bootcamp, the instructional team enabled the "Dyno bot" which is provides user customizable dashboarding, moderation, and interaction tools. For example, as seen in Figure~\ref{fig:dyno_poll}, an instructor is able to quickly generate an interactive poll in the text channel for the main video room by using Dyno's markup syntax.
The poll dynamically updates and allows for users to change their response through time, allowing for instructors to gauge real time progress of the entire class and be able to direct instructor time to students that have questions or are behind without having to single individuals out. While they were not needed or used for the bootcamp, \gls{Discord} has a robust selection of extensible moderation tools that have been well vetted and improved upon given \gls{Discord}'s use as a community interaction platform.

Areas of potential concern or consideration regarding \gls{Discord} do exist. \gls{Discord} is publicly available, but has additional features as paid services.
The free tier of the service has limited video streaming quality, though for a trivial price the bootcamp was able to purchase a bandwidth extension that delivered video quality that was as good, if not superior, to \gls{Zoom}'s. Similarly, the free tier has a limited number of simultaneous video streams per room, which mean that only one speaker in the general ``auditorium'' room could share their video at a time, or in a student and instructor breakout room. Additionally, in-platform video recording is not supported unless users are using Open Broadcaster Software (OBS), which can be problematic if users are not experienced with setting up video recording software. Self hosting of the servers for \gls{Discord} is also currently not supported, which could be problematic for use in some countries given the Mozilla Foundation rates \gls{Discord}'s data privacy practices as not being particularly robust~\cite{Mozilla-Foundation:Discord}.

\subsubsection{Summary}\label{subsec:computing_bootcamp_summary}

The use of a \gls{Discord} server for the 2020 US-ATLAS and Canada ATLAS Computing Bootcamp had a large positive impact on the success of the bootcamp. In addition to creating a centralized virtual location for all discussion and instruction across text, audio, and video, the lower barrier for rapid interaction between students and instructors when students had questions was critical to students learning. Having a centralized location for all interactions also kept the bootcamp focused and avoided overwhelming participants with application fatigue. The server is persistent as well, so the discussions and resources added to the \gls{Discord} server are still useful references at the time of writing in 2021.

It is worth noting though that the use of \gls{Discord} for bootcamps or workshops is not contingent of the events being fully remote. It is possible to still have student peer interactions and discussions that leverage the \gls{Discord} server's strengths, and possibly serve as a bridge in hybrid in-person and remote workshops between the remote and local participants. While the experience of debugging code with a student while physically next to them is preferable and not replicated in a fully remote environment, the shared rooms and tools on the \gls{Discord} server made the experience significantly better than attempting to have the interactions be over a text chat only.

\subsection{Neutrino 2020}
\label{sec:experiences_Neutrino}

\subsubsection{Neutrino 2020 Virtual Conference}

The Neutrino conference series is the largest conference in neutrino physics. It is an international conference which has been hosted around the world every other year since 1972. The Neutrino 2020~\cite{Neutrino2020} conference was slated to be held in the U.S. in June 2020 and planning had begun for a 5-day in-person conference in downtown Chicago, Illinois. In March of that year, given the realities of the COVID-19 pandemic, the conference was completely retooled to an online conference. The Neutrino 2020 team, including 3 conference co-chairs, 11 local organizing committee members, 5 session chairs, and additional support from Fermilab staff had less than 3 months to prepare for a virtual conference. The conference was hosted online by Fermilab and the University of Minnesota on eight 1/2 days over the course of two weeks from June 22 - July 2, 2020. Conference registration was handled via Fermilab \gls{Indico}. Conference content was archived both on the Neutrino 2020 website~\cite{Neutrino2020} and using \gls{Zenodo}~\cite{nu2020-zenodo}.

\begin{figure}[ht!]
    \centering
    \includegraphics[width=0.4\linewidth]{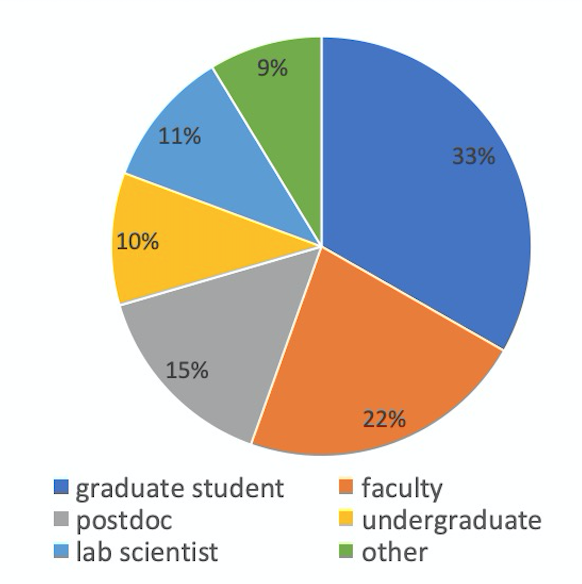}
    \caption{Demographics of Neutrino 2020 conference participants by career stage. More than half of the participants were students and postdocs. The "other" category includes the general public, members of the press, etc.}
    \label{fig:nu2020-participants}
\end{figure}

Neutrino 2020 was an unexpectedly large success. It brought a very large number of neutrino physicists together during a challenging time. While recent in-person Neutrino conferences have typically had roughly 600 to 900 participants in attendance, the Neutrino 2020 conference had record attendance with 4\,350 people from every continent, including Antarctica, participate in the online conference. The conference reached participants from 67 countries, roughly 60~\% of whom were students or post-docs (Figure~\ref{fig:nu2020-participants}), thus allowing access to a much broader and more diverse set of conference participants.

\subsubsection{Technical Aspects}

The Neutrino 2020 conference included four main components: 

\begin{itemize}
    \item {\bf Plenary talks} were broadcast live via a \gls{Zoom} webinar Monday through Thursday from 7:00 to 11:30 CDT across the two weeks of the online conference. Conference participants were given Fridays and the weekends off during the event. During the conference, 79 plenary talks were delivered by neutrino physicists around the globe. These talks were recorded live and posted daily~\cite{nu2020-talks} so that conference participants who were not able to attend the live presentation sessions could view the talks after they were presented and still be a part of online discussions of their content over \gls{Slack}. There were over 62\,000 visits to the Neutrino 2020 web page to view recorded talks during the conference. Increased support provided by Fermilab was needed to handle the web load.
    \item A web portal~\cite{nu2020-posters} was created to host the conference {\bf poster session}. This web portal was custom built for the conference to host the 532 conference posters (a Neutrino conference series record) as well as an optional 2 minute pre-recorded summary video that was uploaded by poster presenters. This too was popular. There were over 5\,800 views of the posters on \gls{YouTube} throughout the duration of the conference. The web portal also offered an alternative to the virtual reality in case participants were not able to attend the live poster sessions in the virtual reality platform.
    \item Conference interactions took place via \gls{Slack}. During the conference, over 23\,000 communications were posted on \gls{Slack}. Separate \gls{Slack} channels for each of the conference plenary sessions were created in advance as well as a channel for interacting directly with the conference organizers. Conference participants were also allowed to create their own channels. An especially popular channel, for example, was a Jobs Board on \gls{Slack} advertising job openings in neutrino physics.
    \item Perhaps the most popular aspect of Neutrino 2020 was an innovative {\bf Virtual Reality} platform~\cite{nu2020-vr} that was used to host the three live poster sessions during the conference. Over 3\,400 participants visited the virtual reality platform that included $>500$ conference posters in themed rooms.
    See section~\ref{sec:nu2020-vr} for more on the Neutrino 2020 Virtual Reality portal.
\end{itemize}

The Neutrino 2020 conference was promoted daily on social media via both \gls{Facebook} and \gls{Twitter} and conference organizers also hosted a public event on the last day of the conference in the form of an online live Physics Slam~\cite{nu2020-physics-slam}.

\subsubsection{Virtual Reality at Neutrino 2020}
\label{sec:nu2020-vr}

\begin{figure}[h]
    \centering
    \includegraphics[width=0.8\linewidth]{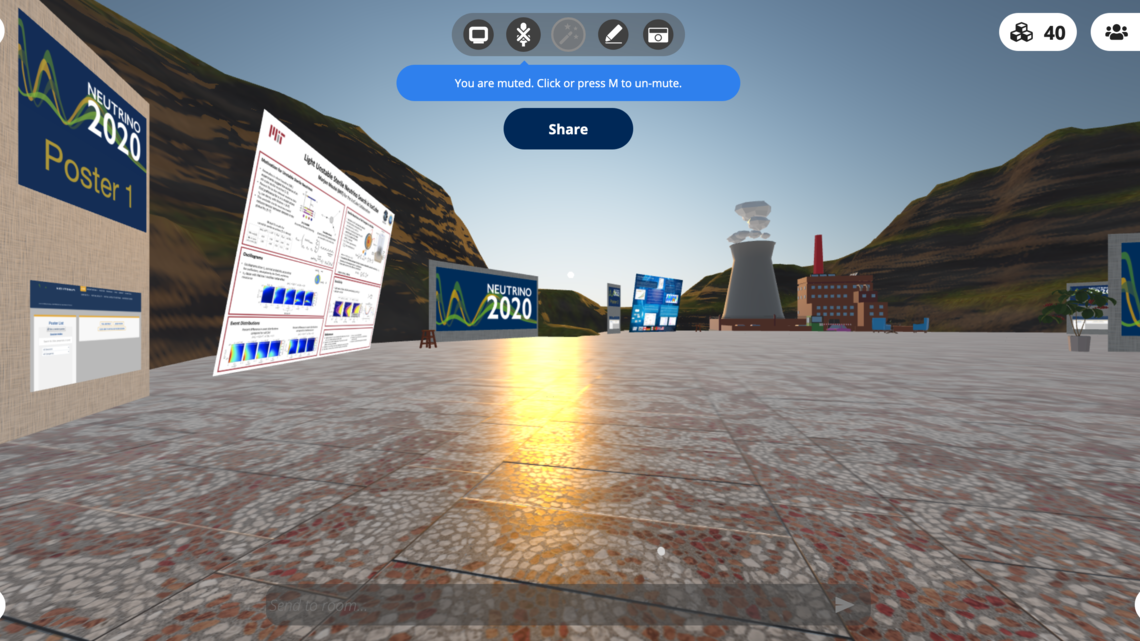}
    \caption{Example view inside the Neutrino 2020 Virtual Reality platform that hosted the 532 neutrino physics posters created by conference participants.}
    \label{fig:nu2020-vr}
\end{figure}

To give conference participants something new to try and to increase personal interaction during the entirely online event, Neutrino 2020 hosted an innovative virtual reality platform~\cite{nu2020-vr} using open source software from Mozilla \gls{mozillahubs}. Conference registrants had access to this virtual reality platform that gave poster presenters an opportunity to share their work and conference participants a chance to interact. The virtual reality platform was available for the duration of the conference, although due to popular demand it was kept live for a few days even after the conference had ended. Using the virtual reality platform, participants had a chance to get immersed in a virtual poster session where they were able to see the posters and interact live with poster presenters during dedicated poster sessions. Visitors were able to create custom avatars and able to move within and between the virtual reality rooms. When a person approached another person's avatar, people could speak and hear each others voices. Participants also had the chance to explore virtual sight-seeing of Fermilab and downtown Chicago as well as visit dedicated virtual social rooms to meet up with colleagues. During the conference, 3,409 conference attendees visited the virtual reality platform, including the 532 early career scientists who presented posters of their research. The virtual reality platform received overwhelmingly positive feedback by those who attended and on social media~\cite{nu2020-vr-twitter}. An example of the space can be seen in Figure~\ref{fig:nu2020-vr}. For more views of the Neutrino 2020 Virtual Reality platform, please also see~\cite{nu2020-press-fnal}.

\subsubsection{What Worked Well}

In the end, the response to the first-ever virtual Neutrino conference was very positive~\cite{nu2020-vr-twitter,nu2020-press-fnal,nu2020-press-cern}. There are several aspects that helped make this large online conference a success:

\begin{itemize}
    \item Dress rehearsals for the local organizing committee were held weekly in the two months leading up to the conference to give conference organizers experience running a \gls{Zoom} webinar and in working together as a team.
    \item Written job descriptions were created for each task needed to run the conference. These assignments included hosting the webinar, speaker assistance, talk recording and posting, session chairing, Q$\&$A handling in the webinar, poster assistance, \gls{Slack} monitoring, virtual reality platform assistance, email list monitoring, social media content posting, and \gls{Zoom} webinar technical support.
    \item An attempt was made to anticipate possible problems in advance. A written trouble-shooting guide was created by the local organizing committee that included a list of possible things that could go wrong during the conference, who would respond, and how one should respond.
    \item At the end of each day during the conference, a regrouping session was held for the local organizing committee so that the group could discuss and solve any issues that arose during the conference that day.
    \item A Code of Conduct~\cite{nu2020-coc} set expectations for behavior during the conference. Conference participants could not register for the conference or receive connection details without agreeing to the Code of Conduct. An online reporting system was also made available during the conference. Reports of violations of the Code of Conduct could be made anonymously.
    \item Conference organizers asked plenary speakers to provide recordings of each of them pronouncing their own names. These audio recordings were made available to session chairs to ensure correct pronunciation of presenters' names.
    \item Pulling together advice from a variety of sources, detailed instructions were provided to speakers including a checklist and tips for giving a talk over \gls{Zoom}. Other conferences may also find these speaker tips helpful~\cite{nu2020-speaker-tips}.
    \item Rehearsal sessions for plenary talk presenters were arranged during the weeks leading up to the conference so that speakers could practice sharing their slides and test their audio and video. Practice sessions were also set up during the first 1/2 hour before the conference on each day of the event in case speakers wanted to do a last minute check of their setup.
    \item Conference organizers set up multiple ways for conference participants to communicate with the organizers during the conference including over email, \gls{Slack}, and in the virtual reality platform so as to be able to ask direct questions of the conference team. A Frequently Asked Questions (FAQ) page~\cite{nu2020-faq} was also posted online for conference participants and updated daily.
\end{itemize}

\subsubsection{Lessons Learned}

Lessons learned from Neutrino 2020 were twofold. First, it would, of course, had been better had the conference organizers had more than 2.5 months to pull together this large scale online event. Unfortunately, due to the timing and uncertainty related to COVID-19, more planning time was not possible. Second, the conference would have benefited from a larger support staff to handle real-time conference registrations, as registration was allowed during the event and many participants expected to be able to attend the conference on the same day as having registered. The latter was an unanticipated large task with upwards of 100 people per day registering for the conference as word got out about the event on social media and in the community.

\subsubsection{Summary}

Neutrino 2020 was hosted as an entirely online event from June 22 to July 2, 2020. This virtual conference was put together by a small team of $\approx15$ people in 2.5 months and included plenary talks, a web portal for posters, a popular virtual reality platform, and an online chat forum via \gls{Slack}. The conference had record attendance: 4\,350 attendees from 67 countries in all 7 continents, $60~\%$ of whom were students and postdocs. This reach included many scientists who might not have otherwise been able to attend an in-person conference because of funding, visas, family responsibilities, or other issues. Given this extensive and diverse participation, some aspects of Neutrino 2020 may well affect the planning and organization of future in-person and online conferences. 

\subsection{Common Themes and Key Findings}
\label{sec:experiences_common}


Several themes from the experience talks and subsequent discussion emerged:
\begin{itemize}
\item Virtual meetings come with several benefits such as increased participation by early-career scientists and international collaborators, smaller overall budgets are required to run an event, decreased cost of attendance and reduced carbon footprint associated with the event (due to a lack of required travel).
\item Roughly 25\%-50\% of registered participants are connected at a given time and this decreases over the event, while the vast majority of registrants attend the event at some point over its duration (see Figures~\ref{fig:cpm_stats} and~\ref{fig:osgahm21}, for example).
\item Challenges include a lack of personal interaction opportunities and diminished quality of those interactions, “\gls{Zoom} fatigue”, attention capture and retention, and synchronous participation over wide range of time zones. 
\item \gls{Zoom} is an adequate current tool for “traditional” presentation-style, but other tools are needed to be able to approach in-person style interactivity (many great ideas shown at the workshop!).
\end{itemize}


\section{Tools and Techniques}
\label{sec:tools_techniques}


For virtual conferences at a minimum we need a way for all participants to hear each other. This could be as simple as a conference call. More subtle communications can occur if the voice call is supplemented with video feeds of the speaker and other participants. This is provided by many of the different platforms. Slides and online demonstrations of software can add additional context and information. Most of the video conferencing systems can allow for someone to share their screen.

In this way, the simple video conferencing systems can provide the basics of what people experience in an in-person conference. There is additional tooling that can be brought to bear that can extend this conference experience and, in some ways, take participants further than has been possible in an in-person environment.

\subsection{Virtual Whiteboards}
Visual communication is a very powerful medium. It is common for workshop presenters to include diagrams on their slides and for organizers to use \glspl{GoogleDoc} for collaborative note taking during the meeting which often also includes diagrams. These diagrams are static and cannot reflect new knowledge gained during the session or the collective wisdom of the gathered participants. 

A useful tool is \gls{Miro}, which is an online whiteboard built for collaboration. It has a number of features that make it easy to share a link to the whiteboard and to manage even a large number of contributors who may wish to view the whiteboard, follow a user around as they discuss or modify the board. Anyone can add items to the board or cleanup diagrams as a discussion proceeds. The \gls{Miro} whiteboard offers infinite space so groups can break up and even work on their own ideas an a separate area of the board. Current alternatives to \gls{Miro} include \gls{Mural} and Google's \gls{Jamboard}.

\subsection{Q and A Management}
In traditional in-person meetings, managing questions from the audience can be difficult and often favor the boldest. For virtual or hybrid meetings this issue is can be harder unless additional tools are brought in, given the need for participants to unmute, and the larger audience sizes. Several tools offer the chance to add more sophistication to the process of submitting and selecting questions. In particular, \gls{Slido}, which is a tool that allows participants to pose questions. The whole audience can review and up-vote questions that interest them. Using this tool makes is easy for people to submit questions even if they are uncomfortable speaking to the assembled group and insures that precious conference time is used for questions that are of general interest.

Live polling is a way to increase participant engagement and expose additional background on the audience as a whole. Results are often presented in interactive ways that evolve in real time as results come in, often taking the form of word clouds or bar and pie charts. While \gls{Zoom} offers this built-in, tools like \gls{Slido} and \gls{Mentimeter} have a wider variety in question types, outcome displays, and interactivity.

Although particularly pertinent for online and hybrid events, there is nothing to stop these tools being used at fully in-person conferences as well.

\subsection{Increased Social Interactions}
As has been described in this report, a common challenge for virtual events over the last year has been giving participants the same opportunity to meet new people and reconnect with old acquaintances. Several tools exist which can help in this regard with many being trialed in the different events described in Section~\ref{sec:experiences}. One such approach is to give participants a common map which can, for example, represent the conference ``venue''. By moving around this map participants can interact with each other based on their proximity. In particular, \gls{Gather.Town} uses a 2D map reminiscent of old 8-bit computer games while \gls{mozillahubs} offers a 3D world that can be used with virtual reality headsets.  Both of these have been used successfully for poster sessions and other more interactive pieces of typical conferences.  
An alternative approach is to offer participants small rooms which allow for a handful of people to join at a time.  Platforms like \gls{Airmeet} and \gls{Remo} allow audience members to choose a table to sit down at and begin discussions, similar to finding a table in a lunch canteen or classroom. An alternative to this is guided networking, where participants are less directly involved in choosing who they meet, such as offered by \gls{RemotelyGreen}.


\section{Diversity, Inclusion and Accessibility Considerations}
\label{sec:DIA}

The shift to fully virtual meetings, conferences, and workshops during the pandemic, and subsequent discussions on the likelihood of being able to maintain such a strong virtual component in the future has highlighted many areas of inequality in our current model of in-person meetings. In this section, we highlight the benefits as well as potential pitfalls in terms of diversity, inclusion, and accessibility that virtual meetings have over in-person ones. Given the lessons learned over the past year it is imperative that we continue to take these considerations into account post-pandemic, and ensure we do not simply return to what we were doing before.

One of the largest impacts of online meetings is in the reduced travel costs for attendees. This directly allows for participation by a larger and more diverse group of people around the world, and especially benefits people in less well-funded groups, newly-established research groups, and those located far away from the meeting location. This also has a strong benefit for early career researchers which often have challenging economic circumstances, given that many institutions require one to cover all the bills in advance and only reimburse after the meeting, sometimes many weeks later. Finally, many groups have a funding limit of one in-person event per year, but researchers would participate in many more events remotely. This is again especially useful for early career researchers (or those who are new to the field).  

On the other hand, network reliability in many parts of the world - perhaps most especially those who would benefit most from the reduced travel costs - may be a hindrance to being able to fully participate in and get the most out of the meeting. In such a situation, a balance between live and pre-recorded talks may be helpful, as well as having the meeting recorded and available to watch at a later time.

Travel itself also has challenges for many people who may benefit from online meetings. Travel time is a luxury that many, such as those with family or other personal commitments, cannot afford. Some individuals may be unable to travel due to disabilities, health limitations, or, especially in near-term post-COVID times, those who have been unable to be vaccinated. Many U.S. National Laboratories have travel caps or bans in place, limiting researchers' travel to in-person meetings and certain countries. And while visa restrictions are not generally a concern for most Northern American or European passport holders, the costs, administration load, and time involved in visa applications is significant for many researchers around the world. Additionally, while all effort should be made to ensure the safety of attendees at all times, meetings may be held in locations that are potentially life-threatening to participants, given their race, sexual or gender orientation, or religion.

A particular feature of online meetings that has accessibility benefits is the relative ease of availability of captioning for presentations. While this does have additional cost implications for the organisers, it should be considered up-front as default rather than an after-thought. As automatic captioning in discussions with many technical terms can be complicated and render inaccuracies, it is highly preferable to accurately caption the content in as close to real-time as possible through professional captioning services or other similarly effective approaches. Captioning benefits not only those with hearing-impairment, but also non-native speakers, and even new students and researchers who may be unfamiliar with all the technical jargon in the field.

Organising social activities for fully online meetings over the past year has required some innovation, but in the cases where it was done successfully has contributed positively to the inclusivity of the event. The social component is a very important part of in-person meetings, but can be alienating if one does not already know a number of people attending. One of the things to think about going into the future will be how to maintain this inclusivity and avoid a fabricated "class divide" in meetings with both online and in-person components.

A Code of Conduct must exist and be agreed upon by all parties in advance to explicitly outline the rules for participation and the values associated with the event. A carefully crafted Code of Conduct can go a long way towards fostering diversity and inclusion by 
establishing acceptable behavior and promoting inclusive thinking. However, having an agreed upon Code of Conduct is not enough -- clearly stated protocols for reporting and handling of violations, including expectations for anonymity, data privacy and repercussions for violators must be associated with any Code of Conduct policy advertisement otherwise it risks being ineffectual. Virtual events bring additional challenges regarding conduct issues as compared with in-person events given the impersonal and sometimes quasi-anonymous nature of remote interactions.

Finally, it is important for event organizers to be proactive about establishing a diverse set of speakers, panelists and participants (and organizers!). There is much more to say on this important topic, especially regarding effective and ineffective approaches to improving diversity in meetings, workshops and conferences (and our research itself!). However, most points are not specific to the virtual event format which is the topic of this report. Still, we would be remiss in not mentioning this critically important aspect of event organization, so we conclude this section with a reminder on this point.  


\section{Best Practices}
\label{sec:best_practices}


With participants spread out around the globe, the most important goal for any virtual meeting is to most efficiently and effectively use the time that is available. \textit{Synchronous time} - where all attendees are available and able to be attentive - is to be considered as \textit{precious}, and careful and thorough planning is necessary to make optimal use of this commodity. Drawing from the experience gained from the conferences and workshops described in Section~\ref{sec:experiences}, we list some non-exhaustive suggestions for virtual meeting organizers and attendees to help with smooth running, and to make the meeting as productive as possible.

\subsection{Timetable Planning}
\begin{itemize}
    \item The most important initial step is to consider the aim and goals of the meeting or conference, as this will dictate the way the timetable is planned. 
    \item For large conferences that aim to show latest results, organizers should carefully consider the fraction of synchronous time dedicated to plenary talks (whether live or pre-recorded), compared to the time spent on questions and discussions, and ensure these are in line with the conference goals.
    \item For workshops where the goal is to have discussions or brainstorming of solutions, the majority of synchronous time should be dedicated to these activities. Any necessary material should be distributed and reviewed beforehand, though highlights or focal points could be summarised at the start of each session.
    \item A well planned agenda that includes links to any tools that may be used during the meeting will go a long way in ensuring smooth running and optimal use of time.
    \item Realistic planning of the timetable is important; meeting chairs should contact speakers in advance to agree on the amount of time allocated for presentations, questions, or any other activities. 
\end{itemize}

\subsection{Before the meeting}
\begin{itemize}
    \item For large conferences, dress rehearsals for the local organizing committee in the weeks leading up to the conference can give the organizers experience running the various tools used in the conference, and in working together as a team.
    \item Written job descriptions should be created for each task needed to run the conference and assigned to individuals. These may include hosting the webinar, speaker assistance, talk recording and posting, session chairing, Q$\&$A handling in the webinar, poster assistance, chat program monitoring, virtual reality platform assistance, email list monitoring, social media content posting, and webinar technical support.
    \item It is helpful to try to anticipate possible problems in advance. A written trouble-shooting guide should be created that includes a list of possible things that could go wrong during the conference, how one should respond, and the person responsible for handling the problem.
    \item A Code of Conduct should be a non-negotiable aspect of every conference, workshop, or meeting, to set expectations for behavior during the event. Where registration is required, participants should not be able to register for the conference or receive connection details without agreeing to the Code of Conduct.
    \item Clearly stated protocols for the reporting and handling of conduct violations should be associated with the Code of Conduct itself. An online reporting system should be made available during the conference. Reports of violations of the Code of Conduct must be able to be made anonymously. Organizers should consider assigning someone outside of the organizing committee to serve as an at-the-ready ombudsperson to investigate complaints and strive for a satisfactory resolution.
    \item Conference organizers should ask all speakers for their preferred pronouns (provided voluntarily, of course), and to provide audio recordings of each of them pronouncing their own names. The organizers should provide the means (a specific service or more likely instructions for a common computing platforms) for speakers to easily produce these audio recordings. These audio recordings should be made available to session chairs to ensure correct pronunciation of presenters' names. Presenter pronouns should always be respected.
    \item Conference organizers should consider asking participants to provide, on a voluntary basis, information beyond the usual registration content to facilitate interactions, such as research interests. For example, I might be interested in knowing if there are other participants interested in machine learning approaches to neutrino event reconstruction. A searchable database of keywords from participants research interests could prove useful to increase interactions. Something similar could be achieved by allowing participants to create topical channels on the event's Slack workspace.
    \item Detailed instructions should be provided to speakers including a checklist and tips for giving a talk over the conference platform of choice. As an example, see the speaker tips provided by Neutrino2020~\cite{nu2020-speaker-tips}.
    \item Rehearsal sessions for plenary talk presenters should be arranged during the weeks leading up to the conference, so that speakers can practice sharing their slides and test their audio and video. Practice sessions could also be set up during the first half hour before the conference on each day of the event, in case speakers want to do a last-minute check of their setup.
    \item Conference organizers can set up multiple ways for conference participants to communicate with the organizers during the conference including over email, collaboration workspaces such as \gls{Slack}, and in a virtual reality platform (if available). Be sure to allocate someone responsible for overseeing each.
    \item A Frequently Asked Questions (FAQ) page (see for example~\cite{nu2020-faq}) should be posted online for conference participants and updated daily.
\end{itemize}

\subsection{During the Meeting}
\begin{itemize}
    \item Staying on time is possibly the biggest challenge for any meeting or conference - virtual or not. Chairs should have the authority to stop a talk, for example by muting the speaker, if they run overtime. They should also be able to show the speaker how much time they have remaining at any time. 
    \item Have a back channel, instant communication mechanism between key organizers during the meeting to troubleshoot on the fly.
    \item At the end of each day during the conference, a regrouping session should be held for the local organizing committee to discuss and solve any issues that arose during the conference that day.
\end{itemize}

\subsection{Discussion sessions}
\begin{itemize}
    \item Chairs should communicate their preferred method for taking questions, whether by ``raising one's hand'', or by comments in the chat panel or a dedicated workspace. Each of these should, of course, be monitored.
    \item Dedicated workspaces are a useful tool for extending discussions after the allocated time is up, or for additional questions there may not have been time for.
\end{itemize}

\subsection{For Presenters}
\begin{itemize}
    \item Speakers, whether live or pre-recorded, must do their best to stick to their allocated time. To run overtime in your presentation shows a lack of preparation at best and can be construed as disrespect for everyone else involved.
    \item Speakers should make sure they attend any rehearsal sessions held for them to familiarise themselves with the conference platform.
\end{itemize}

\subsection{Socialising}
\begin{itemize}
    \item Section~\ref{sec:tools_techniques} lists a number of tools available to improve the social aspect of fully virtual meetings. While this is no substitute for in-person interactions, use of one or more of these tools is encouraged.
    \item In some of the smaller meetings, a nice touch by the organizing committee has been sending out small ``care packages'' which may include things like local snacks, branded merchandise (such as stickers), etc. Whilst this is an additional financial and administration load, it is well received by participants and helps to foster a feeling of community and can also give a little bit of local flair.
\end{itemize}


\section{Looking Forward}
\label{sec:looking_forward}

It has become exceedingly clear over the past year that online meetings have a variety of strong advantages, and as we move to planning post-pandemic meetings we should strive to maintain as many of these advantages as possible. However, the in-person interactions, informal chats and networking have been greatly missed over the past year, and this is perhaps the strongest motivator for researchers to want to return to in-person meetings as soon as it is allowed.

A possible approach to this is to avoid the binary choice between in-person and online, and work toward creating a new, hybrid event format that pulls from the benefits of each. Such an event would host a sufficient number of in-person attendees to be able to cover most of the costs of the conference, whilst also providing the ability to attend online at a reduced cost if the attendee so wishes. The additional logistical requirements for successfully including this virtual component are certainly not negligible, but have already been demonstrated a number of times in events such as the large LHC experiments' collaboration weeks where remote participants are able to attend talks and participate in post-talk discussions via a teleconferencing tool such as \gls{Zoom}. 

In such meetings one has previously been required to be physically present if one is giving a talk, a requirement that will need to be relaxed to maintain a more inclusive approach to the meeting, and to avoid a ``class divide'' that favours those with privileged access to in-person events. Technological challenges such as unreliable internet connection is of course a concern, but one possibility to overcome this has already been seen in virtual conferences over the past year where talks are pre-recorded and uploaded ahead of time.

On its own this version of hybrid event still lacks the social and networking opportunities for the virtual participants, however, and a way to overcome this could be to host local conference hubs at strategic locations around the world, where the virtual attendees could get together for discussions amongst themselves. The hubs could be located at a local university with potentially low overhead costs, and could also help to solve some of the technological challenges for more remote participants. 

To maintain some level of equity between the meeting host location and the hubs, it should be encouraged that a fraction of the speakers broadcast their talks from the hubs themselves. For those with the means to travel, this could be considered an outreach opportunity, or an chance for a researcher to return to and meet with local researchers from their home country.

A large amount of work goes into organising any conference, and a hybrid conference will certainly prove an even bigger challenge to conference organisers than those that are either fully in-person or fully virtual. Additional costs will be involved, to cover the technological and room requirements necessary for a good online event, while the risk of lower in-person attendance also has financial considerations, as many conference budgets are developed based on a minimum number of (in-person) attendees. An initial approach here could be to charge the usual fee for in-person attendees, and a smaller fee (up to US\$50 per person seems to be considered reasonable based on discussions during this workshop) for the online attendees, to cover the additional costs involved with taking the conference online. Local hubs should be discouraged from charging additional fees for attendance at the hub.

Ultimately, conference organisers want their attendees - especially those attending in-person - to have the best experience possible. A particular challenge for the ``hybrid with hubs'' approach is with time zones, trying to find time slots for the talks and discussions that work for most attendees worldwide. Whilst this is impossible to solve completely, many conferences over the past year have already done a good job at managing their schedule for maximum participation across the globe, providing session recordings for others to be able to catch up on at their convenience, or with pre-recorded talks available shortly before the conference combined with a live discussion session between the speakers during the event. In most of these cases, the actual time spent in live talks has been reduced to half a day instead of a full day, to account for the variation in time zones. This should be considered a positive development, with the other half of the day available for discussions, poster sessions, local outreach activities, or local tours. These supplementary events could be organised in the local hubs as well as the main conference venue, providing part of the social aspect of the conference that virtual attendees would otherwise be missing. Finally, while the conference organisers should be encouraged to make the timetable as accessible as possible for all, in the ``hybrid with hubs'' approach perhaps a good incentive to attend in-person would be to be able to attend the live talks in the most convenient timezone.

For workshops and conferences that are organized through funded projects or projects yet to be proposed, creative thinking about how to make better use of funds nominally slated for participant costs and project personnel travel to create and curate workshop content would be prudent in the post-pandemic "new normal". For example, these funds could be used to improve accessibility to event content through professional captioning or to pay for services that can produce high-quality training materials to provide more in-depth understanding and amplify the impact after completion of the event.

It would be really encouraging if within the next two years one or more major conferences takes a lead on trialling the ``hybrid with hubs'' approach and implements many of the suggestions outlined in this report.  

Given what our community has experienced with virtual meetings over the last year due to this terrible pandemic, it is hard to imagine the HEP community returning to a pre-pandemic approach of "fully" in-person meetings. Through this ongoing journey, we have learned the value of a substantial virtual component to meetings organization as well as many associated challenges. This is a genie that will not easily go back into the bottle. New approaches such as the "hybrid with hubs" approach explored in our workshop and described in this report as well as technical innovations that continue to blur the lines of experience between virtual presence and physical presence will shape the future of meetings in HEP and beyond. 


\section*{Acknowledgments}

We thank the attendees for their active participation in the workshop to discuss these issues and suggest paths forward. This workshop was partially supported through the U.S. National Science Foundation (NSF) under Cooperative Agreement OAC-1836650 (IRIS-HEP).


\section*{Disclaimer}

Certain commercial software platforms or services are identified in this paper in order to specify the conference formats adequately. This does not imply a recommendation or endorsement by IRIS-HEP, the individual conference organizers or sponsors, the National Institute of Standards and Technology or the National Science Foundation that the software platforms or services identified are necessarily the best available for the purpose.

\newpage
\printglossaries

\newpage
\bibliography{main}

\end{document}